\documentclass[11pt, onecolumn]{IEEEtran}
\IEEEoverridecommandlockouts \usepackage{hyperref}
\usepackage{makecell}
\usepackage{bbm}
\usepackage{graphicx}
\usepackage{subfigure}
\usepackage[usenames,dvipsnames]{color}
\usepackage{epsfig}
\usepackage{amssymb}
\usepackage{amsmath}
\usepackage{amsthm}
\usepackage{latexsym}
\usepackage{setspace}
\usepackage{bbm}
\usepackage{flushend}
\usepackage[top=1in, bottom=1in, left=1in, right=1in]{geometry}
\usepackage{float}
\usepackage{dsfont}
\usepackage{caption}
\usepackage[ruled,vlined]{algorithm2e}
\usepackage{xspace}
\usepackage[normalem]{ulem}

\usepackage{stackrel}

\newcommand{\htc}[1]{{ #1 }}

\newcommand{\dP}{\mathrm{P}}
\newcommand{\dQ}{\mathrm{Q}}
\newcommand{\bP}[2]{\mathrm{P}_{#1}\left(#2\right)}
\newcommand{\bQ}[2]{\mathrm{Q}_{#1}\left(#2\right)}
\newcommand{\bPP}[1]{\mathrm{P}_{#1}}
\newcommand{\bQQ}[1]{\mathrm{Q}_{#1}}
\newcommand{\bPr}[1]{{\mathrm{Pr}}\left(#1\right)}

\newcommand{\bE}[2]{{\mathbb{E}}_{#1}\left\{{#2}\right\}}
\newcommand{\bEE}[1]{{\mathbb{E}}\left[#1\right]}

\newcommand{\cA}{{\mathcal A}}
\newcommand{\cB}{{\mathcal B}}
\newcommand{\cC}{{\mathcal C}}
\newcommand{\cD}{{\mathcal D}}

\newcommand{\cF}{{\mathcal F}}
\newcommand{\cG}{{\mathcal G}}
\newcommand{\cH}{{\mathcal H}}

\newcommand{\cK}{{\mathcal K}}

\newcommand{\mN}{{\mathbbm N}}
\newcommand{\mR}{{\mathbbm R}}
\newcommand{\cO}{{\mathcal O}}
\newcommand{\cP}{{\mathcal P}}

\newcommand{\cR}{{\mathcal R}}
\newcommand{\cS}{{\mathcal S}}
\newcommand{\cT}{{\mathcal T}}
\newcommand{\cU}{{\mathcal U}}
\newcommand{\cV}{{\mathcal V}}

\newcommand{\cX}{{\mathcal X}}
\newcommand{\cY}{{\mathcal Y}}
\newcommand{\cZ}{{\mathcal Z}}

\makeatletter
\newtheorem*{rep@theorem}{\rep@title}
\newcommand{\newreptheorem}[2]{%
\newenvironment{rep#1}[1]{%
 \def\rep@title{#2 \ref{##1}}%
 \begin{rep@theorem}}%
 {\end{rep@theorem}}}
\makeatother

\newtheorem{theorem}{Theorem}[section]

\newtheorem{proposition}[theorem]{Proposition}
\newtheorem{corollary}[theorem]{Corollary}
\newtheorem*{corollary*}{Corollary}

\newtheorem{lemma}[theorem]{Lemma}
\newtheorem*{lemma*}{Lemma}

\newreptheorem{theorem}{Theorem}
\newreptheorem{lemma}{Lemma}

\theoremstyle{remark}
\newtheorem{remark}{Remark}
\newtheorem*{remark*}{Remark}
\newtheorem*{remarks*}{Remarks}

\theoremstyle{definition}
\newtheorem{definition}{Definition}[section]
\newtheorem{example}{Example}[section]

\newcommand{\ed}{\mathbin{:=}}

\def\undertilde#1{\mathord{\vtop{\ialign{##\crcr
$\hfil\displaystyle{#1}\hfil$\crcr\noalign{\kern1.5pt\nointerlineskip}
$\hfil\tilde{}\hfil$\crcr\noalign{\kern1.5pt}}}}}

\newcommand{\ep}{\varepsilon}

\newcommand{\ttlvrn}[2]{d\left( #1 , #2\right)}

\newcommand{\ICe}{{\tt IC}_{\tt e}}
\newcommand{\ICi}{{\tt IC}_{\tt i}}
\newcommand{\varp}{\mathtt{V}(\pi| \bPP{X_1, X_2})}
\newcommand{\ic}{\mathtt{ic}}
\newcommand{\icp}{\ic(\Pi; X_1,X_2)}

\newcommand{\indicator}{{\mathds{1}}}
\newcommand{\mc}{-\!\!\!\!\circ\!\!\!\!-}

\allowdisplaybreaks
\newcommand{\order}{o}

\newcommand{\Z}{\mathbb{Z}}

\begin{document}
\title{Communication for Generating Correlation: \\ A Unifying Survey}

\author{\IEEEauthorblockN{Madhu Sudan\,\,\,} \and
  \IEEEauthorblockN{Himanshu Tyagi\,\,\,} \and \IEEEauthorblockN{Shun
    Watanabe} }

\maketitle 

{\renewcommand{\thefootnote}{}\footnotetext{
%\hspace*{-.11in}\rule{24ex}{.05em}
M. Sudan is with the Harvard John A. Paulson School of Engineering and
Applied Sciences, 33 Oxford Street, Cambridge, Massachusetts 02138,
USA.  (email: madhu@cs.harvard.edu).

H. Tyagi is with the Department of Electrical Communication Engineering,
Indian Institute of Science, Bangalore 560012, India.  (email: htyagi@iisc.ac.in).

S. Watanabe is with the Department of Computer and Information Sciences,
Tokyo University of Agriculture and Technology, Tokyo 184-8588, Japan.
(email: shunwata@cc.tuat.ac.jp). }}

%\maketitle

\renewcommand{\thefootnote}{\arabic{footnote}}
\setcounter{footnote}{0}

\newcommand{\textchangeMS}[1]{{#1}}
\newcommand{\textchangeHT}[1]{{#1}} %for the submission
\newcommand{\textchangeSW}[1]{{#1}}
\newcommand{\textchangeSWtwo}[1]{{#1}}
%\newcommand{\textchange}[1]{#1}

%%%%%%%%%%%%%%%%%%%%%%%%%%%%%%%%%%%%%%%%%%%%%%%
\begin{abstract}
The task of manipulating correlated random variables in a distributed
setting has received attention in the fields of both Information
Theory and Computer Science. 
Often shared correlations can be converted, using a little amount of communication, into
perfectly shared uniform random variables. Such perfect shared randomness, in turn, enables the solutions of many tasks. Even the reverse conversion of perfectly shared uniform
randomness into variables with a desired form of correlation 
turns out to be  insightful and technically useful. In this article, we
describe progress-to-date on such problems and lay out pertinent
measures, achievability results, limits of performance, and point to new directions.
\end{abstract}
\newpage
%%%%%%%%%%%%%%%%%%%%%%%%%%%%%%%%%%%%%%%%%%%%%%
\tableofcontents
%%%%%%%%%%%%%%%%%%%%%%%%%%%%%%%%%%%%%%%%%%%%%%
\section{Introduction}
The ability to harness and work with randomness has been at the heart
of information theory and modern computer science.  Randomness has
been used to model the ``unknown'' in information theory, be it a
message produced by a source or the error introduced by a channel. In
computer science, randomness is a resource that enables simpler,
faster, and sometimes the only solution to many central problems. Of
particular interest in this article is the use of randomness by a set
of parties that are spread out geographically. Randomness enables such
parties to \htc{generate secret keys and transmit messages
  securely~\cite{Sha49}. It allows them to complete distributed computation tasks
  such as comparison of two strings using very few bits of
  communication~\cite{CarWeg79,Yao79}. In  Shannon theory, shared
  randomness is necessary to attain positive-rate
  channel codes for the arbitrary varying channel \cite{BlackwellBT60,Ahlswede78}.  
Beyond these small sampling of applications which are directly related 
to the problems we shall consider in this article, there are many more
applications of shared randomness, including synchronization, leader
election, and consensus which are known to have no deterministic
solutions for most multiparty settings of interest.}

In many applications, it suffices to have a
weak form of shared randomness instead of perfect uniform
shared randomness. This leads to a quest for understanding weak
forms of randomness and its limitations. A classic example of such a
quest goes back to 
von Neumann (cf.~\cite{Neumann51}) who asked whether one could
simulate an unbiased coin (uniform distribution over $\{0,1\}$) using
a coin of unknown bias.  His solution involves a sequence of tossing of
pairs of coins till they show different outcomes. The first
coin of this final pair is tantamount to a perfectly unbiased coin. If the biased
coin has expected value $p$, then his solution tosses $1/(p(1-p))$
coins in expectation to get one unbiased coin. Improving the rate of
usage of the biased coin and extending solutions to more general
settings of imperfection, including correlations among coins, has lead to
a rich theory of randomness extractors
(cf.~\cite{SanthaVazirani86,BenBraRob88,ImpLevLub89, ImpZuc89, HastadILL99}).

\htc{In this paper, we shall be concerned with a different notion of
imperfectness in randomness, namely that arising from distributed
nature of problems. Many of the applications of randomness we listed
earlier rely on, or can be interpreted as, a conversion of one form
of such imperfect randomness into another, sometimes using
communication between the parties. Focusing on two-party scenarios,
we review research of this nature in this article and  
 describe some of the unifying themes. We have divided these problems into four broad
categories. We list these categories below, mention the sections where
they appear, and highlight some interesting results in each category. 
}
%% To set the context, we point out that several of the solutions to
%% challenges in distributed settings 
%% involve parties sharing identical copies of some random variable. For
%% instance when parties share copies of a uniform random string, it
%% enables information-theoretically secure exchange of secrets. What
%% happens to such solutions when the parties share only some correlated
%% variables? Then, is there an analog of the von Neumann solution here that
%% allows the parties to transform their random variables into identical
%% ones with possibly less entropy? Does this require communication? If
%% so, how much? Questions of this nature were raised in the seminal
%% works of Maurer~\cite{Mau93} and Ahlswede and
%% Csisz\'ar~\cite{AhlCsi93, AhlCsi98} and continue to be the subject of
%% active investigation.

%% We also shall be interested in questions of a ``reverse'' nature,
%% where the parties do have access to identical copies of a random
%% variable in form on shared randomness, but now seek is to generate variables that have a 
%% specified correlation. Interestingly this direction turns out to be of
%% technical interest as well, and protocols that achieve it have played
%% a crucial role in several developments in computer science.

\paragraph{Generating common randomness using correlated observations
  and interactive communication (Section~\ref{s:CR})} The common
  randomness generation problem entails generating shared (almost)
  uniformly distributed bits at the parties, using initial correlated
  observations and interactive communication between the parties.  
When parties share copies of a uniform random string, they can 
accomplish several distributed tasks such as information-theoretically
secure exchange of secrets. The study of common randomness generation
is motivated by questions such as: what happens to such solutions when
the parties share only some correlated variables? Is there an
analog of the von Neumann solution that allows the parties to
transform their random variables into identical ones with possibly less entropy? Does this require communication? If so, how much? Questions of this nature were raised in the seminal works of Maurer~\cite{Mau93} and Ahlswede and Csisz\'ar~\cite{AhlCsi93, AhlCsi98} and continue to be the subject of active investigation.

A result of G\'acs and  K\"orner~\cite{GacKor73} says that unless the
parties shared bits to begin with, the number of bits of common
randomness they can generate per observed independent sample without
communicating goes to $0$. In fact, Witsenhausen~\cite{Wit75} showed
that the parties cannot even agree on a single bit without communicating. However, we shall see that by communicating the parties agree on more bits than they communicate \textchangeSW{\cite{AhlCsi98}}. These extra bits can be extracted as a secret key that is independent of the communication. 

\paragraph{Generating secure common randomness, namely the problem of 
  secret key agreement (Section~\ref{s:SK_agreement})} The secret
key agreement problem is closely related to common randomness
generation and imposes an additional security requirement on the
generated common randomness. Specifically, it requires that the
generated common randomness be almost independent of the communication
used to generate it. Such a common randomness will constitute a secret
key that is information theoretically secure and can be used for
cryptographic tasks such as secure message transmission and message
authentication. The main result in this section says, roughly, that
the rate of secret key that can be generated is given by the rate of
common randomness minus the rate of communication. In fact, the two
problems are intertwined and a complete characterization 
of communication-common randomness rate tradeoff will lead to a complete
characterization of communication-secret key rate tradeoff.
 
\paragraph{Generating samples from a joint distribution without communicating (Section~\ref{s:sim_no_comm})}
In the next class of problems we consider,
two parties observe correlated samples from a distribution and seek
to generate samples from another. We consider the basic problems of
{\em approximation of output statistics}, where the goal is to
generate samples from a fixed distribution at the output of the
channel by using a uniformly distributed input, and {\em Wyner common
  information}, where the goal is to generate samples from a fixed
joint distribution using as few bits of shared randomness as
possible. Another important problem in this class is that of {\em
  correlated sampling} where the knowledge of the joint distribution
is not completely available to any single party.

From the many interesting results covered in this section, we
highlight the 
following to pique the reader's interest.  A well-known result in
probability theory and optimal transport  theory states that given two distributions $\dP$ and $\dQ$, one can find a joint distribution $\bPP{XY}$ such that $\bPP{X} = \dP$,
  $\bPP{Y}=\dQ$, and $\bPr{X\neq Y} = \ttlvrn{\dP}{\dQ}$, where $\ttlvrn{\cdot}{\cdot}$ denotes variational distance. In fact, this is the least probability of disagreement possible for any such joint
  distribution $\bPP{XY}$, \textchangeSWtwo{which is known as the optimal coupling.} 
  We shall see that even when the knowledge of $\dP$ and $\dQ$ is only local, namely Party 1  knows $\dP$ and Party 2 knows $\dQ$, the same probability of
  disagreement \textchangeSWtwo{as the optimal coupling} can be attained up to a factor of $2$.

\paragraph{Generating samples from a distribution using
  communication (Section~\ref{s:sim_comm})} The final class of
problems we consider is similar to the previous one, except that now
the parties are allowed to communicate. Specific instances include the {\em reverse Shannon theorem} and 
{\em interactive channel simulation}, 
where the parties seek to simulate a given conditional distribution using as
few bits of shared randomness and (noiseless) communication as possible;
and {\em simulation of interactive protocols}, where the parties seek to
simulate the distribution of transcripts of a given interactive
protocol using minimum communication. 

Many of these results have
driven recent advances in communication complexity and even quantum
information theory, and are also of independent interest. 
In particular, the reverse Shannon theorem says that, when the parties have access to shared randomness, they can simulate a channel by communicating at rate roughly equal to the mutual
  information between the input and the output, thereby establishing a
  ``reverse'' of Shannon's classic channel capacity theorem. The
  interactive channel simulation problem is an abstraction that
  includes as a special case almost all problems we cover in this
  article. Thus, the reader might temper expectations for very
  general results for this problem. 
  \htc{Note that the simulation problem is related closely to
    the (data) compression problem, which is 
  well-studied in information theory. But there are some distinctions.
  While the
  latter necessitates obtaining an estimate for a given realization of
  a random variable, the former merely requires producing a copy
  of a random variable with a prescribed distribution.} As a
  consequence, simulation of noisy channels typically requires less
  communication  than compression; in simulation a part of
  communication can be realized from the shared randomness.

In addition to these topics, in Section~\ref{s:preliminaries} we review the basic tools from probability
and randomness extraction that will be used throughout. Also, in
Section~\ref{s:app} we discuss some nonstandard applications of correlated
sampling in approximate nearest neighbor search (locality sensitive
hashing) and in showing hardness of approximation (the parallel repetition
theorem). We conclude with pointers to extensions involving multiple parties
and quantum correlation.

Many of the topics we cover are already the subjects of excellent review articles and
monographs. See, for instance, ~\cite{Vadhan12} for a
review of information theoretic randomness extraction and its
extension to the computational setting; \htc{~\cite{CsiKor11} for a
  chapter on information theoretic secret key agreement and wiretap
  channel;} ~\cite{TyagiNarayan16} for common randomness and 
secret key generation by multiple parties; and the online
manuscript~\cite{RaoYehudayoff} for simulation of protocols and its
application to communication complexity. Our goal here is to present
unifying themes underlying these diverse topics, with the hope of
providing a treatment that is appealing to the information theorist as
well as the computer 
scientist. To that end, we have reworked the presentation of some of
the original proofs to bring out connections between various
formulations.

{\it Notation.} All random variables are denoted by capital letters
$X$, $Y$, etc., their realizations by $x$, $y$, etc., and their range
sets by the corresponding calligraphic letters $\cX$, $\cY$, etc.. The
probability distribution of random variable $X$ is denoted by $\bPP X$. 
The {\em variational distance} between $\dP$ and $\dQ$, denoted
$\ttlvrn{\dP}{\dQ}$, is given by $\sup_{A\subseteq \cX} \dP(A) - \dQ(A)$, which
equals $1/2 \sum_x |\dP(x)- \dQ(x)|$ for discrete distributions. The
{\em Kullback-Leibler} (KL) divergence $D(\dP\|\dQ)$ for discrete
distributions $\dP$ and $\dQ$ equals $\sum_x \dP(x)\log
(\dP(x)/\dQ(x))$ when ${\tt supp}(\dP) \subseteq {\tt supp}(\dQ)$, and
infinity otherwise. For random variables $X$ and $Y$, $I(X\wedge Y)=
D(\bPP{XY}\| \bPP{X}\times \bPP{Y})$ denotes the mutual information
between $X$ and $Y$; we write $I(X\wedge (Y,Z))$ as $I(X\wedge Y,
Z)$. The conditional mutual information $I(X\wedge Y|Z)$ denotes the
conditional KL divergence $D(\bPP{XY|Z}\| \bPP{X|Z}\times
\bPP{Y|Z}|\bPP{Z})=\bE{Z}{D(\bPP{XY|Z}\| \bPP{X|Z}\times \bPP{Y|Z})}$
and equals $I(X\wedge Y,Z) - I(X\wedge Z)$ (see~\cite{CsiKor11} for
further elaboration on our notation).  Instead of instrumenting a
consistent notation for the varied problems we consider, we abuse the
notation $C$ and use it for expressing the fundamental limits in
different contexts. The exact meaning will be clear from the context
and the sub- and super-scripts used. \htc{Throughout, we shall denote
  asymptotic optimal quantities by $C$ and single-shot optimal
  quantities by $L$}.

\section{Preliminaries}\label{s:preliminaries}
In this section, we review some basic results and definitions that
will be used throughout. Specifically, we review 
the leftover hash lemma, the maximal coupling lemma, and the basic
measures of correlation such as maximal correlation and
hypercontractivity. Further, we give a definition of interactive
communication protocols with public coins and private coins, which
will be used throughout. Finally, we provide a brief description of 
some informal terms that are common in information theory, but may not
be familiar to a general reader. 
The presentation is brisk and introductory,
and can be skipped if the reader is aware of these basic notions.
%%%%%%%%%%%%%%%%%%%%%%%%
\subsection{Leftover hash and maximal coupling}\label{s:lhl}
We review two basic tools that underlie several proofs in this
area. The leftover hash lemma allows us to extract uniform randomness
that is a function of a given random variable $X$ and is almost
independent of another random variable $Z$ correlated with $X$.
\textchangeSW{Heuristically, the length of extractable uniform
  randomness is characterized by a measure of ``leftover randomness''
in $X$ given $Z$, such as
  the conditional entropy $H(X|Z)$.  Even though this leftover randomness can be characterized using conditional entropy in the asymptotic setting, it turns out
  that a more relevant quantity in the non-asymptotic setting is the
  conditional min-entropy, to be defined below.}  The second result, termed the maximal coupling 
lemma, is classic in probability theory as well as
analysis. Specifically, given two distributions $\dP$ and $\dQ$ on the
same alphabet $\cX$, the maximal coupling lemma yields a joint
distribution $\bPP{XY}$ with marginals $\dP$ and $\dQ$ such that
$\bPr{X\neq Y}= \ttlvrn{\dP}{\dQ}$ (which is the least possible value
of $\bPr{X\neq Y}$).

Traditionally, achievability proofs in information theory that use
random binning arguments involved a randomly selected mapping
from the set of all mappings with a given finite-size range. In fact,
many of those proofs can be completed using a more economical
construction \htc{that uses
  randomization over families of mappings 
  with much smaller cardinality, termed a {\em $2$-universal hash family}, than the set of all mappings. This
  construction} arose in the computer science literature in
\cite{CarWeg79} and has gained popularity in information theory
over the last decade. The leftover hash lemma uses $2$-universal hash
families as well. We review their definition below. 
%is essential to the treatment of leftover hash lemma.
\begin{definition}[$2$-Universal hash family]\label{d:2UHF}
A class of functions $\cF$ from $\cX$ to $\{0,1\}^l$ is called
a $2$-{\it universal hash family} ($2$-UHF) if $\bP{}{F(x) =
  F(x^\prime)} \le 2^{-l}$ for every $x\neq x^\prime \in \cX$, where $F$
is distributed uniformly over the family $\cF$.
%\footnote{The uniformity of $\bPP{F}$ is not necessary.}
\end{definition}
Also, given random variables $(X,Z)$, we need a notion of residual
randomness that will play a role in the leftover hash
lemma and, at a high level, will constitute a single-shot variant
of the conditional entropy $H(X|Z)$.
\begin{definition}[Min-entropy and conditional min-entropy]
The {\it min-entropy} of $\dP$ is defined as
\begin{align*}
H_{\min}(\dP) \ed \min_{x \in {\cal X}} \log \frac{1}{\dP(x)}.
\end{align*}
For distributions $\bPP{XZ}$ and $\bQQ{Z}$, the {\em conditional
  min-entropy} of $\bPP{XZ}$ given $\bQQ{Z}$ is defined as
\begin{align*}
H_{\min}(\bPP{XZ} | \bQQ{Z}) \ed \min_{x \in {\cal X},z \in
  \mathtt{supp}(\bQQ{Z})} \log \frac{\bQ{Z}{z}}{\bP{XZ}{x,z}}.
\end{align*}
Then, the conditional min-entropy of $\bPP{XZ}$ given $Z$ is defined
as
\begin{align}
H_{\min}(\bPP{XZ}|Z) \ed \max_{\bQQ{Z}} H_{\min}(\bPP{XZ}|\bQQ{Z}),
\end{align}
\textchangeSW{where the maximization is taken over $\bQQ{Z}$
  satisfying $\mathtt{supp}(\bPP{Z}) \subseteq
  \mathtt{supp}(\bQQ{Z})$.\footnote{\textchangeSW{In fact, the maximum
      is attained by $\bQ{Z}{z} \propto \bP{Z}{z} \max_x
      \bP{X|Z}{x|z}$ \cite{KonigRS09, IwamotoS13}.}}}
\end{definition}
While simple to define and operationally relevant
(see~\cite{KonigRS09}), the conditional min-entropy defined above is
not easily amenable to theoretical analysis.
%usually difficult to handle. 
It is more convenient to use its ``smooth'' variant defined next.
\begin{definition}[Smooth conditional min-entropy]
For distributions $\bPP{XZ}$ and $\bQQ{Z}$, and smoothing parameter $0
\le \ep <1$, the smooth conditional min-entropy of $\bPP{XZ}$ given
$\bQQ{Z}$ is defined as
\begin{align}
H^\ep_{\min}(\bPP{XZ} | \bQQ{Z}) \ed \max_{\tilde{\mathrm{P}}_{XZ} \in
  {\cal B}_\ep(\bPP{XZ})} H_{\min}(\tilde{\mathrm{P}}_{XZ} | \bQQ{Z}),
\label{e:smoothing}
\end{align}
where
\begin{align*}
{\cal B}_\ep(\bPP{XZ}) \ed \big\{ \tilde{\mathrm{P}}_{XZ} \in \cP_{\tt
  sub}(\cX\times \cZ): d(\tilde{\mathrm{P}}_{XZ}, \bPP{XZ}) \le \ep
\big\},
\end{align*}
and $\cP_{\tt sub}(\cX\times \cZ)$ is the set of subnormalized
distributions on $\cX\times \cZ$, namely all nonnegative $\tilde{\mathrm{P}}_{XZ}$ such that $\sum_{x,z} \tilde{\mathrm{P}}_{XZ}(x,z)\leq 1$. The smooth
conditional min-entropy of $\bPP{XZ}$ given $Z$ is defined as
\begin{align}
H^\ep_{\min}(\bPP{XZ}|Z) \ed \max_{\bQQ{Z}}
H^\ep_{\min}(\bPP{XZ}|\bQQ{Z}),  \nonumber
\end{align}
where the maximization is taken over $\bQQ{Z}$ satisfying $\mathtt{supp}(\bPP{Z}) \subseteq
  \mathtt{supp}(\bQQ{Z})$.
\end{definition}
Note that the maximum in \eqref{e:smoothing} is taken over all
subnormalized distributions, instead of just all distributions. This
is simply for technical convenience; often, a smoothing over all
distributions will suffice, but handling it requires more work.

An early variant of leftover hash lemma appeared in
\cite{BenBraRob88}. A version of the lemma closer to that stated
below appeared in \cite{ImpLevLub89};\footnote{For variants of
  leftover hash lemma using other notions of leakage,
  see~\cite{BenBraCreMau95, Hay11}.}  the appellation ``leftover
hash'' was given in \cite{ImpZuc89}, \htc{perhaps motivated by
  the heuristic interpreation that the lemma provides a hash of
  ``leftover randomness'' in $X$ that is independent of side
  information.}  The form we present below, which is an extension
of that in \cite{HastadILL99}, is from \cite{HayTyaWat14ii} and can
be proved using the treatment in
\cite{Ren05}.\footnote{\textchangeSW{In the course of proving the
    leftover hash lemma with min-entropy, we can derive a leftover
    hash lemma with collision entropy \cite{Ren05}; it is known that
    the version with collision entropy provides tighter bound than the
    one with min-entropy \cite{Hay16, YanSchPoo17}.  For another
    variant of leftover hash lemma with collision entropy, see
    \cite{FehBer14}.}} This form involves additional side-information
$V$ which takes values in a set $\cV$ of finite cardinality $|\cV|$.

\begin{theorem}[Leftover Hash Lemma] \label{t:LHL}
\textchangeSW{Let $K=F(X)$ be a
  key of length $l$ generated by a mapping $F$ chosen uniformly at
  random from a $2$-UHF $\cF$ and independently of $(X,Z,V)$.  Then,
  it holds that}
\begin{align}
\lefteqn{ \ttlvrn{\bPP{K Z VF}}{ \mathrm{P}_{\cK}^{\mathtt{unif}} \times
  \bPP{ZV} \times \bPP{F}} } \nonumber \\
%% &\le
%% 2\ep + \frac{1}{2} \sqrt{2^{l-
%%     H^\ep_{2}(\bPP{XZ}|Z)}}. 
%%  \label{eq:smooth-conditional-LHL-H2}  
%% \\
&\leq 2\ep + \frac{1}{2} \sqrt{2^{l- H^\ep_{\min}(\bPP{XZ}|Z) + \log
    |\cV|}},
\label{eq:smooth-conditional-LHL-Hmin}  
\nonumber
\end{align}
\textchangeSW{where $\mathrm{P}_{\cK}^{\mathtt{unif}}$ is the uniform
  distribution on the range $\cK$ of $F$.}
\end{theorem}
\htc{Noting that $\ttlvrn{\bPP{K Z VF}}{
    \mathrm{P}_{\cK}^{\mathtt{unif}}\times \bPP{ZV} \times \bPP{F}}$
  equals
  $\bE{F}{\ttlvrn{\bPP{F(X)ZV}}{\mathrm{P}_{\cK}^{\mathtt{unif}}\times
      \bPP{ZV}}}$, for a given source $\bPP{XZV}$ we can derandomize
  the left-side and obtain a fixed mapping $f$ in $\cF$ such that}
\[
\ttlvrn{\bPP{f(X)ZV}}{\mathrm{P}_{\cK}^{\mathtt{unif}}\times \bPP{ZV}}
\leq 2\ep + \frac{1}{2} \sqrt{2^{l- H^\ep_{\min}(\bPP{XZ}|Z) +
    \log|\cV|}}.
     \]
\textchangeSWtwo{As a special case, we can find a fixed mapping $f$ for iid distribution
  $\bPP{X^nZ^n}$ and $V$ given by a fixed mapping $g(X^n, Z^n)$. As a consequence, in our applications of leftover hash lemma below to
  common randomness generation and secret key agreement, we can attain
  the optimal rate using deterministic mappings.}
  
\textchangeSWtwo{In fact, we need not fix the distribution and the same
  derandomization can be extended to the case when the distribution
  $\bPP{XZV}$ comes from a family $\cP$ that is not too large. 
  Interestingly, deterministic extractors even with one bit output
  do not exist for the broader class of sources with bounded
  min-entropy, i.e., $\cP = \{ \bPP{X} : H_{\min}(\bPP{X}) \ge k\}$ for some threshold $k$,
  but one-bit deterministic extractors are possible for sources comprising two independent
  components each with min-entropy greater than a
  threshold~\cite{ChoGol88}.  A recent breakthrough result in this
  direction shows that we can find a deterministic extractor as long
  as the sum of min-entropies of two components is logarithmic in the
  input length (in bits)~\cite{ChaZuc16}. Review of this exciting
  topic is beyond the scope of this review article.  }

%\textchange
In a typical application, the random variable $Z$ represents the
initial observation of an eavesdropper while the random variable $V$
represents an additional message revealed during the execution of a
protocol. \textchangeSW{The result above roughly says that a secret
  key of length
  \begin{align*}
  l \simeq H^\ep_{\min}(\bPP{XZ}|Z) - \log |\cV|
  \end{align*}
  can be generated securely. The additional message $V$ could be
  included in the conditional side of the smooth min-entropy along
  with $Z$. But, the form above is more convenient since it does
  not depend on how $V$ is correlated to $(X,Z)$; an additional
  message of length $m$ decreases the key length by at most $m$ bit.}
  
We remark that the smooth version is much easier to apply than the standard version
with $\ep=0$. Also, the proof of the smooth version is almost 
the same as that of the standard version and only applies triangle
inequality in the first step additionally.  
 
%[Comment (SW): leftover hashing with additional message is needed!]

Next, we state the maximum coupling lemma, which in a more general
form was shown by Strassen in \cite{Str65} (see, also, \cite[Lemma
  11.3]{mitzenmacher:book}).
\begin{definition}
Given two probability measures $\dP$ and $\dQ$ on the same 
alphabet $\cX$, a coupling of $\dP$ and $\dQ$ is a pair of random
variables $(X,Y)$ (or their joint distribution $\bPP{XY}$) taking
values in $\cX \times \cX$ such that the marginal of $X$ is $\dP$ and
of $Y$ is $\dQ$. The set of all couplings of $\dP$ and $\dQ$ is
denoted by $\cP(\dP, \dQ)$.
\end{definition}

\begin{lemma}[Maximal coupling lemma]\label{l:MCL}
Given two probability measures $\dP$ and $\dQ$ on the same alphabet, for every $(X,Y) \in \cP(\dP, \dQ)$,
\[
\bPr{X\neq Y} \geq \ttlvrn{\dP}{\dQ}.
\]
Furthermore, there exists a coupling which attains equality.
(The equality-attaining coupling in the bound above is called a
{\it maximal coupling}.)
\end{lemma}
%%%%%%%%%%%%%%%%%%%%%%%%
\subsection{Maximal correlation and hypercontractivity}\label{s:maxcorr_hypcont}
The notion of ``correlation'' lies at the heart of the topic of this
paper. Various measures of correlation will be applied in presenting
the results as well as in their proofs. One such measure is 
mutual information, which appears most prominently in our
treatment. In this section, we review two other measures of
correlation that are standard but perhaps are not known as widely.

The first measure captures roughly the maximum linear correlation that
can be extracted from $X$ and $Y$. Specifically, given $\bPP{XY}$, the
{\it maximal correlation} $\rho_m(X,Y)$ between $X$ and $Y$ is defined
as~\cite{Renyi59} (see, also, ~\cite{Hirschfeld35, Gebelein41})
\begin{align*}
\rho_m(X,Y) = \max_{\substack{f,g\,:\, \bEE{f(X)} = \bEE{g(Y)}=0
    \\\qquad\,\bEE{f^2(X)} =\bEE{g^2(Y)} =1}} \bEE{f(X)g(Y)}.
\end{align*}
%\textchange{[SW: In the above formula, I don't know how to change the size of subscript]}
As an example, consider a binary symmetric source, denoted
$BSS(\rho)$, {$-1\le \rho \le 1$}, comprising $(X_1, X_2)$ taking
values in $\{0,1\}$ such that $\bP{X_1}{1} = 1/2$ and
\[
\bP{X_1, X_2}{0,0}= \bP{X_1, X_2}{1,1}= \frac 14(1+\rho).
\]
For this source, the maximal correlation $\rho_m(X_1, X_2) = \rho$,
and the functions $f$ and $g$ that achieve the maximum in the
definition of $\rho_m$ are given by $f(x) = g(x) = (-1)^x$, $x\in
\{0,1\}$.  As another example, consider a Gaussian symmetric source,
denoted $GSS(\rho)$, comprising jointly Gaussian $(X_1, X_2)$ with
zero mean and covariance matrix given by
\[
\begin{bmatrix}
1 & \rho\\ \rho & 1
\end{bmatrix}.
\]
For this source the maximal correlation is attained by identity
functions and is given by $\rho_m(X_1, X_2) = \rho$.

The second measure we describe, which is closely related to maximal
correlation, is based on {\it hypercontractivity} (see~\cite{Bonami70,
  Gross75, Beckner75, AhlswedeGacs76} for initial results
and~\cite{Donnell14} for a historical review). Specifically, a
distribution $\bPP{XY}$ is $(p,q)$-hypercontractive for $1\leq q \leq
p < \infty$ if for every bounded measurable function $f$ of $X$, the
following holds:
\begin{align} \label{eq:hypercontractivity}
\bEE{|\bEE{f(X)|Y}|^p}^{\frac 1 p} \leq \bEE{|f(X)|^q}^{\frac 1q}.
\end{align}
\textchangeSW{When $p=q$, \eqref{eq:hypercontractivity} holds because
  of concavity of $p$-norm, which is sometimes known as contraction
  property of the conditional expectation operator. Since $p$-norm is
  monotonically non-decreasing in $p$, the $(p,q)$-hypercontractivity
  characterizes how much the contraction inequality can be
  strengthened.}  The condition above can be replaced equivalently by
the following ``H\"older'' form: For all bounded measurable functions
$f$ of $X$ and $g$ of $Y$
\begin{align} \label{eq:hypercontractivity-holder-form}
\bEE{f(X)g(Y)} \leq \mathbb{E}\big[|f(X)|^{p'}\big]^{\frac 1
  {p'}}\mathbb{E}\big[|g(Y)|^q\big]^{\frac 1 q},
\end{align}
\textchangeSW{where $p'=p/(p-1)$ is the H\"older conjugate of $p\geq
  1$.}  \textchangeSW{Again, when $p=q$, i.e., $p^\prime$ is the
  H\"older conjugate of $q$, \eqref{eq:hypercontractivity-holder-form}
  holds because of the H\"older inequality; the
  $(p,q)$-hypercontractivity also characterizes how much the H\"older
  inequality can be strengthened.}  The set of all $(p,q)$ satisfying
the condition above is sometimes referred to as the hypercontractivity
ribbon of $\bPP{XY}$ and is denoted $\cR(\bPP{XY})$.

For $\bPP{XY}$ given by a $BSS(\rho)$ or a $GSS(\rho)$, the following classic result of Bonami
\cite{Bonami70} (see, also,~\cite{ Beckner75, Gross75}) characterizes
the set of $p,q$ for which $\bPP{XY}$ is $(p,q)$-hypercontractive.
\begin{theorem}\label{t:BB_ineq}
Suppose that $\bPP{XY}$ corresponds to a $BSS(\rho)$ or a
$GSS(\rho)$. Then, $\bPP{XY}$ is $(p,q)$-hypercontractive if and only
if
\[
\frac{q-1}{p-1}\geq \rho^2.
\]
\end{theorem}
%% As a corollary, we can obtain a similar result for $GSS(\rho)$.
%% \begin{corollary}\label{t:GH_ineq}
%% For a $GSS(\rho)$ $\bPP{XY}$ is $(p,q)$-hypercontractive if and only
%% if
%% \[
%% \frac{q-1}{p-1}\geq \rho^2.
%% \]
%% \end{corollary}
Therefore, for $BSS(\rho)$ and $GSS(\rho)$ there is a close
connection between hypercontractivity and maximal correlation
$\rho_m(X,Y)$.

Ahlswede and G\'acs~\cite{AhlswedeGacs76} highlighted a special
parameter related to the hypercontractivity ribbon defined as
\[
s^*(X,Y) = \lim_{p \rightarrow \infty} \inf_{q\,:\, (p, q)\in
  \cR(\bPP{XY})} \frac q p.
\]
They showed in \cite{AhlswedeGacs76} that $s^*(X,Y)\geq
\rho_m(X,Y)^2$. In fact, this inequality holds with equality for the
cases of BSS and GSS (for further elaboration,
see~\cite{AnantharamGKN13}).

The quantities $\rho_m(X,Y)$ and $s^*(X,Y)$ satisfy the following {\it
  tensorization} property: For independent $(X_i, Y_i)$, $1\leq i \leq
n$,
\begin{align*}
\rho_m(X^n, Y^n) &= \max_{1\leq i \leq n} \rho_m(X_i, Y_i), \\
s^*(X^n, Y^n) &= \max_{1\leq i \leq n} s^*(X_i, Y_i).
\end{align*}
Interestingly, the entire hypercontractivity ribbon tensorizes, $i.e.$,
\begin{align}
\cR(X^n, Y^n) = \bigcap_{i=1}^n\cR(X_i, Y_i).
\label{e:tensorization}
\end{align}
Note that information quantities such as mutual information satisfy an
additivity (or subaddivity) property for independent $(X^n, Y^n)$,
$i.e.$, $I(X^n \wedge Y^n) = \sum_{i=1}^n I(X_i\wedge Y_i)$.  In fact,
hypercontractivity has an information theoretic characterization
(see~\cite{AnantharamGKN13}, \cite{Nair15}), which also suggests a
duality between additivity and tensorization
(see~\cite{BeigiGohari15}).  \textchangeSW{An information theoretic
  characterization of the Brascamp-Lieb inequality, which includes the
  hypercontractivity bound as a special case, has been known earlier in the
  context of functional analysis \cite{CarCor09}; for a recent
  treatment of the Brascamp-Lieb inequality in the context of
  information theory, see \cite{BeigiNair16, LiuCouCufVer18}.}

%%%%%%%%%%%%%%%%%%%%%%%%
\subsection{Communication protocols}\label{s:comm_prot}
The final concept we review in this section on preliminaries is that of
interactive communication protocols, the key enabler of the tasks we
consider in this paper. The reader may already have a
heuristic notion of interactive communication in her mind, but a formal definition
is necessary to specify the scope of our results, particularly of our
converse bounds. Note that throughout we assume that the communication
channel is noiseless which circumvents issues of synchronization
that arise in interactive communication over noisy channels, and allows
us to restrict ourselves to a simpler notion of interactive communication.

\begin{figure}[t]
\centering \includegraphics[scale=0.3]{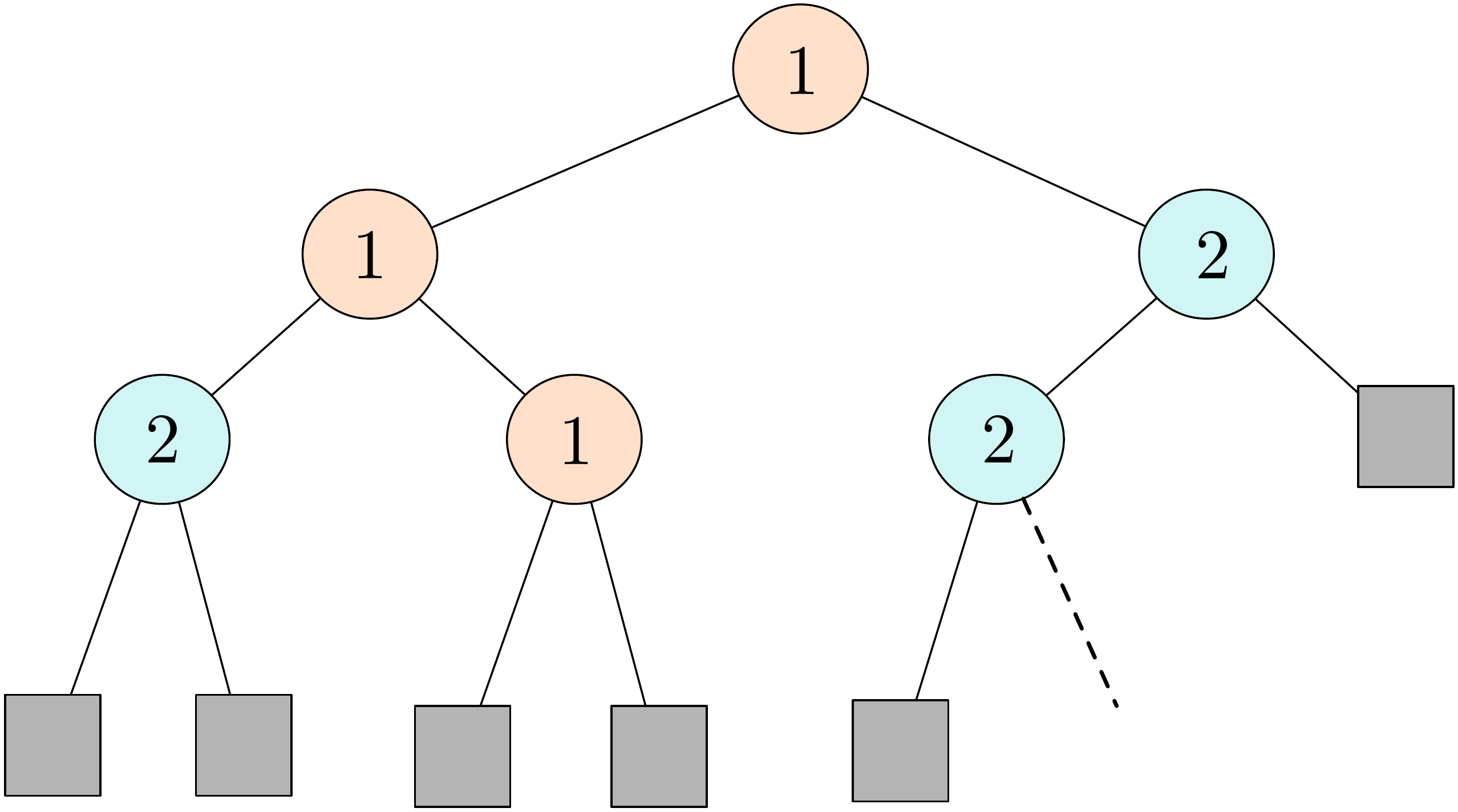}
\caption{A two-party interactive protocol tree.}
\label{f:tree_protocols}
\end{figure}

We restrict attention to {\it tree protocols} for interactive communication,
which were introduced in the work of Yao \cite{Yao79}. Parties $\cP_1$ and $\cP_2$
observe input $X_1$ and  $X_2$ generated from $\bPP{X_1X_2}$, with $\cP_i$
given access to $X_i$.  Additionally, $\cP_i$ has access to local
randomness (private coins) $R_{{\tt pvt}, i}$, $i\in \{1,2\}$, and both
parties have access to shared randomness (public coins) $R_{{\tt
    pub}}$. We assume that random variables $R_{{\tt pvt}, 1}, R_{{\tt pvt}, 2},
R_{{\tt pub}}$ are mutually independent and are independent jointly
of $(X_1, X_2)$.  An interactive communication protocol $\pi$ is
described by a labeled binary tree, where each node has a label from
the set $\{1,2\}$. Starting from the root node, when the protocol
reaches a node $v$ labeled $i$, $\cP_i$ transmits a bit $b_v =
b_v(X_i, R_{{\tt pvt}, i}, R_{{\tt pub}})$, and the protocol proceeds
to the left- or right-child of $v$ when $b_v$ is $1$ or $2$,
respectively. The communication protocol terminates when a leaf node
is reached, at which point each party declares an output. The (random)
bit sequence representing the path from root to leaf is called the
transcript of the protocol and is denoted by $\Pi$. Further, denoting
by $O_i = O_i(X_i, R_{{\tt pvt}, i}, R_{{\tt pub}}, \Pi)$ the output
of $\cP_i$, $i\in \{1,2\}$, we say that the protocol $\pi$ has input
$(X_1, X_2)$ and output $(O_1, O_2)$. The length of a protocol $\pi$,
denoted $|\pi|$, is the maximum number of bits transmitted in any
execution of the protocol and is given by the depth of the protocol
tree for $\pi$.  Figure~\ref{f:tree_protocols} provides an
illustration of a tree protocol. Note that the root is labeled
  $1$ denoting that $\cP_1$ initiates the communication. \textchangeSWtwo{Without loss of generality}, this will be
  our assumption throughout the paper.
The protocols that allow a nonconstant shared randomness $R_{\tt pub}$
are referred to as {\it public coin protocols}. When shared randomness
is not allowed, but local randomness is allowed, the protocols are
referred to as {\it private coin protocols}. Finally, the protocols
that do not allow private or shared randomness are called {\it
  deterministic protocols}.

The definition above allows the labels to switch arbitrarily
along a path from the root to a leaf. A restricted class, termed {\it
  $r$-round protocols}, consists of protocols where the maximum number
of times the label can switch along a path from the root to a leaf is
$r$.
   
While the tree protocol structure described above is seemingly
restrictive, typical lower bound proofs rely on some simple properties
of such protocols.
%Below we review these properties for $m=2$.
\begin{enumerate}
\item {\it Monotonicity of correlation. (cf.~\cite{Mau93,AhlCsi93})}
  For a private coin protocol $\pi$ with input $(X_1, X_2)$,
\begin{align}
I(X_1\wedge X_2) \geq I(X_1\wedge X_2| \Pi).  \label{eq:monotonicity-correlation}
\end{align}
In particular, if $X_1$ and $X_2$ are independent, they remain so upon
conditioning on $\Pi$. 

\item {\it Rectangle property. (cf.~\cite{Yao79,
    KushilevitzNisan97})} For a private coin protocol $\pi$, denote by
  $p(\tau|x_1, x_2)$ the probability of $\Pi= \tau$ given that the
  input is $(x_1, x_2)$. Then, there exist functions $f_\tau$ and
  $g_\tau$ such that
\begin{align}
p(\tau|x_1, x_2) = f_\tau(x_1)g_\tau(x_2), \quad \forall\, x_1\in
\cX_1, x_2\in \cX_2.  \nonumber
\end{align}
For deterministic protocols, this implies that if a transcript $\tau$
appears for inputs $(x_1, x_2)$ and $(x_1', x_2')$, then it must
appear for $(x_1, x_2')$ and $(x_1', x_2)$ as well. In other words,
the set $\Pi^{-1}(\tau)$ constitutes a rectangle.
\end{enumerate}
Both results above are, in essence, observations about the correlation
we can build using tree protocols and are easy to prove. However, note
that they are valid only for private coin protocols. When shared
randomness (public coin) $R_{{\tt pub}}$ is used, the results above do
not hold and the correlation has a more complicated
structure. Nevertheless, even in this case, the results are
recovered on conditioning additionally on $R_{{\tt pub}}$.

Closely related to the length of a protocol, namely the amount of
information communicated in any execution of the protocol, is the
so-called {\it information cost} of the protocol. Heuristically,
information cost captures the number of bits of information that is
revealed by the protocol. We recall two variants of information cost
for private coin protocols.

The {\it external information cost} of a private coin protocol $\pi$
with inputs $(X_1, X_2)$ is given by \cite{ChakrabartiSWY01}
\begin{align}
\ICe(\pi| X_1, X_2) = I(\Pi \wedge X_1, X_2), \nonumber
\end{align}
and its {\it internal information cost} is given by \cite{BarakBCR13}
(an early conference version appeared as~\cite{BarakBCR10})
\begin{align}
\ICi(\pi| X_1, X_2) = I(\Pi \wedge X_1| X_2) + I(\Pi \wedge X_2| X_1).
\nonumber
\end{align}
Since 
\begin{align}
\lefteqn{ \ICe(\pi| X_1, X_2) - \ICi(\pi| X_1, X_2) } \nonumber \\
&= I(X_1 \wedge X_2) - I(X_1 \wedge X_2|\Pi),
\label{e:identity-ie-ic}
\end{align}
the following observation is equivalent to the monotonicity of
correlation property (see, for instance, \cite{CsiNar04, CsiNar08,
  BarakBCR13}):
\begin{align}
\ICi(\pi| X_1, X_2) \leq \ICe(\pi| X_1, X_2).  
\label{e:monotonicity-correlation-cost-form}
\end{align}
\textchangeSW{The internal cost can be regarded as the amount of
  information conveyed between the parties, and the external cost can
  be regarded as the amount of information conveyed to an
  external observer of the transcripts. Thus, the  
inequality above says that parties share less information 
with each other than with an external observer, which is perhaps
natural to expect since the inputs of the parties are correlated, and therefore, they have prior knowledge of each other's input.}

%%%%%%%%%%%%%%%%%%%%%%%%
\subsection{Information theory parlance}
\htc{In our narrative in this article, we shall be using the standard language
  of information theory. Some of the terms used are informal, but are
  standard occurrences in information theory parlance. Here we provide
  a quick listing of these terms for the benefit of the 
  reader.}

\htc{Several quantities in information theory are defined
  operationally as the optimal cost for a problem (such as minimum
  communication or maximum length of a secret key). These optimal
  costs are often characterized by a closed form formula which often
  finds applications beyond the original \emph{operational
    significance}. The foremost example is that of channel capacity,
  which is an operationally defined quantity and is characterized as
  mutual information optimized over input distributions, but it has found use-cases beyond channel coding. Throughout this article, we endow information theoretic
  quantities with operational significance.}

\htc{Another term that often shows up in Shannon theory is the
  so-called {\em single-letter characterization}, which we only describe informally here. Usually, operational quantities mentioned above can
  be easily characterized in terms of information theoretic quantities
  such as entropy, \textchangeSWtwo{but involved random variables may take infinitely many
  values}. Several open problems in information theory seek to express
  these quantities in terms of random variables taking finitely many
  values. Such expressions are called single-letter expressions in 
  information theory. In the computer science literature, similar
  questions have been underlying the so-called {\em direct-sum theorems}
  where one seeks to solve a single instance of a problem using a
  protocol that solves multiple instances simultaneously.}

\textchangeSWtwo{Also, we take recourse to the notion of {\em typical sets} at
  several places. A formal definition of this notion can be found in
  the seminal textbook~\cite{CsiKor11}. In particular, in the proof outline for
  Theorem~\ref{t:CR_cap} we use the standard notion of
  $\bPP{X}$-typical sets from~\cite{CsiKor11}, sometimes referred to
  as {\em strongly typical sets}, which is roughly the set of
  $n$-length sequences with normalized frequencies of each element $x$
  close to $n\bP X x$.}

%%%%%%%%%%%%%%%%%%%%%%%%%%%%%%%%%%%%%%%%%%%%%%%
\section{Common randomness generation}\label{s:CR} 
We begin with the common randomness (CR) generation problem.
%% Note that the problem is
%% nontrivial only when private coin protocols are allowed; if public
%% coin protocols are allowed then the shared randomness directly yields
%% CR.
For simplicity, we restrict ourselves to private coin protocols in
this section\footnote{In principle, shared randomness can be included
  as a part of the input $(X_1, X_2)$.}. As another simplifying
assumption, we consider only the protocols that start at $\cP_1$, namely the
root of the protocol tree is labeled $1$. We also assume that the
cardinalities $|\cX_1|$ and $|\cX_2|$ are finite; results for the
Gaussian case are also highlighted whenever available.

Throughout this section, we restrict ourselves to source models where the
parties are given correlated observations from a joint distribution. A
richer model is a channel model where $\cP_1$ can select an input
$x\in \cX$ for a channel $W: \cX \to \cY$ whose output $Y$ is observed
by $\cP_2$. In addition, the parties can communicate over an error
free channel using interactive protocols. We do not cover the results
for this interesting setting; see \cite{AhlCsi98, VenAna98, CsiNar08} for
initial results.

We present two variants of the CR generation problem: In the first, the
amount of communication is fixed and the largest possible amount of CR
that can be generated is characterized; and in the second, the
amount of CR is fixed and the minimum amount of communication required
is characterized. In principle, both variants above are closely
related and studying one should shed light on the other. In
practice, however, the specific formulations and the techniques
considered in one case are difficult to transform to the
other. \htc{Furthermore, the result we present in the second setting
  looks at very small probability of agreement, exponentially small in the
  CR length, and characterizes the minimum communication needed for
  generating a fixed length CR.}

%%%%%%%%%%%%%%%%%%%%%%%%
\subsection{CR using limited communication}
 The fundamental quantity of interest here is the following.
\begin{definition}
For jointly distributed random variables $(X_1, X_2)$, an integer $l$ is
an $(\ep,c,r)$-achievable CR length if there exists an $r$-round
private coin protocol $\pi$ of length less than $c$ bits and with outputs
$(S_1, S_2)$ such that, for a random string $S$ distributed uniformly
over $\{0, 1\}^l$,
\begin{align}
\bPr{S_1 = S_2 = S} \geq 1-\ep.  \nonumber
\end{align}
The supremum over all $(\ep, c,r)$-achievable CR lengths $l$ is
denoted by $L_{\ep, r}(c|X_1, X_2)$.
%% Further, denote the supremum
%% of $L_{\ep, r}(c|X_1, X_2)$ over $r$ as $L_\ep(c|X_1, X_2)$. 
\end{definition}
The random variable $S$ is referred to as an $(\ep, c,r)$-CR of length
$l$ using $\pi$; we omit the dependence on the parameters when it is
clear from the context.

The formulation above was introduced by \cite{AhlCsi93, AhlCsi98}
where they studied an asymptotic, capacity version of the quantity
$L_{\ep,r}(c|X_1,X_2)$.
\begin{definition}[Common randomness capacity]
 For $R\geq 0$, $r\in \mN$, and an iid sequence $\{X_{1,i},
 X_{2,i}\}_{i=1}^\infty$, the $(\ep, r)$-CR capacity for communication
 rate $R$, denoted
%\footnote{For brevity, we omit the dependence of
%  $C_\ep(R)$ on $(X_1,X_2)$ from our notation.} 
$C_{\ep,r}(R)$, is given by
\begin{align*}
C_{\ep, r}(R) = \lim_{n \rightarrow \infty} \frac 1 n
L_{\ep,r}\big(nR| X_1^n, X_2^n\big).
\end{align*}
Further, the $r$-round CR capacity for communication rate $R$, denoted
$C_r(R)$, is given by $C_r(R)=\lim_{\ep \rightarrow 0}C_{\ep,
  r}(R)$. Finally, denote by $C(R)$ the supremum of $C_r(R)$ over
$r\in \mN$.
\end{definition}
The formulation of CR capacity in \cite{AhlCsi98} allowed only two
rounds of interaction, namely $\cP_1$ upon observing $X_1^n$ sends
$\Pi_1= f_1(X_1^n)$ to $\cP_2$, who in turn responds with $\Pi_2=
f_2(\Pi_1, X_2^n)$. Furthermore, while \cite{AhlCsi98} considered
private coin $1$-round protocols, the extension to $2$ rounds was
restricted to deterministic protocols.  We denote this restricted
notion of CR capacity using a $2$-round deterministic protocol of rate
$R$ by $C_2^d(R)$; it is characterized as follows.
\begin{theorem}[\cite{AhlCsi98}]\label{t:AhlCsi}
For $R\geq 0$ and random variable $(X_1, X_2)$ taking values in a
finite set $\cX_1\times \cX_2$, we have
\begin{align}
C_2^d(R) = \max_{\bPP{UVX_1X_2}\in \cP(R)} I(U\wedge X_1) + I(V\wedge
X_2|U)
\label{e:CR2_char}
\end{align}
where $\cP(R)$ denotes the set of joint pmf $\bPP{UVX_1X_2}$ such that
the following conditions hold:
\begin{enumerate}
\item[(i)] $\bPP{UVX_1X_2} = \bPP{X_1X_2}\bPP{U|X_1}\bPP{V|X_2U}$;
\item[(ii)] $U$ and $V$ take values in finite sets $\cU$ and $\cV$
  such that $|\cU| \leq |\cX_1|+1$ and $|\cV|\leq |\cX_2||\cU| + 1$;
  and
\item[(iii)] $I(U\wedge X_1|X_2) + I(V\wedge X_2|X_1, U)\leq R$.
\end{enumerate}
\end{theorem}
The expression on the right-side of \eqref{e:CR2_char} entails two
interesting quantities: The first, $I(U\wedge X_1) + I(V\wedge X_2|U)$
which in the view of the Markov relations $U \mc X_1 \mc X_2$ and $V
\mc (X_2, U) \mc X_1$ equals $I(U,V \wedge X_1, X_2)$, and the second
$I(U\wedge X_1|X_2) + I(V\wedge X_2|X_1, U)$ which appears in the
constraints set and equals $I(U,V \wedge X_1|X_2) + I(U,V\wedge X_2|X_1)$. Both these quantities have a long history in the literature on
network information theory; see, for instance, \cite{Wyn75iii,
  AhlKor75, WynZiv76, Kas85}. In the computer science literature,
these quantities have been rediscovered in a slightly different
operational role, namely that of external and internal information
costs defined in Section~\ref{s:preliminaries}. 
Specifically, consider
a $2$-round protocol where $\cP_1$ uses its private coin to sample
$U$ using $\bPP{U|X_1}$ and sends it to $\cP_2$. Next, $\cP_2$ samples
$V$ using $\bPP{V|X_2,U}$ and sends it to $\cP_1$. The overall joint
distribution $\bPP{UVX_1X_2}$ is maintained since $U \mc X_1\mc X_2$
and $V\mc (X_2, U)\mc X_1$. Also, the protocol is a tree protocol
since $|\cU|$ and $|\cV|$ can be restricted to be finite. It is easy
to see that any $2$-round private coin protocol can be expressed
similarly in terms of $U$ and $V$. Thus, \eqref{e:CR2_char} can be
restated as follows:
\begin{align}
C_2^d(R) = \max_{\pi: \ICi(\pi| {X_1, X_2}) \leq R} \ICe(\pi| X_1,
X_2),
\label{e:CR2_char_ic}
\end{align}
where the maximum is restricted to $2$-round private coin protocols
$\pi$. Note that while the ``$n$-fold'' problem for $C_2^d(R)$ did not
allow randomization, the ``single-letter'' characterization above
entails optimization over private coin protocols. 

The result can be extended to the case when arbitrary (but fixed) round private
coin protocols are allowed. This extension and its proof are perhaps    
known to specialists in this area, but it has not been formally
reported anywhere. However, glimpses of this result can be seen, for
instance, in \cite{YeTh}, \cite{GohAna10}, \cite{Tya13},
\cite{LiuCuffVerdu17}, and \cite{GhaziJayram17}.

In fact, it is
interesting to track the history of this result in information theory 
and computer science. Following the work of Ahlswede and
Csisz\'ar~\cite{AhlCsi93, AhlCsi98}, a multiparty extension of the CR
agreement problem appeared in a specialized model appeared in~\cite{CsiNar00}
and for the related problem of secret key agreement
in~\cite{CsiNar04}. A result very similar to Theorem~\ref{t:CR_cap}
seems to have appeared first in~\cite[Theorem 5.3]{YeTh}, albeit without a complete
proof. The schemes in all these works in the information theory literature are based
on the classic binning technique of Wyner and
Ziv~\cite{WynZiv76}, which was also used for function computation
in~\cite{OrlRoc01}. This is where the intersection with the computer
science literature first appears. Specifically, the rates achieved by Wyner-Ziv
binning entail terms of the form $I(U\wedge X|Y)$, namely internal
information complexity of one round protocols. This quantity was used
as a measure of information complexity for function computation
in~\cite{BarakBCR10, BraRao11}, following the pioneering
works~\cite{ChakrabartiSWY01,BarJKS04}. Interestingly, the same result
as~\cite{BraRao11} was obtained independently in~\cite{MaIsh08,MaIsh11} in the
information theory literature, where the information complexity
quantities facilitated Wyner-Ziv binning in the scheme; the converse proof
in~\cite{MaIsh11} used a method introduced in~\cite{Kas85}, which was
slightly different from the ``embedding'' used in~\cite{BraRao11}.\footnote{The scheme
  proposed in~\cite{BraRao11} was much more general and was also valid
in the single-shot setting.} Till this point, these two lines of works
bringing in information complexity emerged independently. This seems
to have changed after a workshop at Banff on Interactive Information
Theory in 2012 where the authors of~\cite{BraRao11}
and~\cite{MaIsh11} participated and learnt of these two views on the
same results. Subsequently, review articles such as~\cite{Braverman12ii}
appeared, but still the application of information complexity to CR
generation was not explicitly mentioned anywhere. This connection was
exploited in works such as~\cite{TyagiVVW17} (for instance,~\cite[Eqn. 14]{TyagiVVW17}
is a single-shot counterpart of~\eqref{e:identity-ie-ic}), but the
first instance where this connection was explicitly mentioned
is~\cite{GhaziJayram17}.

Let $C^d(R)$ be the analog of $C(R)$ 
for deterministic communication protocols. Our characterization\footnote{Thanks to Noah Golowich for detecting an error in a previous version of our characterization and suggesting a fix.} of $C^d(R)$ and $C(R)$
involves a function $f(R)$, which is an extension of the function on the right-side of \eqref{e:CR2_char_ic}
to multiple rounds, given by
\begin{align}
f(R) := \sup_{\pi: \ICi(\pi|{X_1,X_2})\leq R}
  \ICe(\pi|{X_1,X_2}), \label{e:CR-multiple-round-expression}
\end{align}  
where the supremum is taken over all protocols with arbitrary (finite) number of rounds.
%%%%%%
\begin{theorem}\label{t:CR_cap}
For $R\ge 0$ and random variable $(X_1, X_2)$ taking values in a finite
set $\cX_1\times \cX_2$, we have
\begin{align}
  C^d(R) = f(R)
 \label{e:CR_char_ic_d}
\end{align}
and 
\begin{align}
C(R) = \sup_{t \ge 0} f(R-t) + t.
\label{e:CR_char_ic_r}
\end{align}
\end{theorem}
%%%%%%%
\paragraph{Proof of Theorem~\ref{t:CR_cap} for deterministic protocols}
We first prove the lower bound (converse) part of~\eqref{e:CR_char_ic_d}, i.e.,
   $C^d(R)\geq f(R)$. Consider an
$(\ep, c, r)$-CR $S$ of length $l$ that can be recovered using an
$r$-round deterministic communication protocol $\pi_n$ with input
$(X_1^n, X_2^n)$ \htc{where $X_i^n=(X_{i1}, ..., X_{in})$ denotes the
  observation of $\cP_i$, $i=1,2$.}  For simplicity, we assume that
$\cP_1$ is the last party to communicate; the other case can be
handled similarly. Denote by $\Pi_i$ the communication sent in round
$i$ and by $S_1$ the estimate of $S$ formed at $\cP_1$. Since $S$ is
uniformly distributed, by Fano's inequality we have
\begin{align}
l \leq H(S_1) + \ep l +1.
\label{e:bound_fano1}
\end{align}
Thus, it suffices to bound $H(S_1)$. We show that there exists an
$r$-round private coin protocol $\pi_{1}$ with input $(X_1, X_2)$ such
that $H(S_1) \leq n \ICe(\pi_1|X_1, X_2)$ and $n\ICi(\pi_1|X_1,X_2) \lesssim c$. Specifically, abbreviating $X_{ij}^k = X_{ij}, ..., X_{ik}$,
$i=1,2$ and with $J$ distributed uniformly over $\{1, ..., n\}$
independently of $(\Pi, X_1^n, X_2^n)$, let $U_1 = (X_{11}, ...,
X_{1(J-1)}, X_{2(J+1)} ,..., X_{2n}, \Pi_1, J)$, $U_i = \Pi_i$ for $1<
i< r$, and $U_r = (\Pi_r, S_1)$. The following Markov relation can be
shown to hold: For $0\leq i\leq r$
\begin{align}
&U_{1} \mc X_{1J} \mc X_{2J}, \nonumber \\ &U_{i+1}\mc (U^i,
  X_{1J})\mc X_{2J},\quad i\text{ even}, i\geq 2, \nonumber
  \\ &U_{i+1}\mc (U^i, X_{2J})\mc X_{1J},\quad i\text{ odd}, i\geq 1.
\label{e:mc1}
\end{align}
\textchangeSWtwo{The Markov relations can be obtained as a consequence of monotonicity of correlation property 
of interactive communication (see Section~\ref{s:comm_prot}).  We outline the proof
  for $U_{i+1} \mc (U^i, X_{1J}) \mc X_{2J}$ for even $i<r$; the
  remaining can be derived similarly. Consider a hypothetical
  situation in which $\cP_1$ observes $(X_{1(j+1)}^n, X_{2(j+1)}^n)$ and
  $\cP_2$ observes $(X_{11}^j,X_{21}^j)$, which are independent.
  First, $\cP_1$ and $\cP_2$ exchange $X_{2(j+1)}^n$ and $X_{11}^j$ so that $\cP_1$ obtains $X_1^n$ and $\cP_2$
  obtains $X_2^n$, which enables them to simulate the transcript $\Pi^i$ via interactive communication.
  Then, using the monotonicity of correlation property, we get
\begin{align*}
0 
&=I(X_{1(j+1)}^n, X_{2(j+1)}^n \wedge X_{11}^j,X_{21}^j ) \\
&\ge I(X_{1(j+1)}^n, X_{2(j+1)}^n \wedge X_{11}^j,X_{21}^j | X_{11}^j, X_{2(j+1)}^n, \Pi^i ) \\
&= I(X_{1(j+1)}^n, X_{2(j+1)}^n, \Pi_{i+1} \wedge X_{11}^j,X_{21}^j | X_{11}^{j}, X_{2(j+1)}^n, \Pi^i ) \\
&\ge I(\Pi_{i+1} \wedge X_{2j} | X_{1j}, X_{11}^{j-1}, X_{2(j+1)}^n, \Pi^i )
\end{align*}
where the second identity uses the fact that $\Pi_{i+1}$ is a
  function of $X_1^n$ given $\Pi^i$. Thus, $I(\Pi_{i+1} \wedge X_{2J} | U^i)=0$, which given \eqref{e:mc1}.}

\textchangeSWtwo{By noting the Markov relation $\Pi_i \mc (X_1^n,\Pi^{i-1}) \mc X_2^n$
for odd $i$ and $\Pi_i \mc (X_2^n,\Pi^{i-1}) \mc X_1^n$ for even $i$,\footnote{We assume that $r$ is odd, but the case with even $r$ can
be handled similarly with $S_2$ in the role of $S_1$.} we can get
\begin{align*}
\lefteqn{ H(S_1) } \\
&\leq I(S_1, \Pi \wedge X_1^n, X_2^n) \\ 
&= \sum_{i=1: \mathrm{odd}}^{r-2} I(\Pi_i \wedge X_1^n | \Pi^{i-1}) \\
&~~~+ \sum_{i=2: \mathrm{even}}^{r-1} I(\Pi_i \wedge X_2^n| \Pi^{i-1}) + I(S_1,\Pi_r \wedge X_1^n | \Pi^{r-1}) \\
&= H(X_1^n) + \sum_{i=1: \mathrm{odd}}^{r-2} \big[ H(X_2^n | \Pi^i) - H(X_1^n | \Pi^i) \big] \\
&~~~ + \sum_{i=2: \mathrm{even}}^{r-1} \big[ H(X_1^n | \Pi^i) - H(X_2^n | \Pi^i) \big] - H(X_1^n | S_1,\Pi^r) \\
&\le n H(X_{1 J}) +\sum_{i=1: \mathrm{odd}}^{r-2} \big[ H(X_2^n | \Pi^i) - H(X_1^n | \Pi^i) \big] \\
&~~~ + \sum_{i=2: \mathrm{even}}^{r-1} \big[ H(X_1^n | \Pi^i) - H(X_2^n | \Pi^i) \big]  - n H(X_{1,J} | U^r) \\
&= n \bigg[ H(X_{1 J}) + \sum_{i=1: \mathrm{odd}}^{r-2} \big[ H(X_{2 J} | U^i) - H(X_{1 J} | U^i) \big] \\
&~~~ + \sum_{i=2: \mathrm{even}}^{r-1} \big[ H(X_{1 J} | U^i) - H(X_{2 J} | U^i) \big] - H(X_{1 J} | U^r) \bigg] \\
&= n I(U^r \wedge X_{1 J}, X_{2 J}),
\end{align*}}
where the second last equality uses\footnote{\textchangeSWtwo{The identity \eqref{eq:csiszar-identity} is very popular in multiterminal
  information theory and sometimes referred to as the Csisz\'ar
  identity (see \cite{ElGKim}). But perhaps it can be best attributed  to
  Csisz\'ar, K\"orner, and Marton; see~\cite{CsiKor11}. To the best of our knowledge,
  the earliest use of this identity appears in \cite{KorMar77}.}}
\begin{align} 
\lefteqn{ H(X_1^n | \Pi^i)-H(X_2^n|\Pi^i) } \nonumber \\
&=  n\big[ H\big(X_{1J}|X_{11}^{J-1},X_{2(J+1)}^n, \Pi^i,J \big) \nonumber \\
&~~~  - H\big(X_{2J}|X_{11}^{J-1},X_{2(J+1)}^n,\Pi^i,J \big) \big].
\label{eq:csiszar-identity}
\end{align}
\htc{Also, for deterministic protocols, the monotonicity of
  correlation property is the same as }
\[
H(\Pi) \geq H(\Pi\mid X_1^n) + H(\Pi\mid X_2^n),
\]
\textchangeSWtwo{which together with the Fano inequality gives
\begin{align*}
H(\Pi) &\geq I(\Pi \wedge X_1^n|X_2^n) + I(\Pi \wedge X_2^n|X_1^n) \\
&\geq  I(S_1, \Pi \wedge X_1^n|X_2^n) + I(S_1, \Pi \wedge X_2^n|X_1^n) - l\ep-1.
\end{align*}
In order to derive the single-letter characterization, again by noting the Markov relation $\Pi_i \mc (X_1^n,\Pi^{i-1}) \mc X_2^n$
for odd $i$ and $\Pi_i \mc (X_2^n,\Pi^{i-1}) \mc X_1^n$ for even $i$, we can get
\begin{align*}
\lefteqn{ I(S_1, \Pi \wedge X_1^n|X_2^n) + I(S_1, \Pi \wedge X_2^n|X_1^n) } \\
&=  \sum_{i=1: \mathrm{odd}}^{r-2} I(\Pi_i \wedge X_1^n | X_2^n, \Pi^{i-1}) \\
&~~~ + \sum_{i=2: \mathrm{even}}^{r-1} I(\Pi_i \wedge X_1^n | X_2^n, \Pi^{i-1}) \\
&~~~ + I(S_1,\Pi_r \wedge X_1^n | X_2^n,\Pi^{r-1}) \\
&= H(X_1^n | X_2^n) + \sum_{i=1: \mathrm{odd}}^{r-2} \big[ H(X_2^n | X_1^n,\Pi^i) - H(X_1^n | X_2^n,\Pi^i) \big] \\
&~~~ + \sum_{i=2: \mathrm{even}}^{r-1} \big[ H(X_1^n | X_2^n,\Pi^i) - H(X_2^n | X_1^n, \Pi^i) \big] \\
&~~~ - H(X_1^n | X_2^n, S_1,\Pi^r) \\
&\ge n H(X_{1 J} | X_{2,J}) \\
&~~~ + \sum_{i=1: \mathrm{odd}}^{r-2} \big[ H(X_2^n | X_1^n,\Pi^i) - H(X_1^n | X_2^n,\Pi^i) \big] \\
&~~~ + \sum_{i=2: \mathrm{even}}^{r-1} \big[ H(X_1^n | X_2^n,\Pi^i) - H(X_2^n | X_1^n, \Pi^i) \big] \\
&~~~ - n H(X_{1,J} | X_{2,J}, U^r) \\
&= n\bigg[ H(X_{1 J} | X_{2,J}) \\
&~~~ + \sum_{i=1: \mathrm{odd}}^{r-2} \big[ H(X_{2 J} | X_{1 J},U^i) - H(X_{1 J} | X_{2 J},U^i) \big] \\
&~~~ + \sum_{i=2: \mathrm{even}}^{r-1} \big[ H(X_{1 J} | X_{2 J},U^i) - H(X_{2 J} | X_{1 J}, U^i) \big] \\
&~~~ - n H(X_{1 J} | X_{2 J}, U^r) \bigg] \\
&= n\big[ I(U^r \wedge X_{1J}|X_{2J}) + I(U^r \wedge X_{2J}|X_{1J}) \big],
\end{align*}
where the second last inequality again uses \eqref{eq:csiszar-identity} along with the identity
\begin{align*}
\lefteqn{ H(X_2^n | X_1^n, \Pi^i) - H(X_1^n | X_2^n,\Pi^i) } \\
& = H(X_1^n | \Pi^i ) - H(X_2^n | \Pi^i).
\end{align*}}

Therefore, noting that $(X_{1J}, X_{2J})$ has the same distribution as
$(X_1, X_2)$, in the limits as $n$ goes to infinity and $\ep$ goes to
zero (in that order), we get by~\eqref{e:bound_fano1} and the bounds
above that the rate of CR is bounded above by $f(R)$ defined in
\eqref{e:CR-multiple-round-expression}.  Also, to claim that $U^r$
correspond to a protocol, we need to bound the cardinalities of their
support sets. Under our assumption of finite $|\cX_1|$ and $|\cX_2|$,
we can restrict the cardinalities of $\cU_i$ to be finite using the
support lemma \cite[Lemma 15.4]{CsiKor11}.

For the proof of the upper bound (achievability) of the deterministic
case, we begin with an outline of the proof for achieving $f(R)$
restricted to $r=1$, namely achieving
\begin{align}
\max_{\substack{U: U\mc X_1 \mc X_2 \\\qquad I(U\wedge X_1|X_2)\leq R}
} I(U \wedge X_1).
\label{e:CR_r1}
\end{align}
We use standard typical set arguments to complete the proof
\cite{CovTho06, Csi98, CsiKor11}.  Consider a random codebook
\[
\cC=\{U^n(i,j), 1\leq i \leq \lceil 2^{nR}\rceil, 1\leq j \leq \lceil
2^{n\gamma} \rceil \},
\]
where $U^n(i,j)\in \cU^n$ are iid (for different $i,j$) and uniformly
distributed over the $\bPP{U}$-typical set. Consider the following
protocol:
\begin{enumerate}
\item[1.] $\cP_1$ finds the smallest $i$ for which there exists a $j$
  such that $(X_1^n, U^n(i,j))$ is $\bPP{UX_1}$-typical. Let $\Pi_1$
  denote this smallest index $i$ and $Y_1$ denote the sequence
  $U^n(i,j)$ identified above.

\item[2.] $\cP_1$ sends $\Pi_1$ to $\cP_2$.
\item[3.] $\cP_2$ searches for the smallest index $j$ such that
  $(X_2^n, U^n(\Pi_1, j))$ is $\bPP{UX_2}$-typical. Denote by $Y_2$
  the sequence $U^n(\Pi_1, j)$.
\end{enumerate}
The standard covering and packing arguments in multiterminal
information theory (cf.~\cite{CsiKor11}) imply that for $\gamma =
I(U\wedge X_2) -2\delta$ and $R=I(U\wedge X_1|X_2)+3\delta$, the
protocol above yields $Y_1$ and $Y_2$ that agree with large
probability (over the random input and the random
codebook). Furthermore, for every fixed realization of the codebook
$\cC$, it can be shown using standard typical set arguments that with
$\cT$ denoting the $\bPP{UX_1}$-typical set
\begin{align*}
\bPr{Y_1 = u^n(i,j)} &\leq \bP{X_1^n}{\{x^n: (u^n,x^n)\in\cT\}}
\\ &\leq \exp\left( - nI(U\wedge X_1) + o(n)\right).
\end{align*}
Therefore, using the leftover hash lemma (see Theorem~\ref{t:LHL})
\htc{with $Y_1$ in the role of $X$ and $(Z,V)$ set to constants, and
  noting that the min-entropy of $Y_1$ is roughly $nI(U\wedge X_1)$ by
  the previous bound,} there exists a fixed function $g$ of $Y_1$ and
$S$ consisting of roughly $nI(U\wedge X_1)$ uniformly distributed
random bits such that $S_1=g(Y_1)$ satisfies
$\ttlvrn{\bPP{S_1}}{\bPP{S}}\leq \ep/2$. Using the maximal coupling
lemma (see Lemma~\ref{l:MCL}), there exists a joint distribution
$\bPP{S_1S}$ with the same marginals as the original $S_1$ and $S$
such that $\bPr{S_1 \neq S}\leq \ep/2$. Therefore, for $\bPP{X_1^n
  X_2^n S_1 S} = \bPP{X_1^n X_2^n S_1}\bPP{S|S_1}$, we have
\begin{align*}
\bPr{S= S_1 = S_2} &\geq \bPr{S=S_1} + \bPr{S_1 = S_2} - 1 \\
& \geq 1-\ep,
\end{align*}
for all $n$ sufficiently large. The final step in our protocol is now
simple:
\begin{enumerate}
\item[4.] $\cP_i$ outputs $S_i=g(Y_i)$, $i=1,2$.
\end{enumerate}
Note that the rate of communication is no less than $R= I(U\wedge
X_1|X_2) + 3\delta$ and the rate of CR generated is $I(U\wedge
X_1)$. Furthermore, since the mapping $g$ can be found for any fixed
realization of the codebook $\cC$, we can derandomize the argument
above to obtain a deterministic scheme. This completes the
achievability proof for the rate in \eqref{e:CR_r1}. To extend the
proof to $r=2$, we repeat the construction above but conditioned on
the previously generated CR, namely the sequence $U^n$ found in Step
1. The analysis will remain the same in essence, except that the
mutual information quantities will be replaced by conditional mutual
information given $U$; leftover hash will be applied to the overall
shared sequence pair. Extension to further higher number of rounds is
obtained by repeating this argument recursively.  \qed

%%%%%%%
\paragraph{Extending the proof to private coin protocols}
We now move to the general case where private coin protocols are
allowed and prove~\eqref{e:CR_char_ic_r}. Achievability follows from
the time sharing between a (deterministic) scheme attaining CR rate
$f(R-t)$ with communication rate $R-t$ and a trivial private coin
scheme attaining CR rate $t$ with communication rate $t$ which simply
shares $t$ random bits over the communication channel.  To extend the
proof of converse to private coin protocols, we assume without loss of
generality that private randomness of $\cP_1$ and $\cP_2$,
respectively, are given by iid sequences $W_1^n$ and $W_2^n$. We can
then use the proof for the deterministic case to get a single-shot
protocol $\pi_1$ such that the rate of CR is less than $\ICe(\pi_1|
(X_1, W_1), (X_2, W_2))$ and the rate of communication is more than
$\ICi(\pi_1| (X_1, W_1), (X_2, W_2))$. To complete the proof, we show
that there exists another private coin protocol $\pi_1'$ and
nonnegative $t$ such that
\begin{align} \label{eq:randomized-deterministic-ICe}
\ICe(\pi_1| (X_1, W_1), (X_2, W_2)) = \ICe(\pi_1'| X_1, X_2) + t,
\end{align}
and
\begin{align} \label{eq:randomized-deterministic-ICi}
\ICi(\pi_1| (X_1, W_1), (X_2, W_2)) = \ICi(\pi_1'| X_1, X_2) + t,
\end{align}
whereby it follows that $C(R) \leq f(R-t) + t$ for some $t \ge 0$.

To see~\eqref{eq:randomized-deterministic-ICe}
and~\eqref{eq:randomized-deterministic-ICi}, for transcript $U^r$ of
protocol $\pi_1$ and for every $1 \le i \le r$, by the monotonicity of
correlation we get
\begin{align*}
I(W_1 \wedge X_2, W_2 | X_1, U^i) &= 0, \\ I(W_2 \wedge X_1, W_1 |
X_2, U^i) &= 0.
\end{align*}
Indeed, the first relation follows by noting that $(X_1,U^i)$ is an
interactive communication protocol for two parties where $\cP_1$
observes $W_1$ and $\cP_2$ observes $(X_1,X_2,W_2)$; the second one
can be obtained similarly.  Using these conditional independence
relations, for odd $i$ we have
\begin{align*}
\lefteqn{ I(U_i \wedge X_2 | X_1, U^{i-1}) } \\
&\le I(U_i, W_1 \wedge X_2, W_2 | X_1, U^{i-1}) \\ 
&= I(W_1 \wedge X_2, W_2 | X_1, U^{i-1}) \\
&~~~+ I(U_i \wedge X_2,W_2 | X_1, W_1, U^{i-1}) 
\\ &= 0.
\end{align*}
Similarly, for even $i$, we have
\begin{align}
I(U_i \wedge X_1 | X_2, U^{i-1}) = 0.  \nonumber
\end{align}
Thus, we find that $U^r$ constitutes transcript of an interactive
protocol $\pi_1^\prime$ with observation $(X_1,X_2)$. Furthermore, we
can expand the information costs of $\pi_1$ as follows:
\begin{align*}
\lefteqn{ \ICe(\pi_1| (X_1, W_1), (X_2, W_2)) } \\ &= I(U^r \wedge
X_1,W_1,X_2,W_2) \\ &= \sum_{i: \mathrm{odd}} I(U_i \wedge X_1,W_1 |
U^{i-1}) + \sum_{i : \mathrm{even}} I(U_i \wedge X_2, W_2 | U^{i-1})
\\ &= \sum_{i: \mathrm{odd}} \big[ I(U_i \wedge X_1 | U^{i-1}) + I(U_i
  \wedge W_1 | X_1, U^{i-1}) \big] \\ &~~~ + \sum_{i: \mathrm{even}}
\big[ I(U_i \wedge X_2 | U^{i-1}) + I(U_i \wedge W_2 | X_2, U^{i-1})
  \big] \\ &= \ICe(\pi_1'| X_1, X_2) + \sum_{i: \mathrm{odd}} I(U_i
\wedge W_1 | X_1, U^{i-1}) \\
&~~~ + \sum_{i: \mathrm{even}} I(U_i \wedge W_2
| X_2, U^{i-1})
\end{align*}
and
\begin{align*}
\lefteqn{ \ICi(\pi_1| (X_1, W_1), (X_2, W_2)) } \\ 
&= I(U^r \wedge X_1,W_1 | X_2, W_2) + I(U^r \wedge X_2, W_2 | X_1, W_1) \\ 
&= \sum_{i:
  \mathrm{odd}} \big[ I(U_i \wedge X_1, W_1 | U^{i-1}) - I(U_i \wedge X_2, W_2 | U^{i-1}) \big] \\ 
  &~~~ + \sum_{i: \mathrm{even}} \big[
  I(U_i \wedge X_2, W_2 |U^{i-1}) - I(U_i \wedge X_1, W_1 | U^{i-1}) \big] \\ 
  &= \sum_{i: \mathrm{odd}} \big[ I(U_i \wedge X_1| U^{i-1})
  - I(U_i \wedge X_2| U^{i-1}) \big]  \\
  &~~~ + \sum_{i: \mathrm{even}} \big[
  I(U_i \wedge X_2|U^{i-1}) - I(U_i \wedge X_1 | U^{i-1}) \big]
\\ &~~~ + \sum_{i: \mathrm{odd}} \big[ I(U_i \wedge W_1 | X_1,
  U^{i-1}) - I(U_i \wedge W_2 | X_2, U^{i-1}) \big] \\ &~~~ + \sum_{i:
  \mathrm{even}} \big[ I(U_i \wedge W_2 | X_2, U^{i-1}) - I(U_i \wedge
  W_1 | X_1, U^{i-1}) \big] \\ 
  &= \ICi(\pi_1'| X_1, X_2) + \sum_{i:
  \mathrm{odd}} I(U_i \wedge W_1 | X_1, U^{i-1}) \\
  &~~~ + \sum_{i:  \mathrm{even}} I(U_i \wedge W_2 | X_2, U^{i-1}),
\end{align*}
where the last identity holds since for odd $i$, we have
\begin{align*}
\lefteqn{ I(U_i \wedge W_2 | X_2, U^{i-1}) } \\
&\le I(U_i, X_1, W_1 \wedge W_2 | X_2, U^{i-1}) \\ 
&= I(X_1, W_1 \wedge W_2 | X_2, U^{i-1}) \\
&~~~ + I(U_i \wedge W_2 | X_1, W_1, X_2, U^{i-1}) \\ 
&= 0,
\end{align*}
and similarly for even $i$, $I(U_i \wedge W_1 | X_1, U^{i-1}) = 0$.
The required bounds \eqref{eq:randomized-deterministic-ICe} and
\eqref{eq:randomized-deterministic-ICi} follow upon setting
\begin{align*}
t = \sum_{i: \mathrm{odd}} I(U_i \wedge W_1 | X_1, U^{i-1}) + \sum_{i:
  \mathrm{even}} I(U_i \wedge W_2 | X_2, U^{i-1}).
\end{align*} 
\qed

\begin{remark} The proof of converse we have presented is very similar to the
  proof for $r=2$ given in \cite{AhlCsi98} and uses a standard recipe
  in network information theory. The exact choice of ``auxiliary''
  random variables $U^r$ that enable our proof is from
  \cite{Kas85}. In contrast, in the computer science literature, the
  standard approach has been to ``embed'' a single instance of a
  problem in an $n$-fold instance. Specifically, in the case above,
  the approach is to extract a protocol for generating CR from $(X_1,
  X_2)$ given a protocol for extracting CR from $(X_1^n, X_2^n)$ (see, for instance, ~\cite{BarakBCR13,BraRao14}). \textchangeHT{Our proof can be interpretted similarly as follows. We can view $J$ in our proof above
    as the location where the single input must be fed and the rest of
    the inputs $(X_{11}^{J-1}, X_{21}^{J-1}, X_{1(J+1)}^n,
    X_{2(J+1)}^n)$ can be sampled from private and shared randomness in the manner of~\cite[Theorem 3.17]{BraRao14}. Our
    proof shows that we can find random variables $U_1, ..., U_r$ that
    constitute an interactive communication protocol for single inputs
    with external and internal information costs equal to $(1/n)$
    times the information costs of the original protocol $\Pi$ (see
    Section~\ref{s:comm_prot} for the definition of information
    cost).}
\end{remark}

\begin{remark} The proof of achievability is, in essence, from \cite{AhlCsi98}; the
  extension to higher number of rounds is straightforward. While the
  arguments have been presented in an asymptotic form which uses
  typical sets, we can use an information spectrum approach to define
  typical sets and give single-shot arguments \cite{Han03}. Such
  arguments were given, for instance, in \cite{RenWol03, RenWol05,
    Ren05}. The challenge lies in analyzing and establishing
  optimality (in an appropriate sense) of the resulting single-shot
  bounds.
\end{remark}
%%%%%%%%%%%%%%%%%%%%%
\paragraph{Shape of $C(R)$}
We first examine the shape of the function $f(R)$ on the right-side
of~\eqref{e:CR_char_ic_d}.  For a fixed number of rounds $r$, denote
by $C_r^d(R)$ the maximum rate of common randomness that can be
generated using $r$-round deterministic protocols. It is easy to see
that $C_r^d(R)$ is a nondecreasing function of $R$. Also, it can be
argued using a time-sharing argument that $C_r(R)$ is concave in
$R$. Therefore, $C^d(R) = \sup_{r}C_r^d(R)$ must be concave and
nondecreasing function of $R$ as well, and so must be the right-side
of~\eqref{e:CR_char_ic_d}. Note that we can directly verify these
properties by analysing $f(R)$ instead of using the operational
definition of $C^d(R)$, but we find the proof above more illuminating.

Note that since $f(R)$ is a nonnegative, concave, and nondecreasing
function of $R$, if $f(R^\prime)\geq R^\prime $, then $f(R)\geq R$ for
every $R\leq R^\prime$. Furthermore, for $R^\prime = H(X_1|X_2)$, we
can see by setting $\pi$ as the one round protocol with $\Pi_1=X_1$
that $f(R^\prime)\geq H(X_1)\geq R^\prime$. Thus, $f(R)\geq R$ for all
$R\leq H(X_1|X_2)$. This further implies that the slope $f^\prime(R)$
of $f(R)$ is greater than $1$ for every $R\leq R^\prime$. Denote by
$R^*$ the least $R$ for which $f^\prime(R)$ equals $1$.   

We claim that for $R\leq R^*$, $C(R)=f(R)$. Indeed, since $f$ is
concave, $f^\prime(R)\geq 1$ for $R\leq R^*$, and so, $f(R-t)+t\leq
f(R)$ for every $t$, which yields the claim
by~\eqref{e:CR_char_ic_r}. 
Note that using the same arguments as above, $C(R)$, too, is a concave and
nondecreasing function of $R$. Further, since $C(R)$ equals $f(R)$ for
$R\leq R^*$,  $R^*$ must also be the least $R$ for which the slope of
$C(R)$ equals $1$. We have thus characterized the shape of $C(R)$ for
$R\leq R^*$: It is a concave increasing function with slope at least
$1$. 

It remains to characterize the shape of $C(R)$ for $R>R^*$. For that, noting that
\[
\ICe(\pi|X_1, X_2)-\ICi(\pi|X_1, X_2)\leq I(X_1\wedge X_2),
\]
we have $f(R)\leq g(R):=I(X_1\wedge X_2)+R$. Therefore, using~\eqref{e:CR_char_ic_r}, $C(R)\leq \sup_{t\geq 0}g(R-t)+t= g(R)$. Also, graphs of both $f(R)$ and $g(R)$ pass through the point $(H(X_1|X_2), H(X_1))$, whereby $R^*$ is also the least $R$ for which $f(R)=g(R)$. Thus, for $R>R^*$, we can simply attain $C(R)=g(R)$ by using $t=R-R^*$. 

\begin{figure}[t]
\centering \includegraphics[scale=0.4]{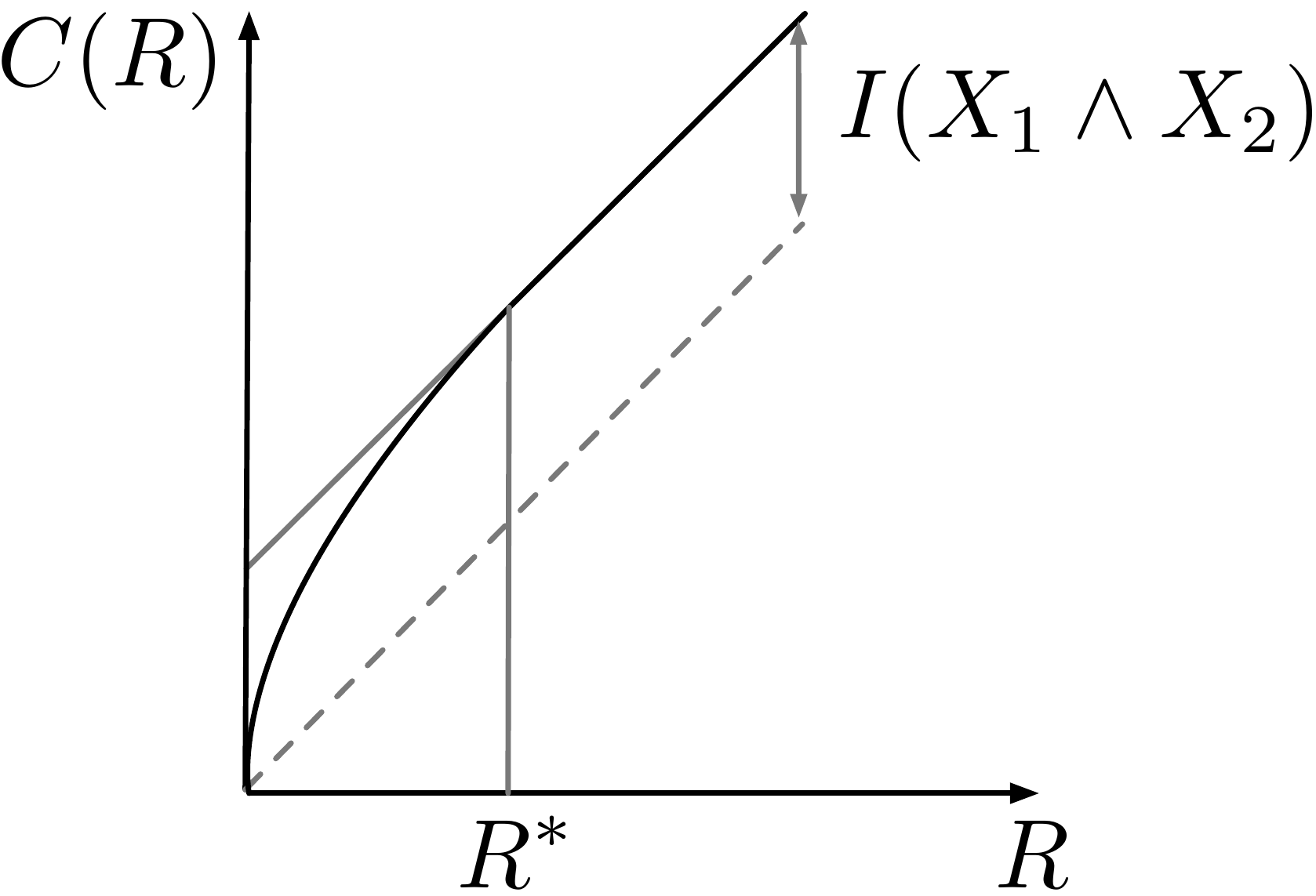}
\caption{$C(R)$ as a function of $R$. $R^\ast$ denotes the point where
  $C(R)$ curve attains the slope $1$.}
\label{f:CR}
\end{figure}

We summarize these observations in the following corollary of
Theorem~\ref{t:CR_cap}; see Figure~\ref{f:CR} for an illustration.  

%%%
\begin{corollary}
For $R\ge 0$ and random variable $(X_1, X_2)$ taking values in a finite
set $\cX_1\times \cX_2$, we have
\begin{align*}
C(R) = \left\{
\begin{array}{ll}
f(R) & \mbox{if } R \le R^* \\
I(X_1 \wedge X_2) + R & \mbox{if } R > R^*
\end{array}
\right..
\end{align*}
\end{corollary}

Another interesting point in the $C(R)$ curve is the slope at $R=0$, namely
the amount of CR that can be generated per bit of communication. A
related problem studied in \cite[Proposition 2]{LiuCuffVerdu17} gives
a characterization of this quantity (with minor changes in the
proof). If we restrict ourselves to $1$-round protocols, $i.e.$, the slope of
$C_1(R)$ for $R=0$, it is given by $1/(1-s^*(X_1,X_2))$ where $s^*(X_1,X_2)$
is defined in Section~\ref{s:maxcorr_hypcont}.

%%%%%%%%%%%%%%%%%%%%%%%%
\subsection{Communication for a fixed-length CR}
The variant of the CR agreement problem that we describe in this
section has been proposed recently, and the literature on it is thin
in comparison with the classic formulation of the previous section. In
fact, most of our treatment is based on a recent
paper~\cite{GR16}. Nevertheless, the techniques used and the results
are interesting. Furthermore, a comprehensive understanding of the CR
problem requires a unified treatment that will yield both variants of
the CR generation problem as special cases.

We are interested in the following quantity.
\begin{definition}
For jointly distributed random variables $(X_1, X_2)$,  $c\geq 0$ is
an $(\ep, r)$-achievable communication length for CR of length $l$ if
there exists an $r$-round private coin protocol $\pi$ of length less
than $c$ and with outputs $(S_1, S_2)$ such that, for a random string
$S$ distributed uniformly over $\{0, 1\}^l$,
\begin{align}
\bPr{S_1 = S_2 = S} \geq 1-\ep.  \nonumber
\end{align}
The infimum over all $(\ep,r)$-achievable communication lengths  for CR
of length $l$ is denoted by $C_{\ep, r}(l|X_1, X_2)$. Further, denote
the infimum of $C_{\ep, r}(l|X_1, X_2)$ over $r$ as $C_\ep(l|X_1,
X_2)$.
\end{definition}
As in the previous section, we are interested in understanding the
behavior of $C_{\ep,r}(l|X_1^n, X_2^n)$ as a function of $l$ and $n$;
the dependence on $r$ and $\ep$ is also of interest, but perhaps more
challenging to study. However, no general result characterizing the
trade-off between the communication length, the CR length, and the
number of samples $n$ is available. We shall focus on the limiting
behavior as $n$ goes to infinity. This represents a fundamental
trade-off between communication and CR lengths, regardless of the
number of samples. In fact, we restrict ourselves to one round protocols and
consider the following quantity:
\[
\Gamma_{\ep}(l) \ed \limsup_{n\rightarrow \infty} \, C_{\ep, 1}(l|
X_1^n, X_2^n).
\]
We review a representative result of the treatment in \cite{GR16}
which focuses on a $BSS(\rho)$
\footnote{The paper \cite{GR16} handles symmetric Gaussian sources as
  well as the binary erasure source, in addition to $BSS(\rho)$
  considered here. The techniques used extend to all the
  distributions, but the resulting bounds may not be sharp.}.
The
notion of CR used in \cite{GR16} is slightly different from the one we
described
%\textchange{\sout{in the previous section}
 above.
%[SW: The requirement $H_{\tt min}(S_1)\geq l$ seems more
%relaxed than the condition that $S$ being uniform on $\{0,1\}^l$]}
In particular, the definition of CR in \cite{GR16} requires that the
estimate $S_1$ of $\cP_1$ equals $S$ and replaces the uniformity of
$S$ on $\{0,1\}^l$ with an alternative requirement of $H_{\tt
  min}(S_1)\geq l$. The key technical difference is that this
definition insists that one of the parties gets the exact CR $S$
(unlike our definition where both parties only obtained estimates of
$S$). The following result of \cite{GR16} applies to this restrictive
notion of CR:
\begin{theorem}\label{t:GR}
Given $(X_1, X_2)$ generated by $BSS(\rho)$, $l>0$, and $\theta>0$,
there exists $\ep \leq 1- 2^{-\theta l - \cO(\log l) }$ such that
\[
\Gamma_{\ep}(l)\leq \big((1-\rho^2)(1-\theta) -
2\rho\sqrt{(1-\rho^2)\theta}\big)\cdot l.
\]
Furthermore, for every $\ep\leq 1-2^{-\theta l}$, it holds that
\[
\Gamma_{\ep}(l)\geq \big((1-\rho^2)(1-\theta) -
2\rho\sqrt{(1-\rho^2)\theta}\big)\cdot l.
\]
\end{theorem}
%{\tt[Question for Madhu: Is the following justification okay?]}  
Note that the result above focuses on very small probability of
agreement and is uninteresting when $\ep$ is required to be close to
$0$. This regime is interesting for historical reasons. Specifically,
the problem of generating CR without communicating goes back to the
classic paper of G\'acs and K\"orner \cite{GacKor73} which shows that
(for indecomposable distributions) no positive rate of CR can be
established without communicating, even when a fixed probability of
error is allowed (see also \cite{MakMak12} for an alternative proof).
A companion result was shown by Witsenhausen~\cite{Wit75} establishing
that the parties cannot even agree on a single bit with non-vanishing
(in observation length $n$) probability of agreement. An extension of
this result appears in \cite{BogdanovMossel11} (see~\cite{ChanMN14}
for further refinements) where it is shown that the largest
probability with which the parties can agree on $\ell$ bits without
communication is exponentially small in $\ell$ and the best exponent
is established. Theorem~\ref{t:GR} is in a similar vein and shows that
the two parties can agree on $\ell$ bits with exponentially small probability
using $c\ell$ bits of communication where the constant $c$ depends on
the exponent of the error. In fact, the scheme in \cite{GR16} is
related to \cite{BogdanovMossel11} -- both papers make the point that
simple schemes where the CR is a subset of observed bits is
suboptimal.

\textchangeHT{We remark that it is of interest to consider the problem of common randomness generation when no communication is allowed. We defer this discussion to Section~\ref{s:simulation}, where we consider the problem of generating correlated random variables without communicating. However, for comparison one can consider a simple scheme for the problem of this section where $\cP_1$ and $\cP_2$ simply declare their observed bits $X_1^\ell$ and $X_2^\ell$ as their respective estimates for CR. This scheme does not use any communication and yields $\ep= 1 -(\frac 12 (1-\rho))^\ell$.}

\paragraph{Outline of achievability proof for Theorem~\ref{t:GR}}
The one-way communication scheme proposed in~\cite{GR16} is very
similar to the one we reviewed in the previous section. Note that the
typical set used in our scheme consists, in essence, of sequences
which are correlated in the sense that they are jointly
typical. However, since the focus here is on a simple BSS, a much
simpler notion of correlation and typical sets can be used. In
particular, we can make do with linear correlation. For simplicity, we
assume that $\cP_1$ and $\cP_2$ observe $n$ iid samples $\{(X_{1i},
X_{2i})\}_{i=1}^n$ from $\{-1,1\}$-valued $X_1$ and $X_2$ which have
the same sign with probability $(1+\rho)/2$.

The CR generation protocol we describe below involves parameters $r>0,
\eta>0$, and $c\in (0,1)$, which will be chosen later. Consider a
random codebook comprising $2^\ell$ vectors $U^n(i,j)$, $1\leq i \leq
2^{c\ell}$ and $1\leq j \leq 2^{(1-c)\ell}$. The vectors $U^n(i,j) =
(U_1(i,j), ..., U_n(i,j))$ are iid for different $(i,j)$, each
consisting of a uniformly generated vector from $\{-1, +1\}^n$. The
protocol for CR generation is very similar to the one above:
\begin{enumerate}
\item[1] $\cP_1$ finds $\Pi_1$ which is the smallest $i$ for which there
  exists a $j$ such that the sequence $u^n = U^n(i,j)$ satisfies
\[
\langle X_1^n\,,\, u^n\rangle\ed\sum_{l=1}^n X_{1l}u_l \geq r\sqrt{n}.
\]
Denote by $Y_1$ the sequence $u^n$.
\item[2] $\cP_1$ sends $\Pi_1$ to $\cP_2$.
\item[3] $\cP_2$ searches for the smallest index $j$ such that $v^n =
  U^n(\Pi_1, j)$ satisfies $\langle X_{2}^n,v^n \rangle \geq (1-\eta)
  r\sqrt{n}$. Denote by $Y_2$ the sequence $U^n(\Pi_1, j)$.
\end{enumerate}
The probability that $Y_1$ and $Y_2$ are the same is bounded below by
the probability that the following hold:
\begin{itemize}
\item[(i)] There exists $(i,j)$ such that for $u^n= U^n(i,j)$,
  $\langle X_1^n\,,\, u^n\rangle\geq r\sqrt{n}$ and $\langle X_2^n\,,
  \,u^n\rangle\geq (1-\eta)r\sqrt{n}$;
\item[(ii)] for every other index pair $(i',j')$, $\langle X_1^n\,,\,
  U^n(i', j')\rangle <r\sqrt{n}$;
\item[(iii)] for every other index $j'$, $\langle X_2^n\,,\,U^n(i,
  j')\rangle <(1-\eta)r\sqrt{n}$.
\end{itemize}
For sufficiently large $n$, we can approximate the random variables
$\langle X_1^n\,,\, u^n\rangle$ and $\langle X_2^n\,,\, u^n\rangle$
with Gaussian random variables using the B\'erry-Ess\'een theorem
(cf.~\cite{Fel71}). In particular, $\langle X_1^n\,,\, u^n\rangle$
can be approximated as a Gaussian random variable with mean $0$ and
variance $n$. Therefore, $\bPr{\langle X_1^n\,,\, u^n\rangle \geq
  r\sqrt n} \approx Q(r)$, where $Q(x)= \bPr{G>x}$ and $G$ is the
standard Gaussian random variable.  Furthermore, given a fixed
realization $X_1^n = x_1^n$ such that $\langle x_1^n\,,\, u^n\rangle =
r'\sqrt{n}$ for some $r'\geq r$, $\langle X_2^n\,,\, u^n\rangle$ can
be approximated as a Gaussian random variable with mean $\rho r'
\sqrt{n}$ and variance $(1-\rho^2)n$. Therefore,
\begin{align*}
\bPr{\langle X_2^n\,,\, u^n\rangle \geq \eta r\sqrt{n}| X_1^n =x_1^n}
&\approx Q\left(\frac{\eta r - \rho r'}{\sqrt{1-\rho^2}}\right)
\\ &\geq Q\left(\frac{(\eta - \rho) r}{\sqrt{1-\rho^2}}\right).
\end{align*}
Thus, the probability of agreement can be seen to be bounded below
roughly by
\[
2^{\ell} Q(r)Q\left(\frac{(\eta - \rho) r}{\sqrt{1-\rho^2}}\right)(1-
q_2 -q_3),
\]
where $q_2$ denotes the probability of event $(ii)$ above not
happening given event $(i)$ and $q_3$ for event $(iii)$. Further,
$q_2\leq 2^\ell Q(r)$ and $q_3 \leq 2^{(1-c)\ell} Q((1-\eta)r)$.
Also, note that for every fixed realization of the codebook,
the probability that $Y_1$ equals $u^n=U^n(i,j)$ is bounded above by
$Q(r)$ which yields $H_{\tt min}(Y_1)\geq \ell$ upon choosing
$Q(r)\approx 2^{-\ell}$. This fixes the value of $r$ as
$\theta(\sqrt{\ell})$; the parameter $c$ is chosen as the minimum
possible so that we can find some $\eta$ that yields the required
probability of agreement.

\begin{remark} The scheme proposed in \cite{GR16} uses a slightly
  different (structured) codebook construction, suggested in
  \cite{BogdanovMossel11}, which renders $Y_1$ uniformly distributed
  over $\{0,1\}^\ell$. Our alternative presentation above is aimed at
  pointing out the similarity between the scheme of \cite{GR16} and
  the standard information theoretic approach used in \cite{AhlCsi98}.
\end{remark}

\paragraph{Outline of converse proof for Theorem~\ref{t:GR}}
We have assumed that the CR $S$ equals $S_1$ and is a function, say
$g$, of $X_1^n$, and $H_{\min}(g(X_1^n)) \geq \ell$. The proof of
lower bound we present remains valid for every $n$ by the
tensorization property of hypercontractivity
(cf.~\eqref{e:tensorization}); we fix $n=1$. For simplicity, we
restrict ourselves to deterministic communication protocols $\pi$ of length $t$.
For a fixed $x_2$ and different possible transcripts of the
communication protocol, $\cP_2$ can output different estimates for the
CR; we denote this set of possible estimated CR values by
$\cZ_{x_2}$. Clearly, $|\cZ_{x_2}|\leq 2^t$ for every $x_2\in
\cX_2$. It can be seen that
\[
1-\ep \leq \bEE{\sum_{z\in \cZ_{X_2}} \bPr{g(X_1)=z|X_2}}.
\]
Using H\"older's inequality,
\begin{align*}
1-\ep&\leq\sum_{z}\bPr{z\in \cZ_{X_2}}^\frac{1}{p^\prime}
\bEE{\bPr{g(X_1)=z|X_2}^p}^{1/p} \\ &\leq\sum_{z}\bPr{z\in
  \cZ_{X_2}}^\frac{1}{p^\prime} \bPr{g(X_1)=z}^{\frac 1q} \\ &\leq
2^{-\frac{\ell}q}\sum_{z}\bPr{z\in \cZ_{X_2}}^\frac{1}{p^\prime},
\end{align*}
where the second inequality holds since $\bPP{X_1X_2}$ is
$(p,q)$-hypercontractive and the third by the assumptions that $H_{\tt
  min}(g(X_1))\geq \ell$. The sum on the right-side of the previous
bound can be bounded further as
\begin{align*}
 \sum_z \bPr{z\in \cZ_{X_2}}^{\frac{1}{p^\prime}} &\leq
 \left(\sum_z\bPr{z\in
   \cZ_{X_2}}\right)^{\frac{1}{p^\prime}}|\cZ|^{\frac 1 p} \\ &=
 \bEE{|Z_{X_2}|}^{\frac{1}{p^\prime}}|\cZ|^{\frac 1 p} \\ &\leq
 2^{\frac{t}{p^\prime}+\frac{\ell}p}
\end{align*}
where the first inequality uses H\"older's inequality and the final
uses $|\cZ_{X_2}|\leq 2^{t}$. Finally, using the assumption $1-\ep
\geq2^{-\theta \ell}$, together with the bounds above we get
\[
t \geq \ell \cdot \left[\frac{p-q - \theta pq}{q(p-1)}\right].
\]
Up to this point, our analysis applies to any distribution $\bPP{X_1
  X_2}$. The best bound will be obtained by maximizing the previous
lower bound for $t$ over all $(p,q)$ such that $\bPP{X_1X_2}$ is
$(p,q)$-hypercontractive. In general, this set of $(p,q)$ is not
explicitly characterized. However, for our case of BSS, we can
optimize over $(p,q)$ characterized in Theorem~\ref{t:BB_ineq} to get
the stated result.

%%%%%%%%%%%%%%%%%%%%%%%%
\subsection{Discussion}
In spite of our understanding of the shape of $C(R)$ described above,
several basic questions remain open. Specifically, it remains open if
a finite round protocol can attain $C(R)$, $i.e.$, for a given
$\bPP{X_1X_2}$ and $r\in\mN$, is $C(R) = C_r(R)$? An interesting
machinery for addressing such questions, which also exhibits its
connection to hypercontractivity constants, has been developed
recently in \cite{LiuCuffVerdu17} (see, also,
\cite{MaIsh13,BravermanGPW12}).  \textchangeSW{In another direction,
  it is an important problem to investigate the dependency of CR rate
  on error $\ep$. The first step toward this direction is to prove a
  {\it strong converse}, i.e., $C(R)= C_\ep(R)$ for all $\ep\in
  (0,1)$, where $C_\ep(R)$ is the supremum of $C_{\ep,r}(R)$ over
  $r\in \mN$.  For $r=1$, the strong converse was proved in
  \cite{LiuCuffVerdu15} by using the blowing-up lemma \cite{CsiKor11}.
  More recently, the strong converse for general $r \in \mN$ was
  proved by using a general recipe developed in \cite{TyaWat18}.
  Finer questions such as the second-order asymptotics of CR length in
  $n$ for a fixed allowed error $\ep$ and bounds for $L_{\ep,r}(c|
  X_1,X_2)$ are open; recently, a technique to derive the second-order
  converse bound using reverse hypercontractivity was developed in
  \cite{LiuHanVer17} (see also \cite[Sec. 4.4.4]{LiuThesis}).}

For the fixed-length CR case, the analysis for $BSS(\rho)$ presented
above extends to $GSS(\rho)$ verbatim. But much remains open. For
instance,
%% while the analysis of \cite{GR16} extends to the case of
%% fixed probability of agreement $\ep< 1/2$, it does not yield a precise
%% formula such as Theorem~\ref{t:GR} for $\ep\geq 1/2$. Furthermore, 
the proof of the lower bound in \cite{GR16} requires the nagging
assumption that the CR is a function of only the observations of
$\cP_1$. It is easy to modify the proof to include local randomness,
but it is unclear how to handle CR which depends on both $X_1^n$ and
$X_2^n$.  Perhaps a more interesting problem is the dependence of
communication on the number of rounds; only a partial result is proved
in~\cite{GR16} in this direction which shows that for binary symmetric
sources interaction does not help if the CR is limited to a function
of observation of one of the parties.  Of course, the holy grail here
is a complete trade-off between the communication length, the CR
length, and the number of samples, which is far from understood. Some
recent progress in this direction includes a sample efficient explicit
scheme for CR generation in \cite{GhaziJayram17} \htc{and examples
  establishing lower bounds for number of round for fixed amount of
  communication per round in~\cite{BafnaGGS18}.  Yet several very
  basic questions remain open; perhaps the simplest to state is the
  following: Does interaction help to reduce communication for CR
  agreement for a binary symmetric source? An interested reader can see~\cite{LiuCuffVerdu17} for further discussion on this question.}

\htc{The results we covered above were only for iid sources. We close this section with references to an interesting recent line of work~\cite{BeigiEG17,BeigiBEG17,BeigiBEG18} that extends the results of~\cite{GacKor73, Wit75} (a two-party extension of) the more general class of Santha-Vazirani sources (see~\cite{SanthaVazirani86}.} 

%%%%%%%%%%%%%%%
\section{Secret key agreement}\label{s:SK_agreement}
We now introduce the secret key (SK) agreement problem which entails
generating a CR that is independent of the communication used to
generate it.
% and, perhaps, another correlated random variable.

\subsection{Secret keys using unlimited communication} \label{subsection:SK-no-constraint}
We start with SK agreement when the amount of communication over the
public channel is not restricted. Parties $\cP_1$ and $\cP_2$
observing $X_1$ and $X_2$, respectively, communicate over a noiseless
%Consider the secret key agreement by two arties $\cP_1$
%and $\cP_2$ using interactive public communication.  Upon observing
%$X_1$ and $X_2$, the parties communicate interactively over a 
public communication channel that is accessible by an eavesdropper,
who additionally observes a random variable $Z$ such that the tuple
$(X_1,X_2,Z)$ has a (known) distribution $\bPP{X_1 X_2 Z}$.

 The parties communicate using a private coin protocol $\pi$ to
 generate a CR $K$ taking values in $\cK$ and with $K_1$ and $K_2$
 denoting its estimates at $\cP_1$ and $\cP_2$, respectively. The CR
 $K$ constitutes an $(\varepsilon,\delta)$-SK of length $\log |\cK|$
 if it satisfies the following two conditions:
\begin{align}
\Pr( K_1 = K_2 =K) &\ge 1 - \varepsilon, \label{eq:reliability-SK}
\\ d(\bPP{K \Pi Z}, \bPP{\mathtt{unif}} \times \bPP{\Pi Z}) &\le
\delta, \label{eq:security-SK}
\end{align}
where $\Pi$ denotes the random transcript of $\pi$ and
$\bPP{\mathtt{unif}}$ is the uniform distribution on $\cK$.  The first
condition \eqref{eq:reliability-SK} guarantees the reliability of the
SK and the second condition \eqref{eq:security-SK} guarantees secrecy.

\begin{definition}
Given $\varepsilon,\delta \in [0,1)$, the supremum over the length
  $\log |\cK|$ of $(\varepsilon,\delta)$-SK is denoted by
  $S_{\varepsilon,\delta}(X_1,X_2|Z)$.
\end{definition}
The only interesting case is when $\varepsilon+\delta <1$.  In fact,
when $\varepsilon + \delta \ge 1$, it can be shown that the parties
can share arbitrarily long SK, i.e.,
$S_{\varepsilon,\delta}(X_1,X_2|Z) = \infty$ \cite[Remark
  3]{TyaWat15}.

%%%%%%%%%%%%%%%%%%%%%%%%%%%
%%%%%%%%%%%%%%%%%%%%%%%%%%%
\paragraph{Achievability techniques} 
Loosely speaking, the SK agreement schemes available in the literature
can be divided into two steps: An {\em information reconciliation}
step where the parties generate CR (which is not necessarily uniform
or secure) using public communication; and a {\em privacy
  amplification} step where a SK independent of the observations of
the eavesdropper, i.e $(Z,\Pi)$, is extracted from the CR.

In the CR generation problem of the previous section, the specific
form of the randomness that the parties agreed on was not critical. 
In contrast, typical
schemes for SK agreement generate CR comprising specific random variables such as $X_1$
or $(X_1, X_2)$. This gives us an analytic handle on the amount of
randomness available for extracting the SK. Clearly, agreeing on
$(X_1, X_2)$ makes available a larger amount of randomness for the
parties to extract an SK. However, this will require a larger amount of
public communication, thereby increasing the information leaked to the
eavesdropper. To shed further light on this tradeoff, we review the
details of the scheme where both parties agree on $X_1$ in the
information reconciliation step.

In this case, $\cP_1$ needs to send a message to $\cP_2$ to enable the
latter to recover $X_1$. This problem was studied first by Slepian and
Wolf in their seminal work~\cite{SleWol73} where they characterized
the optimal asymptotic rate required for the case where $(X_1,X_2^n)$ is iid. This asymptotic
result can be recovered by setting $U=X_1$ and $V$ to be a constant in
\eqref{e:CR2_char}. In a single-shot setup, i.e., $n=1$, the Slepian-Wolf scheme
can be described as follows: $\cP_1$ sends the hash value $\Pi_1 =
F(X_1)$ of observation $X_1$ where $F$ is generated uniformly from from a
$2$-UHF.
%\footnote{Using the random binning is a special case of $2$-UHF.}
Then, $\cP_2$ looks for a unique $x_1$ in a guess-list ${\cal L}_{x_2}
\subseteq \cX_1$ given $X_2 = x_2$ that is compatible with the
received message $\Pi_1$.  A usual choice of the guess-list is the
(conditionally) typical set: ${\cal T}_{\bPP{X_1|X_2}} := \{ (x_1,x_2)
: h_{\bPP{X_1|X_2}}(x_1|x_2) \le t - \gamma\}$, where
$h_{\bPP{X_1|X_2}}(x_1|x_2) = - \log \bP{X_1|X_2}{x_1|x_2}$ is the
conditional entropy density, $t$ is the length of the message sent by
$\cP_1$ and $\gamma \ge 0$ is a slack
parameter.\footnote{\textchangeSW{Unlike the notion of typical set  used
    in the classic information theory textbooks \cite{CsiKor11, CovTho06}, the
    typical set ${\cal T}_{\bPP{X_1|X_2}}$ only involves one-sided
    deviation event of conditional entropy density. Such a typical set
    is more convenient in non-asymptotic analysis using the information
    spectrum method \cite{Han03}.}}  \textchangeSW{In this case, the
  size of guess-list can be bounded as $|\{ x_1 : (x_1,x_2) \in {\cal
    T}_{\bPP{X_1|X_2}} \}| \le 2^{t-\gamma}$.  Since $\cP_2$'s recovery
  $\hat{X}_1$ may disagree with $X_1$ when $(X_1,X_2)$ is not included
  in the typical set or there exists $\hat{x}_1 \neq X_1$ such that
  $(\hat{x}_1, X_2) \in {\cal T}_{\bPP{X_1|X_2}}$ and
  $F(X_1)=F(\hat{x}_1)$,} the error probability is bounded as (eg.~see
\cite[Section 7.2]{Han03} for details)
\begin{align*}
\Pr\bigg( X_1 \neq \hat{X}_1 \bigg) \le \bP{X_1X_2}{{\cal
    T}_{\bPP{X_1|X_2}}^c} + 2^{-\gamma}.
\end{align*}
When the observations are iid, by the law of large numbers, the error
probability converges to $0$ as long as the message rate is larger
than the conditional entropy, i.e.,
\begin{align} \label{eq:IR-bound}
t \ge n(H(X_1|X_2)+\nu)
\end{align}
for some $\nu > 0$.

Once the parties agree on $X_1$, the parties generate a SK from $X_1$
by using $2$-UHF.  By an application of the leftover hash lemma
\textchangeSW{with $X_1$ and $\Pi_1$ playing the role of $X$ and $V$
  in Theorem \ref{t:LHL}}, a SK satisfying \eqref{eq:security-SK} can
be generated as long as
\begin{align*}
\log |\cK| \le H^{(\delta-\eta)/2}_{\min}(\bPP{X_1Z}|Z) - t - \log
(1/4\eta^2)
\end{align*}
for some $0 \le \eta < \delta$. A common choice of smoothing is a
truncated distribution
\begin{align*}
\tilde{\mathrm{P}}_{X_1Z}(x_1,z) = \bP{X_1Z}{x_1,z} \mathbf{1}\big[
  h_{\bPP{X_1|Z}}(x_1|z) > r \big]
\end{align*}
for some threshold $r$. \textchangeSW{Then, we have
  $H_{\min}(\tilde{\mathrm{P}}_{X_1Z}|Z) \ge r$.  By adjusting the
  threshold $r$ so that $\tilde{\mathrm{P}}_{X_1Z} \in {\cal
    B}_{(\delta-\eta)/2}(\bPP{X_1Z})$, we have }
\begin{align*}
\lefteqn{ H^{(\delta-\eta)/2}_{\min}(\bPP{X_1Z}|Z) } \\
&\ge \sup\big\{ r : \Pr\big( h_{\bPP{X_1|Z}}(X_1|Z) \le r \big) \le \delta - \eta \big\}.
\end{align*}
When the observations are iid, by the law of large numbers, a secret
key with vanishing security parameter $\delta$ can be generated as
long as
\begin{align} \label{eq:PA-bound}
\log |\cK| \le n (H(X_1|Z) - \nu)
\end{align}
for some $\nu>0$.

By combining the two bounds \eqref{eq:IR-bound} and
\eqref{eq:PA-bound}, for vanishing $\ep,\delta$, we can conclude that
$(\ep,\delta)$-SK of length roughly $n[H(X_1|Z) - H(X_1|X_2)]^+$ can be
generated, where $[t]^+ = \max\{ t, 0\}$.

Alternatively, the parties can agree on $(X_1, X_2)$ in the
information reconciliation step. This is enabled by first
communicating $X_1$ to $\cP_2$ using the scheme outlined above and
then $X_2$ to $\cP_1$ using a standard Shannon-Fano code.\footnote{We can also use the Slepian-Wolf coding for
communication from $\cP_2$ to $\cP_1$ as well; however, since $\cP_2$ has already recovered $X_1$, it is more efficient
to use a standard Shannon-Fano code.} For iid
observations, this will require $n(H(X_1|X_2)+ H(X_2|X_1))$ bits of
communication $(\Pi_1,\Pi_2)$. Furthermore, by using the leftover hash
lemma \textchangeSW{with $(X_1,X_2)$ and $(\Pi_1,\Pi_2)$ playing the
  role of $X$ and $V$ in Theorem \ref{t:LHL}}, we will be able to
extract a SK of length roughly $n[H(X_1X_2|Z) - H(X_1|X_2)-H(X_2|X_1)]^+$, which in general is not comparable with the
rate attained in the previous scheme. However, when $Z$ is constant
the two rates coincide. This observation was made first in
\cite{CsiNar04} where the authors used the latter CR generation, termed
attaining {\em omniscience}, for multiparty SK agreement. In fact, a
remarkable result of \cite{CsiNar04} shows that, when $Z$ is constant, the omniscience leads to an
optimal rate SK even in the multiparty setup with arbitrary number of parties.

%%%%%%%%%%%%%%%%%%%%%%%%%%%%
\paragraph{Converse techniques} 
 Moving now to the converse bounds, we begin with a simple bound based
 on Fano's inequality. For special cases, this bound is asymptotically
 tight for iid observations, when $\ep$ and $\delta$ vanish to $0$.
\begin{theorem} \label{theorem:basic-converse}
For every $0\le \ep,\delta<1$ with $0 \le \ep + \delta < 1$, it holds
that
\begin{align*}
S_{\ep,\delta}(X_1,X_2|Z)  \le \frac{I(X_1\wedge X_2|Z) + h(\ep) +
  h(\delta)}{1-\ep-\delta}
\end{align*}
\end{theorem}
The proof of Theorem \ref{theorem:basic-converse} entails two steps.
First, by using Fano's inequality and the continuity of the Shannon
entropy, an $(\ep,\delta)$-SK with estimates $K_1,K_2$ for
$\cP_1,\cP_2$, respectively, satisfies
\begin{align} \label{eq:mediate-basic-converse}
\log |\cK| \le \frac{I(K_1 \wedge K_2| Z,\Pi) + h(\ep) + h(\delta) }{
  1- \ep-\delta}.
\end{align}
The claimed bound then follows by using the monotonicity of
correlation property of interactive communication (cf.~\eqref{eq:monotonicity-correlation}).

Next, we present a stronger converse bound which, in effect, replaces
the multiplicative loss of $1/(1-\ep-\delta)$ by an additive $\log
1/(1-\ep-\delta)$. The bound relies on a quantity related to binary
hypothesis testing; we review this basic problem first. For
distributions $\mathrm{P}$ and $\mathrm{Q}$ on $\cX$, a test is
described by a (stochastic) mapping $\mathrm{T}:\cX \to
\{0,1\}$. Denote by $\dP[T]$ and $\dQ[T]$, respectively, the size of
the test and the probability of missed detection, $i.e.$,
\begin{align*}
\mathrm{P}[\mathrm{T}] &= \sum_x \mathrm{P}(x) \mathrm{T}(0|x),
\\ \mathrm{Q}[\mathrm{T}] &= \sum_x \mathrm{Q}(x) \mathrm{T}(0|x).
\end{align*}
Of pertinence is the minimum
probability of missed detection for tests of size greater than
$1-\ep$, $i.e.$,
\begin{align*}
\beta_\ep(\mathrm{P}, \mathrm{Q}) := \inf_{\mathrm{T}:
  \mathrm{P}[\mathrm{T}] \ge 1 - \ep} \mathrm{Q}[\mathrm{T}],
\end{align*}
When $\bPP{}^n$ and $\bQQ{}^n$ are iid distributions, Stein's lemma
(cf.~\cite{CsiKor11}) yields
\begin{align*}
\lim_{n\to\infty} - \frac{1}{n} \log \beta_\ep(\bPP{}^n, \bQQ{}^n) =
D(\bPP{} \| \bQQ{}),~\forall 0 < \ep < 1.
\end{align*} 
The following upper bound for SK length from \cite{TyaWat14, TyaWat15}
involves $\beta_\ep$. Heuristically, it relates the length of SK
to the difficulty in statistically distinguishing the
distribution $\bPP{X_1X_2Z}$ from a ``useless'' distribution in which the
observations of the parties are independent when conditioned on the
observations of the eavesdropper.
\begin{theorem} \label{theorem:CIT}
Given $0 \le \ep,\delta < 1$ and $0 < \eta < 1- \ep-\delta$, it holds
that
\begin{align*}
\lefteqn{ S_{\ep,\delta}(X_1,X_2|Z) } \\
&\le - \log \beta_{\ep+\delta+\eta}(\bPP{X_1
  X_2Z}, \bQQ{X_1 X_2Z}) + 2 \log (1/\eta)
\end{align*}
for any $\bQQ{X_1 X_2Z}$ satisfying $\bQQ{X_1 X_2Z}=
\bQQ{X_1|Z}\bQQ{X_2|Z}\bQQ{Z}$.
\end{theorem}
We outline the proof of Theorem \ref{theorem:CIT}.  The first
observation is that the reliability and secrecy conditions for an
$(\ep,\delta)$-SK impliy (and are roughly equivalent to) the following
single condition:
\begin{align} \label{eq:combined-criterion}
d(\bPP{K_1 K_2 Z \Pi }, \mathrm{P}^{(2)}_{\mathtt{unif}}\times \bPP{Z
  \Pi }) \le \ep + \delta,
\end{align}
where
\begin{align*}
\mathrm{P}^{(2)}_{\mathtt{unif}}(k_1,k_2) :=
\frac{\indicator(k_1=k_2)}{|\cK|}.
\end{align*}
Next, note that for a distribution $\bQQ{X_1X_2Z}$ satisfying
$\bQQ{X_1 X_2Z} = \bQQ{X_1|Z}\bQQ{X_2|Z}\bQQ{Z}$, the property in \eqref{eq:monotonicity-correlation} implies that the
distribution $\bQQ{K_1K_2Z\Pi}$ of the ``view of the protocol'' equals
the product $\bQQ{K_1 |Z \Pi } \bQQ{K_2|Z \Pi } \bQQ{Z \Pi }$.

\textchangeSW{We will show that the two observations above yield the
  following bound:} For any $(K_1,K_2)$ satisfying
\eqref{eq:combined-criterion} and any $\bQQ{K_1K_2 Z \Pi }$ of the
form $\bQQ{K_1 |Z \Pi } \bQQ{K_2|Z \Pi } \bQQ{Z \Pi }$, it holds that
\begin{align} \label{eq:mediate-bound}
\log |\cK| \le - \log \beta_{\ep+\delta+\eta}(\bPP{K_1 K_2 Z \Pi },
\bQQ{K_1 K_2 Z \Pi }) + 2 \log (1/\eta).
\end{align}
The bound of Theorem \ref{theorem:CIT} can be obtained by using the
``data-processing inequality'' for $\beta_\ep(\dP\circ W, \dQ \circ W)
\leq \beta_\ep(\dP, \dQ)$, where $(\dP \circ W)(y) = \sum_x \dP(x)W(y|x)$.

For proving \eqref{eq:mediate-bound}, we prove a reduction of
independence testing to SK agreement. In particular, we use a given SK
agreement protocol to construct a hypothesis test between $\bPP{K_1
  K_2 Z \Pi }$ and $\bQQ{K_1 K_2 Z \Pi }$. The constructed test is a
standard likelihood-ratio test, but instead of the likelihood ratio
test between $\bPP{K_1 K_2 Z \Pi}$ and $ \bQQ{K_1 K_2 Z \Pi }$, we
consider the likelihood ratio of
$\mathrm{P}^{(2)}_{\mathtt{unif}}\times \bPP{Z \Pi }$ and $ \bQQ{K_1
  K_2 Z \Pi }$. Specifically, the acceptance region for our test is
given by
\begin{align*}
{\cal A} := \bigg\{ (k_1,k_2, z,\tau) : \log
\frac{\mathrm{P}^{(2)}_{\mathtt{unif}}(k_1,k_2)}{\bQ{K_1
    K_2|Z\Pi}{k_1,k_2|z,\tau}} \ge \lambda \bigg\},
\end{align*}
where $\lambda = \log |{\cal K}| - 2 \log (1/\eta)$.  Then, a
change-of-measure argument of bounding probabilities under $\bQQ{K_1K_2|Z\Pi}$ by those under $\mathrm{P}^{(2)}_{\mathtt{unif}}$
yields the following bound on the type II
error probability:
\begin{align*}
\bQ{K_1 K_2 Z \Pi}{{\cal A}} \le \frac{1}{|\cK| \eta^2}.
\end{align*}
On the other hand, the security condition
(cf.~\eqref{eq:combined-criterion}) yields a bound on the type I
error probability:
\begin{align*}
\lefteqn{ \bP{K_1K_2 Z \Pi}{{\cal A}^c} } \\
&\le d(\bPP{K_1 K_2 Z \Pi},
\mathrm{P}^{(2)}_{\mathtt{unif}}\times \bPP{Z \Pi }) +
\mathrm{P}^{(2)}_{\mathtt{unif}}\times \bPP{Z \Pi }({\cal A}^c) \\ 
&\le \ep + \delta +
\mathrm{P}^{(2)}_{\mathtt{unif}}\times \bPP{Z \Pi }({\cal A}^c),
\end{align*}
where the first inequality follows from the definition of the
variational distance. \htc{For the second term, note that by the
  definition of the set $\cA$ for any $(k, k, z,\tau)\in \cA^c$}
\[
1\leq |\cK|^2\eta^2\bQ{K_1, K_2|Z\Pi}{k,k|z, \tau},
\]
which yields
\begin{align*}
\lefteqn{ \sum_{(k_1, k_2): (k_1,k_2,z,\tau)\in \cA^c}
\frac{\indicator(k_1=k_2)}{|\cK|} } \\
&\leq \eta \sum_{k: (k,k,z,\tau)\in
  \cA^c}\sqrt{\bQ{K_1K_2|Z\Pi}{k,k|z, \tau}} \\ &\leq \eta
\sqrt{\sum_k \bQ{K_1|Z\Pi}{k|z,\tau} \sum_k \bQ{K_2|Z\Pi}{k|z,\tau}} \\
&=\eta,
\end{align*}
\htc{where the last inequality uses the product form of
  $\bQQ{K_1K_2|Z\Pi}$; \eqref{eq:mediate-bound} follows by combining
  the bounds above.}

The proofs of the two converse bounds presented above follow roughly the
same template: First, use the reliability and secrecy conditions to
bound the length of an SK by a measure of correlation (see, for
instance, \eqref{eq:mediate-basic-converse} and
\eqref{eq:mediate-bound}); and next, use properties of interactive
communication and a data-processing inequality for the measure of
correlation used to get the final bounds that entail only the original
distribution. The {\em monotone approach} (cf.~\cite{CerMasSch02,
  RenWol05, GohAna10}) for proving converse bounds is an abstraction
of these two steps. More specifically, the monotone approach seeks to
identify measures of correlation that satisfy properties that enable
the aforementioned bounds. This allows us to convert a problem of
proving converse bounds to that of verifying these properties for
appropriately chosen measures of correlation. The two bounds above, in
essence, result upon choosing $I(X_1\wedge X_2|Z)$ and
$\min_{\bQQ{X_1X_2Z}}\beta_\ep(\bPP{X_1X_2Z}, \bQQ{X_1X_2Z})$ as those
measures of correlation. This approach is  used not only in the SK
agreement analysis, but is also relevant in other related problems of
``generating correlation" such as multiparty secure computation
\cite{WolWul08,PraPra14} and entanglement distillation
\cite{DonHorRud02} in quantum information theory.
%%%%%%%%%%%%%%%%%%%%%%%%%%%%%
%%%%%%%%%%%%%%%%%%%%%%%%%%%%%
\paragraph{Secret key capacity} \label{subsec:asymptotic-SK}
When the observations of the parties and the eavesdropper comprise an
iid sequences $X_1^n, X_2^n, Z^n$, we are interested in examining how
the maximum length of a SK that the parties can generate grows with
$n$.  The first order asymptotic term is the {\it secret key capacity}
defined as
\begin{align} \label{eq:secret-key-capacity-ep-delta}
C^{\mathtt{sk}}_{\varepsilon,\delta}(X_1,X_2|Z) :=
\liminf_{n\to\infty} \frac{1}{n} S_{\varepsilon,\delta}(X_1^n,
X_2^n|Z^n)
\end{align}
and\footnote{\textchangeSWtwo{Note that the secret key capacity is defined for vanishing
error and secrecy instead of exactly zero-error and zero-secrecy. The zero-error secret key capacity was studied in \cite{OrlWig93}.}}
\begin{align} \label{eq:secret-key-capacity}
C^{\mathtt{sk}}(X_1,X_2|Z) := \lim_{\varepsilon,\delta \to 0}
C^{\mathtt{sk}}_{\varepsilon,\delta}(X_1,X_2|Z).
\end{align}
Evaluating the achievability bound derived above for iid observations,
we get
\begin{align} \label{eq:standard-lower-bound-SK}
C^{\mathtt{sk}}(X_1,X_2|Z) &\ge H(X_1|Z) - H(X_1|X_2) \\ &= I(X_1
\wedge X_2) - I(X_1 \wedge Z).\nonumber
\end{align}

For the converse bound, we can evaluate the single-shot result of
Theorem \ref{theorem:CIT} using Stein's Lemma (cf.~\cite{CsiKor11})
to obtain the following:
\begin{theorem} \label{theorem:basic-upper-bound-SK}
For $\varepsilon,\delta \in (0,1)$ with $\varepsilon+\delta < 1$, we
have
\begin{align}
C^{\mathtt{sk}}_{\varepsilon,\delta}(X_1,X_2|Z) \le I(X_1 \wedge X_2
|Z).  \nonumber
\end{align}
The inequality holds with equality when $X_1$, $X_2$, and $Z$ form
Markov chain in any order.\footnote{When $X_1$, $Z$, and $X_2$ form
  Markov chain in this order, the SK capacity is $0$.} In particular,
when $Z$ is constant, the SK capacity is
\begin{align*}
C^{\mathtt{sk}}_{\varepsilon,\delta}(X_1,X_2) = I(X_1 \wedge X_2).
\end{align*}
\end{theorem}
The remarkable feature of this bound is that it holds for every fixed
$0<\ep, \delta$ such that $\ep+\delta<1$.  Note that if we used
Theorem \ref{theorem:basic-converse} in place of Theorem
\ref{theorem:CIT}, we would have only obtained a matching bound when
$\ep$ and $\delta$ vanish to $0$, namely a weak converse
result. Instead, using Theorem \ref{theorem:CIT} leads to a
characterization of capacity with a strong converse. When the Markov
relation $X_1\mc X_2 \mc Z$ holds, \eqref{eq:standard-lower-bound-SK}
and the identity $I(X_1 \wedge X_2|Z) = H(X_1|Z) - H(X_1|X_2)$ yield
the claimed characterization of capacity. 
\textchangeSWtwo{In such a case, Theorem \ref{theorem:basic-converse} claims that
the SK capacity does not depend on $\ep$ and $\delta$ as long as $\ep+\delta<1$. 
%Although $C^{\mathtt{sk}}_{\varepsilon,\delta}(X_1,X_2|Z) \ge C^{\mathtt{sk}}(X_1,X_2|Z)$ holds trivially, it is not clear if the equality hold in general. 
An interesting question is if $C^{\mathtt{sk}}_{\varepsilon,\delta}(X_1,X_2|Z)$ only depends on $\ep+\delta$ in general. In fact, a weaker claim can be proved:
by using an argument to convert a high reliability protocol to a zero-secrecy protocol \cite[Proposition 4]{HayTyaWat14ii},
$C^{\mathtt{sk}}_{\varepsilon^\prime,\delta^\prime}(X_1,X_2|Z) \ge C^{\mathtt{sk}}_{\varepsilon,\delta}(X_1,X_2|Z)$ holds
as long as $\ep^\prime \ge \ep$ and $\ep^\prime + \delta^\prime \ge \ep+\delta$.}

\textchangeSWtwo{In general, a characterization of SK capacity is an open problem. 
It is also difficult to decide whether the SK capacity is positive or not,
but there is a recent progress on this problem in \cite{GohGunKra17}.
In fact, it can be proved that the SK capacity is positive if and only if one-bit secret key
with $\ep+\delta< \frac{3-\sqrt{5}}{8}$ in \eqref{eq:combined-criterion} can be generated.}

For the special case when we restrict ourselves to SK agreement protocols using
\textchangeSW{one-round communication, say from $\cP_1$ to $\cP_2$}, a
characterization of SK capacity $C^{\mathtt{sk}}_1(X_1,X_2|Z)$ was
given in \cite[Theorem 1]{AhlCsi93}.
\begin{theorem} \label{theorem:one-way-formula-SK}
For a pmf $\bPP{X_1X_2Z}$,
\begin{align*}
C^{\mathtt{sk}}_1(X_1,X_2|Z) = \max_{U,V}\big[ I(V \wedge X_2|U) - I(V
  \wedge Z|U) \big],
\end{align*}
where the maximum is taken over auxiliary random variables
$(U,V)$ satisfying $U \mc V \mc X_1 \mc (X_2,Z)$. Moreover, it can be
assumed that $V = (U,V^\prime)$ and both $U$ and $V^\prime$ have range
of size at most $|\cX_1|$.
\end{theorem}
Returning to the general problem of unrestricted interactive protocols,
we note that 
some improvements for the upper
bound of Theorem \ref{theorem:basic-upper-bound-SK} are known
\cite{MauWol99, RenWol03, GohAna10}. The following bound is roughly
the state-of-the-art.
\begin{theorem}\cite{GohAna10} \label{theorem:GA-bound-SK}
For a pmf $\bPP{X_1X_2Z}$,
\begin{align}
C^{\mathtt{sk}}(X_1, X_2|Z) \le \inf \big[I(X_1 \wedge X_2 |U) +
  I(X_1, X_2 \wedge U|Z)\big],
\end{align}
where the infimum is taken over the conditional distributions
$\bPP{U|X_1X_2Z}$.
\end{theorem}
The bound of Theorem~\ref{theorem:GA-bound-SK} is derived by an
application of the monotone approach for proving converse bounds
described earlier. A single-shot version of the asymptotic converse
bound in Theorem \ref{theorem:GA-bound-SK} can be obtained using the
bound in Theorem~\ref{theorem:CIT}; see \cite[Theorem 7]{TyaWat15} for
details.

Moving to the achievability part, the lower bound of
\eqref{eq:standard-lower-bound-SK} is not tight and can be improved in
general.  In fact, even when the right-side of
\eqref{eq:standard-lower-bound-SK} is negative, a positive rate was
shown to be possible in \cite{Mau93}. Interestingly, the scheme
proposed in \cite{Mau93} uses interactive communication\footnote{The
  benefit of feedback in the context of the wiretap channel was pointed
  out in \cite{CheongThesis}.} for information reconciliation, as
opposed to one-way communication scheme that led to
\eqref{eq:standard-lower-bound-SK}. This information reconciliation
technique, termed ``advantage distillation,'' has been studied further
in \cite{MauWol99, MurYosDav06,MurYosDav06b}.  The best known lower
bound for SK capacity appears in \cite{GohAna10} which implies that
\begin{align*}
\lefteqn{ C^{\mathtt{sk}}(X_1, X_2|Z) } \\
&\geq \sup_{\pi} \ICe(\pi| X_1, X_2) -
\ICi(\pi|X_1, X_2) - I(\Pi\wedge Z)
\end{align*}
where the supremum is over all private coin protocols $\pi$.  Note
that the form above is very similar to the one presented in
Theorem~\ref{t:CR_cap} and shows that the rate of the SK is obtained by
subtracting from the rate of CR the rate of communication and the
information leaked to the eavesdropper. In fact, the actual bound
in~\cite{GohAna10}, which we summarize below, is even stronger and
allows us to condition on any initial part of the transcript, thereby
recovering Theorem~\ref{theorem:one-way-formula-SK} as a special case.

\begin{theorem}\label{t:interactive_lb}
For a pmf $\bPP{X_1X_2Z}$,
\begin{align*}
  \lefteqn{ C^{\mathtt{sk}}(X_1, X_2|Z) } \\
  &\geq \sup_{\pi} I(\Pi\wedge X_1, X_2\mid
  \Pi^t) - I(\Pi\wedge X_1|X_2, \Pi^t) \\
  &~~~ - I(\Pi\wedge X_2|X_1, \Pi^t) - I(Z\wedge \Pi \mid \Pi^t), 
\end{align*}
where the supremum is over all private coin protocols $\pi$ and all
$t$ less than $|\pi|$ and $\Pi^t$ denotes the transcript in the first
$t$-rounds of communication.
\end{theorem}
The communication protocol attaining the bound above uses multiple
rounds of interaction for information reconciliation. It was shown in
\cite{GohAna10} that the lower bound of Theorem~\ref{t:interactive_lb}
can strictly outperform\footnote{In \cite{WatMatUyeKaw07}, a SK
  agreement protocol using multiple rounds of communication for
  information reconciliation was proposed in the context of quantum
  key distribution, and it was demonstrated that that the multiround
  communication protocol can outperform the protocol based on
  advantage distillation (cf.~\cite{Mau93}).}  the one-way SK
capacity characterized in
Theorem~\ref{theorem:one-way-formula-SK}. \textchangeSW{However, the
  bound is not tight in general, even for binary symmetric sources
  \cite{GohGunKra17}.}

\textchangeSW{Traditional notion of security used in the information theory literature
is that of {\em weak secrecy} where information leakage is defined using mutual information normalized by block-length $n$ (see, for instance, \cite{Wyn75ii,Mau93,AhlCsi93}). In the past few decades, motivated by cryptography applications a more stringent notion of security with
  unnormalized mutual information, termed {\em strong secrecy}, has
  become popular. In fact, it turns out that the SK capacity under both notions of security coincide \cite{MauWol00}.  In this
  paper, we have employed the security definition with variational
  distance since it is commonly used in the cryptography literature and
   is consistent with other problems treated in this paper. In
  the i.i.d. setting, since the  protocols reviewed above guarantee
  exponentially small secrecy in the variational distance, those
  protocols guarantee strong secrecy as well.}

%%%%%
\subsection{Secret key generation with communication constraint}\label{s:SK_comm_constraint}
In the previous section, there was no explicit constraint placed on
the amount of public communication allowed. Of course, there is an
implicit constraint implied by the secrecy condition. Nevertheless,
not having an explicit constraint on communication facilitated 
schemes where the parties communicated as many bits as required to agree on
$X_1$ or $(X_1,X_2)$ and accounted for the communication rate in the
previous amplification step. We now consider a more demanding problem
where the parties are required to generate a SK using private coin
protocols $\pi$ of length $|\pi|$ no more than $c$.
\begin{definition}
Given $\varepsilon,\delta \in [0,1)$ and $c>0$, the supremum over the
  length $\log |\cK|$ of $(\varepsilon,\delta)$-SK that can be
  generated by a $r$-rounds protocol $\pi$ with $|\pi| \le c$ is
  denoted by $S_{r,\varepsilon,\delta}(X_1,X_2|Z;c)$.
\end{definition}
Given a rate $R>0$, the rate-limited SK capacity is defined as
follows:
\begin{align}
C^\mathtt{sk}_{r,\varepsilon,\delta}(R) := \liminf_{n\to\infty}
\frac{1}{n} S_{r,\varepsilon,\delta}(X_1^n,X_2^n|Z^n;nR)
\end{align}
and
\begin{align}
C_r^\mathtt{sk}(R) := \lim_{\varepsilon,\delta \to 0}
C^{\mathtt{sk}}_{r,\varepsilon,\delta}(R).
\end{align}
The problem of SK agreement using rate-limited communication was first
studied in \cite{CsiNar00}. The general problem of characterizing
$C^{\mathtt{sk}}_r(R)$ remains open. However, a complete
characterization is available for two special cases: First, the rate-limited SK capacity $C^{\mathtt{sk}}_1(R)$ when we restrict ourselves to one-way
communication protocols is known. Second, an exact expression for
$C^{\mathtt{sk}}_r(R)$ is known when $Z$ is constant
(cf.~\cite{Tya13, LiuCuffVerdu17}).  Specifically, for the rate-limited SK capacity with one-way communication,
the following result holds.
\begin{theorem}[\cite{CsiNar00}] \label{theorem:SK-R-one-way}
The rate-limited SK capacity using one-way communication protocols is
given by
\begin{align*}
C^{\mathtt{sk}}_1(R) = \max_{U,V}\big[ I(V \wedge X_2|U) - I(V \wedge
  Z|U) \big],
\end{align*}
where the maximization is taken over auxiliary random variables
$(U,V)$ satisfying $U \mc V \mc X_1 \mc (X_2,Z)$ and
\begin{align*}
I(V \wedge X_1|U) - I(V \wedge X_2|U) \le R.
\end{align*}
Moreover, it may be assumed that $V = (U,V^\prime)$ and both $U$ and
$V^\prime$ have range of size at most $|\cX_1|+1$.
\end{theorem}
\textchangeSW{Theorem \ref{theorem:SK-R-one-way} has the same
  expression as Theorem \ref{theorem:one-way-formula-SK} except that
  there is additional communication rate constraint. Because of this
  additional constraint, unlike the Slepian-Wolf coding used in the
  proof of Theorem \ref{theorem:one-way-formula-SK}, we need to use
  quantize-and-binning scheme \`a la the Wyner-Ziv coding.}

In general, the expression in Theorem \ref{theorem:SK-R-one-way}
involves two auxiliary random variables and is difficult to
compute. However, explicit formulae are available for the specific
cases of Gaussian sources and binary sources \cite{WatOoh11, ChoBlo14,
  LiuCuffVerdu16}.

When the adversary's observation $Z$ is constant, the SK agreement is
closely related to the CR generation problem studied
earlier. Heuristically, if the communication used to generate a SK is
as small as possible, it is almost uniform, and therefore,
$L_{\ep}(c|X_1, X_2)\gtrsim S_{r,\ep,\delta}(X_1, X_2|Z; c) +c$. On
the other hand, using the leftover hash lemma we can show the
following.
\begin{proposition} \label{proposition:connection-SK-CR}
Given $\varepsilon,\delta \in [0,1)$ and $c>0$, we have
\begin{align*}
S_{r,\varepsilon,\delta}(X_1,X_2|Z;c) \ge L_{\varepsilon}(c|X_1,X_2) -
c - 2 \log (1/2\delta) - 1.
\end{align*}
\end{proposition}
 %% (see
%% Proposition \ref{proposition:connection-SK-CR}). In fact, the optimal
%% trade-offs are just parallel transport each other:\footnote{The
%%   inequality $\ge$ of \eqref{eq:transformation-CR-SK} directly follows
%%   from Proposition \ref{proposition:connection-SK-CR}; on the other
%%   hand, the inequality $\le$ is proved via single-letter
%%   characterizations.
%% \begin{align} \label{eq:transformation-CR-SK}
%% C^\mathtt{sk}_r(R| X_1,X_2) = C^\mathtt{cr}_r(R| X_1,X_2) - R
%% \end{align}
%% \begin{theorem} \label{theorem:SK-Rate-Tradeoff}
%% For given $r \ge 1$, we have
%% \begin{align*}
%% C^\mathtt{sk}_r(R| X_1,X_2) = \max_{V^r}\bigg[ \sum_{\stackrel{i=1:}{
%%       \mathrm{odd}}}^r I(V_i \wedge X_2 |V^{i-1}) +
%%   \sum_{\stackrel{i=1:}{ \mathrm{even}}}^r I(V_i \wedge X_1|V^{i-1})
%%   \bigg],
%% \end{align*}
%% where the maximization is taken over auxiliary random variables $V^r =
%% (V_1,\ldots,V_r)$ satisfying
%% \begin{align*}
%% &V_i \mc (X_1,V^{i-1}) \mc X_2,~~~\mbox{for odd } i, \\ & V_i \mc
%%   (X_2,V^{i-1}) \mc X_1,~~~\mbox{for even } i
%% \end{align*}
%% and
%% \begin{align*}
%% \sum_{\stackrel{i=1:}{ \mathrm{odd}}}^r \big[ I(V_i \wedge X_1
%%   |V^{i-1}) - I(V_i \wedge X_2 | V^{i-1}) \big] +
%% \sum_{\stackrel{i=1:}{ \mathrm{even}}}^r \big[ I(V_i \wedge
%%   X_2|V^{i-1}) - I(V_i \wedge X_1 | V^{i-1}) \big] \le R.
%% \end{align*}
%% Moreover, it may be assumed that range sizes are at most $|\cX_1|
%% \prod_{j=1}^{i-1} |\cV_j|+1$ for odd $i$ and $|\cX_2|\prod_{j=1}^i
%% |\cV_j|+1$ for even $i$.
%% \end{theorem}}}
%% \htc{
Using these observations and Theorem~\ref{t:CR_cap}, we get the
following characterization of rate-limited SK capacity.
\begin{theorem}\label{theorem:SK-Rate-Tradeoff}
For $R>0$ and finite-valued $(X_1, X_2)$,
\begin{align}
C^{\tt SK}_r(R) = \sup {\ICe(\pi|X_1, X_2) - \ICi(\pi|X_1, X_2)},
\nonumber
\end{align}
where the supremum is over all $r$-round private coin protocols $\pi$
such that $\ICi(\pi|X_1, X_2)\leq R$.
\end{theorem}

%% %% \begin{figure}[t]
%% %% \begin{center}
%% %% \includegraphics[scale=0.6]{SK-CR.eps}
%% %% \caption{Description of $C_r^{\mathtt{cr}}(R) =
%% %%   C^{\mathtt{cr}}_r(R|X_1,X_2)$ and $C_r^{\mathtt{sk}}(R) =
%% %%   C_r^{\mathtt{sk}}(R|X_1,X_2)$.}
%% %% \label{f:SK-CR-relation}
%% %% \end{center}
%% %% \end{figure}
%% 
In fact, it can be seen that $C_{r}^{\tt sk}(R)$ is $R$ less than the
maximum rate of CR that can be generated using $r$-round communication
protocols of rate less than $R$. Thus, the optimal rate of an SK that can
be generated corresponds to the difference between the $C(R)$ curve
and the slope $1$ line in Figure~\ref{f:CR}. Since the maximum
possible rate of SK is $I(X_1\wedge X_2)$, the quantity $R^*$ depicted
in Figure~\ref{f:CR} corresponds to the minimum rate of communication
needed to generate a SK of rate equal to $I(X_1\wedge X_2)$. This
minimum rate was studied first in \cite{Tya13} where a characterization of
$R^*$ was given. Furthermore, an example was provided where $R^*$
cannot be attained by simple (noninteractive) protocols, which in turn
constitutes an example where two parties can generate a SK of rate
$I(X_1\wedge X_2)$ without agreeing on $X_1$ or $X_2$ or $(X_1, X_2)$.

\subsection{Discussion}
Several basic problems remain open in spite of decades of work in this
area. Perhaps most importantly, a characterization of SK capacity
$C^{\tt sk}(X_1, X_2|Z)$ is open in general. As we pointed out, it is
known that interaction is needed in general to attain this
capacity. Even when $Z$ is a constant, although interaction is not
needed to attain the SK capacity, we saw that it can help reduce the
rate of communication needed to generate an optimal rate SK. In
another direction, \cite{HayTyaWat14ii} studied the second-order
asymptotic term in $S_{\ep, \delta}(X_1^n, X_2^n|Z^n)$ and used an
interactive scheme with $O(n^{1/4})$ rounds of interaction to attain
the optimal term when the Markov relation $X_1\mc X_2 \mc Z$ holds. It
remains open if a noninteractive protocol can attain the optimal
second-order term. One of the difficulties in quantifying the
performance of noninteractive protocols is that the converse bound of
Theorem~\ref{theorem:CIT} allows arbitrary interactive communication
and a general bound which takes the number of rounds of interaction
into account is unavailable.  \textchangeSW{Recently, \cite{TyaWat17}
  provided a universal protocol for generating an SK at rate within a 
  $\cO(\sqrt{n\log n})$ gap to the SK capacity without knowing the
  distribution $\bPP{X_1X_2}$.} Moreover, this protocol is interactive and
uses $O(\sqrt{n})$ rounds of interaction. Studying the role of
interaction in SK agreement is an interesting research direction. A
noteworthy recent work in this direction is \cite{LiuCuffVerdu17}
where a connection between the expression for rate-limited SK capacity
and an extension of the notion of hypercontractivity is used to study
the benefit of interaction for SK agreement.

\htc{ Another refinement that has received attention is the
  exponential decay rate for secrecy parameter $\delta$ 
and the decay exponent for error parameter $\ep$. Bounds for achievable error
  exponents were given in~\cite{CsiNar04} and follow from the classic
  error exponents for Slepian-Wolf coding (cf.~\cite{CsiKor11}). An
  exponential decay rate for $\delta$ was also reported
  in~\cite{CsiNar04} and further refinements can be obtained using
  exponential-leakage refinements for the leftover hash lemma from,
  for instance,~\cite{Hay11, Hay13}. All these works consider error
  and secrecy exponents in isolation, and the problem of
  characterizing exponents for secret key agreement is open.  }

\htc{Also, while our treatment has focused on a simple source model,
  several extensions to other multiterminal source models where one or
  more terminals act as helpers have been studied starting
  with~\cite{CsiNar00}. We have not included a review of this and
  other related setting including channel models for SK agreement
  (cf.~\cite{AhlCsi93, CsiNar08, Cha08, Cha11, GohAna10ii, Cha11,
    Cha12, CsiNar13, TyaWat13ii}). Some of these topics and references
  to related literature can be found in the
  monograph~\cite{TyagiNarayan16}.}

A variant of the two-party SK agreement problem reviewed in this
section has been studied in the computer science literature under the
name of ``fuzzy extractors,'' starting from \cite{DodOstReySmi08}.
Unlike our setup above, in the fuzzy extractors model, the exact
distribution of the observations $(X_1, X_2)$ is not
fixed. Specifically, $\cP_1$ observes a {\em $k$-source} $X_1$,
$i.e.$, a source that has min-entropy at least $k$, and $\cP_2$
observes $X_2$ such that the Hamming distance between $X_1$ and $X_2$
is less than a threshold with probability $1$. This is a two-party
extension of the classic model of randomness extraction introduced in
\cite{ChoGol88}.  Although this direction of research has 
developed independently of the one in the information theory community,
the techniques used are related; for instance, see
\cite{HayTyaWat14ii,FulReySmi16}. Note that the universal setting of
\cite{TyaWat17} constitutes another variant of the problem where the
distribution of the observations is unknown. However, the protocols
proposed are theoretical constructs and, typically, the work on fuzzy
extractors seeks computationally tractable protocols.
%%%%%%%%%%%%%%%%%%%%%%%%%%%%%%%%%%%%%%%%%%%%%%%
\section{Simulation without communication}\label{s:sim_no_comm} 
In this section, we address the distributed simulation (or generation)
of samples from a specified distribution by using samples from another
distribution, but without communicating. We begin with a classic
problem of Wyner where two parties seek to generate samples from a
distribution using shared randomness. Instead of providing the
original treatment of this problem from~\cite{Wyn75}, we recover the
known results using the more powerful framework of {\em approximation
  of output statistics} (AOS) introduced in \cite{HanVer93}. The
latter is reviewed first. We then proceed to the general problem of
simulation, and close with an important ``distributed information
structure'' variant where each party has only partial information
about the distribution to be simulated. In the following section, we
consider variants of simulation problems where communication is
allowed; results of this section will serve as basic tools for the
setting with communication.

%%%%%%%%%%%%%%%
\subsection{Approximation of output statistics} \label{sec:AOS}
A standard simulation step in several applications entails generation
of a given random variable using samples from a uniform
distribution. The AOS problem is an extension where we seek to
generate a given distribution as the output of a noisy channel using a
uniform distribution on a subset of its input.
%% In this section, we shall introduce a problem called {\em
%%   approximation of output statistics} (AOS) also known as {\em channel
%%   resolvability}.  
A version of this problem was originally introduced in \cite{Wyn75} as
a tool to prove the achievability part of ``Wyner common
information,'' which will be discussed in the next section. The
general formulation, also referred to as the {\em channel
  resolvability} problem, was introduced in \cite{HanVer93} in part as
a tool to prove the converse for the identification capacity
theorem~\cite{AhlDue89}.  More recently, it has been used as a tool
for proving the achievability part of the reverse Shannon theorem
(which we will review in Section~\ref{subsec:reverse-Shannon}) and the
wiretap channel capacity \cite{Wyn75ii} (cf.~\cite{Csi96,
  CaiWinYeu04, Hay06, BloLan13}). In the information theory
literature, the AOS problem has emerged as a basic building block for
enabling distributed simulation.

For a given input distribution $\bPP{X}$ and channel $W(y|x)$, our
goal in the AOS problem is to simulate a given output distribution
\begin{align*}
\bP{Y}{y} := \sum_x \bP{X}{x} W(y|x).
\end{align*}
To that end, we construct a code $\cC = \{x_1,\ldots, x_{|\cC|} \}$ so
that the output distribution
\begin{align*}
\bP{\cC}{y} := \sum_{x \in \cC} \frac{1}{|\cC|} W(y|x)
\end{align*}
corresponding to a uniform distribution over the codewords
approximates the target output distribution $\bPP{Y}$. For a given
size $|\cC|$ of input randomness, we seek to make the approximation
error as small as possible. Various measures of ``distance'' have been
used in the literature to evaluate the approximation error: for
instance, Kullback-Leibler divergence, normalized Kullback-Leibler
divergence, and the variational distance\footnote{The normalized
  divergence makes sense only when we consider block coding for a
  given sequence of channels.}. In our treatment here, we use
variational distance to measure error and denote $\rho(\cC,\bPP{Y}) :=
d(\bPP{\cC}, \bPP{Y})$.
\begin{definition}
For a given $\ep \in [0,1)$, the infimum over the length $\log |\cC|$
  of AOS codes satisfying $\rho(\cC,\bPP{Y}) \le \ep$ is denoted by
  $L_\ep(\bPP{X},W)$.
\end{definition}
When the input distribution is iid $\mathrm{P}_X^n$ and the channel
$W^n=\prod_{t=1}^n W$ is discrete memoryless, we consider the
asymptotic limits defined by
\begin{align*}
C^{\mathtt{AOS}}_\ep(\bPP{X},W) := \limsup_{n\to\infty} \frac{1}{n}
L_\ep(\mathrm{P}_X^n, W^n)
\end{align*}
and
\begin{align*}
C^{\mathtt{AOS}}(\bPP{X},W) := \lim_{\ep\to 0} C_\ep(\bPP{X},W).
\end{align*}

\begin{theorem}\cite{Wyn75, HanVer93, HanVer93b, HouKra13, WatHay14} \label{theorem:resolvability-fixed-input}
For a given $\ep \in (0,1)$, we have
\begin{align}
C^{\mathtt{AOS}}_\ep(\bPP{X},W) = C^{\mathtt{AOS}}(\bPP{X},W) =
\min_{\bPP{\tilde{X}}} I(\tilde{X} \wedge \tilde{Y}),
\label{eq:resolvability-fixed-input}
\end{align}
where the minimum is taken over all input distribution
$\bPP{\tilde{X}}$ such that the output distribution $\bP{\tilde{Y}}{y}
= \sum_x \bP{\tilde{X}}{x}W(y|x)$ coincides with the target output
distribution $\bPP{Y}=\bPP{X}\circ W$.
\end{theorem}
In \cite{HanVer93}, the motivation to introduce the AOS problem was to
show the (strong) converse part of the identification capacity
theorem~\cite{AhlDue89}. For that purpose, it is useful to consider
the worst-case with respect to input distributions:
\begin{align}
C^{\mathtt{AOS}}_\varepsilon(W) := \limsup_{n\to\infty}
\sup_{\bPP{X^n}} \frac{1}{n} L_\varepsilon(\bPP{X^n}, W^n)
\label{eq:worst-case-resolvability}
\end{align}
and
\begin{align*}
C^{\mathtt{AOS}}(W) := \lim_{\varepsilon \to 0}
C^{\mathtt{AOS}}_\varepsilon(W),
\end{align*}
where the supremum in \eqref{eq:worst-case-resolvability} is taken
over all input distribution $\bPP{X^n}$ that are not necessarily iid
\htc{(the output distribution we are trying to approximate are given
  by $\bPP{X^n}\circ W^n$).}  Interestingly, this worst-case quantity
coincides with Shannon's channel capacity.
\begin{theorem}[\cite{HanVer93}]
For a given $\ep \in (0,1)$, we have
\begin{align}
C^{\mathtt{AOS}}_\varepsilon(W) = C^{\mathtt{AOS}}(W) = \max_{\bPP{X}}
I(X \wedge Y).
\label{eq:characterization-worst-resolvability}
\end{align}
\end{theorem} 
Even though the worst case AOS is characterized by the maximization of
the single-letter input distribution in
\eqref{eq:characterization-worst-resolvability}, the worst input
distribution attaining the supremum in
\eqref{eq:worst-case-resolvability} may not be iid in general (see
\cite[Example 1]{HanVer93}).

We outline the achievability proof of Theorem
\ref{theorem:resolvability-fixed-input}.  For simplicity, we assume
$\bPP{X}$ itself is the optimal distribution attaining the minimum in
\eqref{eq:resolvability-fixed-input}.  To construct an AOS code, we
randomly generate codewords ${\cal C}_n = \{
\mathbf{x}_1,\ldots,\mathbf{x}_{|{\cal C}_n|}\}$ according to
$\mathrm{P}_X^n$.  Then, the approximation error $\rho(\bPP{{\cal
    C}_n}, \mathrm{P}_Y^n)$ averaged over the random choice of the
code ${\cal C}_n$ can be evaluated by techniques from \cite{HanVer93,
  Hay06, Oohama13, Cuff13}. Specifically, if the rate $\frac{1}{n}
\log |{\cal C}_n|$ of the constructed AOS code is larger than $I(X
\wedge Y)$, the convergence of the approximation error is
guaranteed. A technical tool involved is a bound for the resulting
approximation error; such results have been aptly named {\em soft
  covering} lemmas starting from \cite{Cuff13}. \textchangeSW{The
  traditional covering lemma, proved using combinatorial arguments in
  \cite{CsiKor11}, claims that the typical set of size $2^{nH(Y)}$ in
  the output space can be covered by almost disjoint ``balls'' of size
  $2^{nH(Y|X)}$ each centered around $2^{nI(X\wedge Y)}$
  codewords. Interestingly, the soft covering lemma claims that the
  same number of codewords suffice to cover the output space in the
  sense of approximating the output distribution of a channel.}

Originally, a version of the soft covering lemma was proved in
\cite{HanVer93}.  Later, alternative versions appeared in
\cite[Theorem 2]{Hay06} and \cite[Lemma 3]{Oohama13}; \htc{a general
  version} of the lemma can be found in \cite[Theorem 7.1]{Cuff13}.
The proofs in \cite{Hay06, Oohama13, Cuff13} are all based on a
similar strategy using the Cauchy-Schwarz inequality \textchangeSW{to
  bound the variational distance ($\ell_1$-distance) in terms of the
  $\ell_2$-distance}, which is reminiscent of the proof of the
leftover hash lemma.

The AOS problem has been extended in various directions.  In
fact,~\cite{HanVer93} studied the AOS problem for general channels
that may not be stationary or ergodic.  The convergence speed
(exponent) of the approximation error was studied in \cite{Hay06,
  Oohama13} and \htc{a complete characterization for the random coding
  exponent was derived in~\cite{PariziTM17} (see, also,
  \cite{YagCuf19})}.  The second-order asymptotic rate for this
problem was characterized in \cite{WatHay14} under additional
assumptions.  A general formula for the leading asymptotic term in the
optimal length of AOS codes for general channels and general input
distributions was characterized recently in \cite{Yagi17}.

%%%%%%%%%%%%%%%
\subsection{Wyner common information} \label{sec:wyner-common-information}
In an attempt to define an operational notion of common information of
two random variables, Wyner studied the amount of shared uniform
randomness needed for two parties to generate $n$ independent samples
from a given joint distribution $\bPP{X_1X_2}$. The number of shared
random bits needed per sample is termed {\em Wyner common
  information}~\cite{Wyn75}.

Formally, $\cP_1$ and $\cP_2$ have access to shared randomness $U$
distributed uniformly over a set $\cU$ (constituting public coins) and
unlimited private randomness $U_1$ and $U_2$ (constituting private
coins), respectively. They seek to generate a sample from a fixed
distribution $\bPP{X_1X_2}$. To that end, they execute a simulation
protocol comprising channels $W_1(\cdot |u)$ and $W_2(\cdot | u)$ with
a common input alphabet $\cU$. The output distribution of the protocol
is given by
\begin{align}
\bPP{\cC}(x_1,x_2) = \sum_{u \in \cU} \frac{1}{|\cU|} W_1(x_1|u)
W_2(x_2|u),
\end{align}
and the corresponding simulation error by
\begin{align*}
\rho(\cC, \bPP{X_1 X_2}) = d(\bPP{\cC}, \bPP{X_1 X_2}).
\end{align*}
\begin{definition}
For a given $\ep \in [0,1)$, the infimum over the length $\log |\cU|$
  of simulation protocols satisfying $\rho(\cC, \bPP{X_1X_2}) \le \ep$
  is denoted by $L_\ep(\bPP{X_1X_2})$.
\end{definition}
For iid distribution $\mathrm{P}_{X_1X_2}^n$, the Wyner common
information of $(X_1, X_2)$ is defined as follows:
\begin{align*}
C_\ep^{\mathtt{Wyn}}(X_1,X_2) := \limsup_{n\to\infty} \frac{1}{n}
L_\ep(\mathrm{P}_{X_1X_2}^n)
\end{align*}
and
\begin{align*}
C^{\mathtt{Wyn}}({X_1X_2}) := \lim_{\ep \to 0}
C_\ep^{\mathtt{Wyn}}(\bPP{X_1X_2}).
\end{align*}
A single-letter expression for Wyner common information $C^{\tt
  Wyn}(X_1, X_2)$ was given in \cite{Wyn75}; a strong converse
establishing $C_\ep^{\tt Wyn}(X_1, X_2) = C^{\tt Wyn}(X_1, X_2)$ for
all $0< \ep<1$ has been claimed recently in \cite{YuTan17}. We
summarize both results below.

\begin{theorem}\cite{Wyn75, YuTan17} \label{theorem:wyner}
For a given $\ep \in (0,1)$, we have
\begin{align}
C_\ep^{\mathtt{Wyn}}(X_1,X_2) = C^{\mathtt{Wyn}}({X_1,X_2}) = \min I(V
\wedge X_1,X_2),
\label{eq:wyner-common-info}
\end{align}
where the minimization is taken over all auxiliary random variable $V$
satisfying $X_1 \mc V \mc X_2$.  Moreover, the range of $V$ may be
assumed to be $|\cV| \le |\cX_1||\cX_2|$.
\end{theorem}
This problem is closely related to the AOS problem considered in the
previous section.  In fact, to prove the achievability part of
Theorem~\ref{theorem:wyner}, we construct an AOS code as follows.  For
the optimal joint distribution $\bPP{V X_1 X_2}$ attaining the minimum
in \eqref{eq:wyner-common-info}, note that the distribution
$\bPP{X_1X_2|V}$ can be factorized as $\bPP{X_1|V} \times \bPP{X_2|V}$
using the Markov chain condition. Thus, if we have an AOS code that
approximates the output distribution $\mathrm{P}^n_{X_1X_2}$, which is
the output distribution of channel $\mathrm{P}^n_{X_1X_2|V}$ with
input distribution $\mathrm{P}^n_{V}$, then the parties can simulate
$\mathrm{P}^n_{X_1X_2}$ by using the AOS code as shared randomness and
$\mathrm{P}_{X_1|V}^n$ and $\mathrm{P}^n_{X_2|V}$ as local channels
for the simulation protocol, respectively.

\textchangeSW{In the problem formulation above, we studied the worst-case length
  of common randomness required for generating the target joint distribution
  with vanishing error. Alternatively, we can consider the expected
  length of common randomness required to generate the target joint
  distribution exactly. Such a variant of the problem, termed {\em
    exact common information}, was studied in \cite{KumLiElG14} (see
  also \cite{LiElG17} for a protocol that exactly generates target
  distributions on continuous alphabets).  The exact common
  information is larger than or equal to the Wyner common information
  by definition.  For some sources such as the binary double symmetric
  source, it is known that the former is strictly larger than the
  latter \cite{YuTan18}.}

%%%%%%%%%%%%%%%
\subsection{Simulation of correlated random variables}\label{s:simulation}
One important special case of the CR capacity result given in
Theorem~\ref{t:CR_cap} is when the rate of communication $R=0$. By
Theorem~\ref{t:CR_cap}, this is given by the supremum of $I(U\wedge
X_1)$ such that the Markov relations $U \mc X_1 \mc X_2$ and $U \mc
X_2 \mc X_1$ hold. This double Markov condition enforces $U$ to be a
{\it common function} of $X_1$ and $X_2$, namely a $U$ such that
$H(U|X_1) = H(U|X_2) = 0$; $e.g.$~see \cite[Problem 16.25]{CsiKor11}
and \cite[Lemma 1.1]{CsiNar00} for a slight sharpening of this result.
The maximum of such common functions is referred to as the {\it
  G\'acs-K\"orner common information} of $(X_1,X_2)$, denoted $GK(X_1,
X_2)$ \cite{GacKor73}. G\'acs and K\"orner showed in \cite{GacKor73}
that the maximum rate of CR that two parties observing iid samples
from $\bPP{X_1 X_2}$ can generate without communicating is $GK(X_1,
X_2)$. In fact, when $GK(X_1, X_2) = 0$ and no communication is
allowed, Witsenhausen~\cite{Wit75} showed that parties cannot even
agree on a single unbiased bit.

In this section, we are interested in a generalization of this
question: When can parties observing $(X_1, X_2)$ generate a single
sample from a given distribution $\bQQ{U_1U_2}$ with $\cP_1$ getting
$U_1$ and $\cP_2$ getting $U_2$. Formally, we consider the following
problem.
\begin{definition}[Simulation without communication]
Given distributions $\bPP{X_1X_2}$ and $\bQQ{U_1U_2}$, we say that
$\bPP{X_1X_2}$ can {\it simulate} $\bQQ{U_1U_2}$ if for every $\ep>0$
there exists $n\geq 1$ and functions $f:\cX_1^n\to \cU_1$ and $g:
\cX_2^n \to \cU_2$ such that
$\ttlvrn{\bPP{f(X_1^n)g(X_2^n)}}{\bQQ{U_1U_2}} \leq \ep$. Denote by
$\cS(\bPP{X_1X_2})$ the set of all distributions $\bQQ{U_1 U_2}$ such
that $\bPP{X_1X_2}$ can simulate $\bQQ{U_1U_2}$.
\end{definition}
While private randomness is not allowed in our formulation, it can
easily be extracted using samples from $\bPP{X_1X_2}$. Note that the
common randomness generation problem and the Wyner common information,
respectively, entail simulating a uniformly distributed shared bits
from a given distribution and vice-versa. \htc{Also, a related setting
  where we seek to simulate a given channel using an available channel
  was considered in~\cite{HadYBGA17}. We do not review this problem
  here and restrict ourselves to the simple source model setting
  above.}

An elemental question is to characterize the set $\cS(\bPP{X_1
  X_2})$. Surprisingly, this basic question was formulated only
recently in \cite{AnantharamKamath12} (see, also,
\cite{AnantharamKamath16}). However, several important instances of
this general question appear in the information theory literature and
the treatment of randomness in the computer science literature. In
particular, the following result was shown in \cite{Wit75}:
\begin{theorem}  Every distribution $\bPP{X_1, X_2}$ can simulate $BSS(\rho)$ if
\[
\rho \leq \frac{2}{\pi}\cdot \arcsin(\rho_m(X_1, X_2)).
\]
\end{theorem}
The proof is simple and entails first simulating correlated Gaussian
random variables with correlation $\rho_m(X_1,X_2)$ (using the central
limit theorem) and then declaring their signs. A result of
Borell~\cite{Borell85} shows that for jointly Gaussian vectors
$\bPP{X_1, X_2}$, the maximum of $\rho_m(f_1(X_1), f_2(X_2))$ over
binary-valued $f_1, f_2$ is obtained when $f_1$ and $f_2$ correspond
to half-planes, namely they have the form $f_i(x)=\mathrm{sign}(a_i
\cdot (x-b_i))$. As a corollary of this result \htc{(applied to
  $X_1^n, X_2^n$)} and the theorem above, we obtain the following.
\begin{corollary}
For jointly Gaussian $(X_1, X_2)$ with zero mean and covariance matrix
\[
\begin{bmatrix}
1 &\rho_0\\ \rho_0 &1
\end{bmatrix},
\]
$\bPP{X_1X_2}$ can simulate $BSS(\rho)$ iff $\rho \leq
\frac{2}{\pi}\cdot arcsin(\rho_0)$.
\end{corollary}
The previous result gives precise conditions for $BSS(\rho)$ to be
contained in $\cS(GSS(\rho))$. The characterization of the set
$\cS(GSS(\rho))$, and in general of $\cS(\bPP{X_1X_2})$ for a general
distribution $\bPP{X_1X_2}$, is open. Partial results are available
in~\cite{AnantharamKamath16} which provide general necessary
conditions for a distribution to lie in $\cS(\bPP{X_1X_2})$. We review
this result below, in a 
slightly different form, where a measure of correlation is used to
capture the relation $\bQQ{X_1X_2}\in \cS(\bPP{X_1X_2})$. Instead of
specifying a measure of correlation, the next result provides a
general characterization of such measures of correlation which behave
monotonically along the containment relation (this approach is similar
to that of monotones used in~\cite{RenWol05, GohAna10, GohAna10ii}).
\begin{theorem}\label{t:correlation_bound}
Consider a function $\Gamma(X,Y)$ satisfying the following properties:
\begin{enumerate}
\item {\it Data processing inequality.} $\Gamma(f(X), g(Y)) \leq
  \Gamma(X, Y)$ for all functions $f,g$;

\item {\it Tensorization property.} $\Gamma(X^n, Y^n) = \Gamma(X_1,
  Y_1)$ for iid $(X^n, Y^n)$;

\item {\it Lower semicontinuity.} $\Gamma(X,Y)$ is a lower
  semicontinuous function of $\bPP{XY}$.
\end{enumerate}
If $\bPP{X_1X_2}$ can simulate $\bQQ{U_1, U_2}$, then
$\Gamma(U_1,U_2)\leq \Gamma(X_1, X_2)$.
\end{theorem}
As was pointed-out in \cite{AnantharamKamath16}, both $\rho_m(X,Y)$
and $s^*(X,Y)$ satisfy the conditions required of $\Gamma$ in
Theorem~\ref{t:correlation_bound}. \htc{We note that a more general
  class of measures of correlation satisfying these conditions is
  available~\cite{BeigiGohari18} (see, also,~\cite{BeigiGohari15}).}
As a corollary, we have the following.
\begin{corollary}
$BSS(\rho_1)$ can simulate $BSS(\rho_2)$ iff $\rho_1\geq \rho_2$.
\end{corollary}
Next, we review a few simple properties of $\cS(\bPP{X_1X_2})$ which
have not been reported in literature, but perhaps are well-known.
Clearly, the set $\cS(\bPP{X_1X_2})$ is closed. We note that
the simulation induces a partial order on the set of
distributions. Specifically, denoting by $\bPP{X_1X_2}\succeq
\bQQ{U_1U_2}$ the relation ``$\bPP{X_1X_2}$ can simulate
$\bQQ{U_1U_2}$,'' it can be shown that $\succeq$ is a
preorder. Furthermore, define an equivalence relation $\dP\sim\dQ$ iff
$\dP \succeq \dQ$ and $\dQ\succeq\dP$, and consider the set of
equivalence classes $[\dP]$. Then, the set of equivalence classes is a
poset under the partial order induced by the preorder $\succeq$. It is
easy to see that constants constitute a minimal element for this poset
and distributions $\bPP{X_1X_2}$ with $GK(X_1, X_2)>0$ constitute a
maximal element.

Note that $\succeq$ does not constitute a total order. Indeed,
consider $\dP = BSS(\rho_1)$ and $\dQ = GSS(\rho_2)$ such that
$(2/\pi)arcsin(\rho_2) < \rho_1 < \rho_2$. Then, by
Theorem~\ref{t:correlation_bound} $\dP$ cannot simulate
$\dQ$. Furthermore, the aforementioned result of Borell implies that
$\dQ$ cannot simulate $\dP$. Therefore, one dimensional measures of
correlation such as $\rho_m$ and $s^*$ used
in~\cite{AnantharamKamath16} cannot characterize the simulation
relation. Among the candidate two dimensional measures, the
hypercontractivity ribbon may sound promising as when
$\bPP{X_1X_2}\succeq \bQQ{U_1U_2}$, $\cR(\bPP{X_1X_2}) \subset
\cR(\bQQ{U_1U_2})$ (cf.~\cite{AnantharamKamath16}). But even this
promise is empty since for $\dP$ and $\dQ$ above, $\cR(\dQ) \subset
\cR(\dP)$ but $\dQ$ cannot simulate $\dP$.

While the general problem of characterizing when a distribution
$\bPP{X_1X_2}$ can simulate $\bQQ{U_1U_2}$ remains open, an
algorithmic procedure for testing a ``gap-version'' of the problem
when $U_1$ and $U_2$ are both binary has been proposed recently in
\cite{GhaziKamathSudan16}; it has been extended to the general case in
\cite{DeMosselNeeman17}. At a high level, the procedure is to produce
random variables with as large a maximal correlation as possible from
$\bPP{X_1X_2}$ while maintaining the marginals of the simulated
distribution as close to $\bQQ{U_1}$ and $\bQQ{U_2}$; the algorithm
either produces a sample from a distribution \textchangeSW{such that
  the variational distance with $\bQQ{U_1 U_2}$ is less than error
  parameter $\delta$} or claims that there is no procedure which can
produce a sample with distribution within $O(\delta)$ of
$\bQQ{U_1U_2}$. The key idea is to obtain a finite sample equivalent
of Witsenhausen's construction~\cite{Wit75}, namely claim that using
finitely many samples behavior similar to a Gaussian distribution with
appropriate correlation can be simulated. The treatment is technical
and relies on the {\it invariance principle} shown in
\cite{MosselDonnelOeszkiewicz05, MosselDonnelOeszkiewicz10}.

%%%%%%%%%%%%%%%
\subsection{Correlated sampling}\label{s:correlated_sampling}
The final problem we cover in this section entails a simulation when the
complete knowledge of the target distribution is not
available. Specifically, $\cP_1$ has access to $\dP$ and $\cP_2$ has
access to $\dQ$, where $\dP$ and $\dQ$ are distributions on the same
alphabet $\cX$. Using their shared randomness, $\cP_1$ and $\cP_2$
seek to generate $X\sim \dP$ and $Y\sim \dQ$, respectively, such that
the probability of agreement $\bPr{X=Y}$ under the resulting coupling
is as large as possible. When the marginals $\dP$ and $\dQ$ are
available at the same place, the probability of agreement is maximized
by the maximal coupling and the maximum equals
$\ttlvrn{\dP}{\dQ}$. Interestingly, even in the distributed setup, the
same can be attained up to a multiplicative factor.

The problem of correlated sampling was formally defined in
\cite{Holenstein09} as a tool for providing a simpler proof of the
parallel repetition theorem~\cite{Raz98}. Since the latter is central
to showing several hardness of approximation results, correlated
sampling is one of the foundational tools for randomized computational
complexity theory. On the other hand, variants of correlated sampling
have been applied gainfully in devising efficient randomized
algorithms for data mining; see, for instance, \cite{Broder97,
  KleinbergTardos02}.
 
We begin with a formal definition of the problem.
\begin{definition}[Correlated Sampling]
Given distributions $\dP$ and $\dQ$ on a finite alphabet $\cX$ and a
shared randomness $R_{\tt pub}$, an $\ep$-correlated sample for
$(\dP,\dQ)$ consists of mappings $f_\dP$ and $g_\dQ$ depending only on
$\dP$ and $\dQ$, respectively, such that $X=f_\dP(R_{\tt pub})$ and
$Y=g_\dQ(R_{\tt pub})$ satisfy $\bPP{X} = \dP$, $\bPP{Y}=\dQ$ and
\[
\bPr{X\neq Y}\leq \ep.
\]
We call $(X,Y)$ $\ep$-correlated sample.
\end{definition}
Note that the maximal coupling lemma (Lemma~\ref{l:MCL}) already
characterizes the best $\ep$ that can be attained for a given $\dP$
and $\dQ$ when they are available at the same place. The next basic
result is due to Holenstein~\cite{Holenstein09} and shows that even
when the knowledge of $\dP$ and $\dQ$ is not available at the same
place, roughly the same error $\ep$ can be attained (up to a factor of
$2$).
\begin{theorem}\label{t:correlated_sampling}
Given distributions $\dP$ and $\dQ$ on a finite alphabet $\cX$ such
that $\ttlvrn{\dP}{\dQ}\leq \ep$, there exist an
$2\ep/(1+\ep)$-correlated sample for $(\dP, \dQ)$.
\end{theorem}
{\it Proof sketch.} For the binary case with $\dP\equiv {\tt Ber}(p)$
and $\dQ \equiv {\tt Ber}(q)$, we can simply use the public randomness
to generate $R_{\tt pub}\sim {\tt unif}([0,1])$ to obtain the
correlated sampling as $X= \indicator(R_{\tt pub} \leq p)$ and $Y=
\indicator(R_{\tt pub} \leq q)$. In fact, in this case we obtain a
$\ep$-correlated sample for $(\dP, \dQ)$.

In general, we proceed as follows: Let $R_{\tt pub}$ comprise an iid
sequence $(A_i, B_i)_{i=1}^\infty$ where $A_i\sim {\tt unif}(\cX)$ and
$B_i \sim {\tt unif}([0,1])$. The correlated sample is produced as
below:
\begin{enumerate}
\item $\cP_1$ returns $X= A_i$ where the index $i$ is the least $i$
  such that $\dP(A_i) > B_i$.
\item $\cP_2$ returns $Y = A_j$ where the index $j$ is the least $j$
  such that $\dQ(A_j) > B_j$.
\end{enumerate}
The proof can be completed upon noting that
\begin{align}
\bPr{X=x} = \dP(x), \quad \bPr{Y=y}= \dQ(y), \nonumber
\end{align}
and, denoting by $I$ the smallest index $l$ such that $B_l<
\max[\dP(A_l), \dQ(A_l)]$ ($i.e.$, the smallest index declared by
$\cP_1$ or $\cP_2$),
\begin{align}
\bPr{X=Y} &\geq \bPr{B_I < \min[\dP(A_I), \dQ(A_I)]} \\ &=
\frac{1-\ttlvrn{\dP}{\dQ}}{1 + \ttlvrn{\dP}{\dQ}}.  \nonumber
\end{align}
\qed

The result above has been extended to address various simulation
problems with distributed knowledge of the joint distribution. For
instance, {consider the variant of the simulation problem of the
  previous section where \textchangeHT{the first party observes $X_1$ and the second party observes
    $X_2$ generated
    from a distribution $\bPP{X_1X_2}$. Suppose that a distribution $\bPP{X_1X_2U}$
is known to both the parties, and they}
    seek to generate random variables $(U_1, U_2)$ with $\cP_1$ declaring
$U_1$ and $\cP_2$ declaring $U_2$ and such that $\bPP{X_1X_2U_1}$ and
    $\bPP{X_1 X_2 U_2}$ are both equal to $\bPP{X_1X_2U}$.
    \textchangeHT{Note that since $X_1$ and $X_2$ are known only to $\cP_1$ and $\cP_2$, respectively,
      the conditional distributions 
$\bPP{U|X_1}$ and $\bPP{U|X_2}$, too, are known only to $\cP_1$ and $\cP_2$, respectively.}
    By applying
Theorem~\ref{t:correlated_sampling} twice with $\dP = \bPP{U|X_1=x_1}$
and $\dQ= \bPP{U|X_1=x_1, X_2=x_2}$ and with $\dP = \bPP{U|X_2=x_2}$
and $\dQ= \bPP{U|X_1=x_1, X_2=x_2}$, it was shown
in~\cite{Holenstein09} that the parties can obtain $(U_1, U_2)$ such
that
\begin{align}
\bPr{U_1\neq U_2} 
&\leq 2\big\{ \ttlvrn{\bPP{X_1X_2U}}{\bPP{X_1X_2}\bPP{U|X_1}} \nonumber \\ 
&~~~ +\ttlvrn{\bPP{X_1X_2U}}{\bPP{X_1X_2}\bPP{U|X_2}} \big\}.
\label{e:corr_samp}
\end{align}
Given the key role played by Theorem~\ref{t:correlated_sampling} in
results of hardness of approximation, it is natural to ask if the
result obtained is close to optimal. This question was settled
recently in~\cite{BavarianGHKRS16} where it was shown that for every
$\gamma>0$, there exist $\dP$ and $\dQ$ such that
$\ttlvrn{\dP}{\dQ}\leq \ep$ and any correlated sampling has
probability of error at least $2\ep/(1+\ep) - \gamma$. Thus, the
scheme of Theorem~\ref{t:correlated_sampling} is optimal.

We close by noting that when communication between the parties is
allowed, correlated sampling with arbitrarily small probability of
error can be achieved; we shall revisit this problem in the next
section in the context of simulation of interactive protocols.
%%%%%%%%%%%%%%%
%\subsection{Discussion}

%%%%%%%%%%%%%%%%%%%%%%%%%%%%%%%%%%%%%%%%%%%%%%%
\section{Simulation using communication}\label{s:sim_comm} 
We now move to the more general problem of distributed simulation when
communication is allowed. Unlike the formulation considered in the
previous section, we not only require the parties to generate samples
from a given distribution but seek to emulate a prescribed joint
distribution for the input random variables and the simulated random
variables. The most general problem of {\em interactive channel
  simulation}, formalized in
Section~\ref{s:interactive_channel_simulation} below, encompasses most
of the formulations we have considered in this paper. We begin with a
simpler problem where only one-way communication is allowed; this
restriction is termed the {\em reverse Shannon theorem}. We conclude
with another restriction, namely the {\em protocol simulation
  problem}, where the channel to be simulated has the structure of an
interactive protocol.

%%%%%%%%%%%%%%%
\subsection{The reverse Shannon theorem} \label{subsec:reverse-Shannon}
How many bits must $\cP_1$ observing $X$ communicate noiselessly to
$\cP_2$ to enable $\cP_2$ to output a $\hat{Y}$ such that $\bPP{\hat
  Y|X}$ is close to a given channel $W:\cX\to \cY$? This problem
formulated in~\cite{BenShoSmoTha02} is, in essence, the ``reverse'' of
the Shannon's channel coding problem. The latter states that using a
given noisy channel $W$ with capacity $C(W)=\max_{\bPP X}I(X\wedge
Y)$, we can simulate an $nC(W)+o(n)$-bit noiseless channel. In the
same vein,~\cite{BenShoSmoTha02} posed the question: How many bits
must be sent to simulate $n$ instances of a channel? Remarkably, the
answer will be $I(X\wedge Y)$ as well\footnote{In this
  section, we only review the case with fixed input distribution; the
  case with worst input distribution has been studied
  in~\cite{BenShoSmoTha02, BenDevHarShoWin14}.}.

Formally, for a given channel $W:\cX \to \cY$ and an input
distribution $\bPP{X}$, the parties would like to simulate the joint
distribution $\bP{XY}{x,y} = \bP{X}{x} W(y|x)$.  The parties observe
shared randomness $U$ distributed uniformly over $\{0,1\}^l$, in
addition to private randomness $U_i$ observed by $\cP_i$, $i=1,2$. A
one-way channel simulation protocol entails a $1$-round communication
protocol $\pi$ and the output $\hat Y = \hat{Y}(\Pi,U_2,U)$ produced
by $\cP_2$.  The approximation error for this protocol is given by
\begin{align*}
\rho(\pi) := d(\bPP{X\hat{Y}}, \bPP{XY}).
\end{align*} 

\begin{definition}
For a given $\varepsilon \in [0,1)$ and $\ell\in \mN$, the infimum
  over the length $|\pi|$ of simulation protocols satisfying
  $\rho(\pi) \le \varepsilon$ \htc{and using $\ell$ bits of shared
    randomness} is denoted by $L_\varepsilon(l|\bPP{X}, W)$.
\end{definition}
When the input distribution is iid $\mathrm{P}_X^n$ and the channel is
a discrete memoryless channel, denoted $W^n$, we consider the
asymptotic limits defined by
\begin{align*}
C_\varepsilon^{\mathtt{RS}}(R|\bPP{X},W) := \limsup_{n\to\infty}
\frac{1}{n} L_\varepsilon(nR| \mathrm{P}_X^n,W^n)
\end{align*}
and
\begin{align*}
C^{\mathtt{RS}}(R|\bPP{X},W) := \lim_{\varepsilon \to 0}
C_\varepsilon^{\mathtt{RS}}(R|\bPP{X},W).
\end{align*}
An important corner point is the quantity
\begin{align*}
C_\varepsilon^{\mathtt{RS}}(\bPP{X},W) &:= \inf_{R \ge 0}
C_\varepsilon^{\mathtt{RS}}(R|\bPP{X},W),
\\ C^{\mathtt{RS}}(\bPP{X},W) &:= \inf_{R \ge 0}
C^{\mathtt{RS}}(R|\bPP{X},W).
\end{align*}

\textchangeSWtwo{
  Before we proceed, we illustrate how shared randomness is useful in this problem by an example.
\begin{example}\label{example:RS}
Let $\bPP{XY}$ induced by $\bPP{X}$ and $W$ be such that $X = (B_1,B_2)$ is  two uniform independent bits, and $Y = (J,B_J)$, where $J \in \{1,2\}$ is uniform and independent of $(B_1,B_2)$.
If no shared randomness exists between the two parties, $\cP_1$ has no option but to communicate
both $(B_1,B_2)$ to $\cP_2$, i.e., $C^{\mathtt{RS}}(R|\bPP{X},W) = 2$ for $R=0$. However, in the presence of one bit
of shared randomness, the two parties can take $J$ to be the shared randomness, and only one bit communication $B_J$
is enough, i.e., $C^{\mathtt{RS}}(R|\bPP{X},W) = 1$ for $R \ge 1$.
\end{example}}

In general, we can show the following result, which is an instance of the reverse Shannon theorem.
\htc{Versions of this theorem under various restriction occur
  in~\cite{BenShoSmoTha02, Win02, BenDevHarShoWin14, Cuff13}; the form
  below is from~\cite{Cuff13} where finite rate of shared randomness
  is considered.}
\begin{theorem}\label{theorem:reverse-Shannon}
For discrete random variable $X$ and a discrete memoryless channel
$W:\cX\to \cY$, we have
\begin{align}
C^{\mathtt{RS}}(R|\bPP{X},W) &= \min\big\{ R_\mathtt{c} : \exists
\bPP{V|XY} \mbox{ s.t. } X \mc V \mc Y, \nonumber \\
&~~~ |\cV|\le |\cX||\cY|+1, \nonumber \\ 
&~~~ R_\mathtt{c} \ge I(V \wedge X), \nonumber \\ 
&~~~ R_\mathtt{c}+R \ge
I(V\wedge X,Y) \big\}.  \label{eq:reverse-shannon-tradeoff}
\end{align}
In particular, for every $\varepsilon \in [0,1)$,
\begin{align} \label{eq:reverse-Shannon-unlimited-cr}
C_\varepsilon^{\mathtt{RS}}(\bPP{X},W) = C^{\mathtt{RS}}(\bPP{X},W) =
I(X \wedge Y).
\end{align}
\end{theorem}
One extreme case when unlimited shared randomness is allowed,
highlighted in \eqref{eq:reverse-Shannon-unlimited-cr} above, brings
out the classic mutual information, thereby endowing the latter with
another operational significance. At the other extreme is the case
when no shared randomness is allowed, namely $R=0$. Here, too, the
well-known Wyner's common information (see
Sec.~\ref{sec:wyner-common-information}) appears:
\begin{align*}
C^{\mathtt{RS}}(0|\bPP{X},W) = C^{\mathtt{Wyn}}(\bPP{XY}).
\end{align*}

We briefly outline the proof of achievability for
\eqref{eq:reverse-Shannon-unlimited-cr} in Theorem
\ref{theorem:reverse-Shannon}.  The construction uses AOS codes
described in Section~\ref{sec:AOS}.  Consider the AOS problem for the
reverse channel $\mathrm{P}_{X|Y}^n$. Using a random coding argument,
we construct $2^l$ AOS codes $\cC_u = \{\mathbf{y}_{u
  1},\ldots,\mathbf{y}_{u |\cC_u |}\}$ for each realization of shared
randomness $u \in \{1,\ldots,2^l\}$. Then, upon observing $X^n =
\mathbf{x}$ and $U=u$, $\cP_1$ generates the transcript $\Pi=\tau$ by
using the so-called likelihood encoder \cite{Cuff13}:
\begin{align*}
\bP{\Pi|X^n U}{\tau|\mathbf{x},u} \propto
\mathrm{P}_{X|Y}^n(\mathbf{x}| \mathbf{y}_{u \tau}),
\end{align*} 
where $\propto$ represents equality with the normalized right-side.
On the other hand, $\cP_2$ outputs $\mathbf{y}_{U \Pi}$. By using the
AOS results reviewed in Section~\ref{sec:AOS}, we can show that
$\bPP{X^n \mathbf{y}_{U\Pi}}$ is close to the target joint
distribution $\mathrm{P}_{XY}^n$ as long as the communication rate and
the shared randomness rate satisfy $\frac{|\pi|}{n} > I(X \wedge Y)$
and $\frac{|\pi|+l}{n}>H(Y)$.  The construction for the proof of
\eqref{eq:reverse-shannon-tradeoff} is slightly more involved, but is
based on a similar idea.

The study of the reverse Shannon theorem started in the quantum
information community to investigate the following
question~\cite{BenShoSmoTha02}: Can any two channels of equal capacity
simulate one another with unit asymptotic efficiency, \htc{namely with
  roughly one use of channel per simulated channel instance?} The
answer is in the affirmative if shared randomness is allowed as an
additional resource. The same question for quantum channels has also
been resolved in \cite{BenDevHarShoWin14} (see also \cite{BerChrRen11}
for a proof based on a single-shot approach); in the quantum setting,
an additional resource of entanglement is needed.

Several variants of the reverse Shannon problem have been
considered. In \cite{Win04}, the problem of simulating measurement
outcomes of quantum states was studied.  The cases with
side-information at $\cP_2$ have also been studied \cite{LuoDev09,
  WilHayBusHsi12}.  A similar problem has been studied in the computer
science community as well. Specifically, in \cite{HarJaiMcARad10}, the
average communication complexity of the reverse Shannon theorem with
exact simulation has been studied. The achievability scheme in
\cite{HarJaiMcARad10} uses {\em rejection sampling} which proceeds as follows.
\textchangeSW{The parties $\cP_1$ and $\cP_2$ share
an  infinite dictionary $\{y_1,y_2,\ldots\}$ comprising independent samples from $P_Y$. $\cP_1$
finds (based on a fixed rule) an index $i^*$ so that $\cP_1$'s observation
  $x$ and $y_{i^*}$ are distributed according to the target
  distribution $P_{XY}$. When an efficient encoding for natural
  numbers is used, the expected code length of $i^*$ is roughly
  $I(X\wedge Y)$.  Recently, an alternative proof of the exact
  simulation result was given in \cite{LiElG18} using a strengthened version
  of functional representation lemma \cite{ElGKim}, which is also
  applicable to infinite alphabet.  Furthermore, the trade-off between
  the communication rate and the shared randomness rate for exact
  simulation was studied in \cite{YuTan18b}.}

Another motivation to study the reverse Shannon problem arises in
proving coding theorems in information theory. Specifically, in
problems such as rate-distortion theory or multi-terminal source
coding, the encoder needs to simulate a test channel. The standard method for enabling 
this simulation uses a ``covering lemma''
(cf.~\cite{CsiKor11}).  In place of the covering lemma, we can also
use a reverse Shannon theorem to simulate a test channel, which
simplifies proofs and sometimes provide tighter bounds; see, for
instance,~\cite{Win02, LuoDev09, WilDatHsiWin13, WatKuzTan15,
  SonCufPoo16, HsiWat16}.

Another closely related problem is that of empirical coordination in
\cite{CufPerCov10}. Here, instead of requiring the approximation error
$d(\bPP{X^n\hat{Y}^n}, \mathrm{P}_{XY}^n)$ to be small, we require
that the joint empirical distribution (joint type) of
$(X^n,\hat{Y}^n)$ is close to the target joint distribution $\bPP{XY}$
with high probability. In fact, the latter requirement is known to be
weaker than the former, and the need for shared randomness can be
circumvented, $i.e.$, the communication rate of $I(X\wedge Y)$ is
attainable without using shared randomness.

In our treatment above, we reviewed a construction based on AOS
codes. An alternative approach using leftover hashing (random binning)
has been given in \cite{YasAreGoh14} (see, also, \cite{RenRen11,
  Mur14}).
%%%%%%%%%%%%%%%
\subsection{Interactive channel simulation}\label{s:interactive_channel_simulation}
We now present the interactive channel simulation problem. As
mentioned earlier, this general formulation includes many problems in
the literature as special cases.  For simplicity, we allow the parties
access to unbounded amount of shared randomness; for a more thorough
treatment, see \cite{YasAreGoh15}.

For a given channel $W: \cX_1 \times \cX_2 \to \cY_1 \times \cY_2$ and
an input joint distribution $\bPP{X_1 X_2}$, parties $\cP_1$ and
$\cP_2$ seek to simulate the joint distribution $\bP{X_1 X_2 Y_1
  Y_2}{x_1,x_2,y_1,y_2} = \bP{X_1X_2}{x_1,x_2}
W(y_1,y_2|x_1,x_2)$. \htc{The first party observes $X_1$ and the
  second $X_2$, and} they communicate with each other by using a
public coin protocol $\pi$ with output $(\hat{Y}_1, \hat{Y}_2)$. The
approximation error for the protocol is given by
\begin{align*}
\rho(\pi) := d(\bPP{X_1 X_2 \hat{Y}_1 \hat{Y}_2}, \bPP{X_1X_2
  Y_1Y_2}).
\end{align*}

\begin{definition}
For a given $\ep \in [0,1)$ and $r \ge 1$, the infimum over the length
  $|\pi|$ of $r$-rounds simulation protocols satisfying $\rho(\pi) \le
  \ep$ is denoted by $L_{r,\ep}(W|\bPP{X_1X_2})$.
\end{definition}
When the input distribution is iid $\mathrm{P}_{X_1 X_2}^n$ and the
channel is discrete memoryless channel $W^n$, we consider the
asymptotic limits defined by
\begin{align*}
C_{r,\ep}^{\mathtt{ICS}}(W|\bPP{X_1X_2}) := \limsup_{n\to\infty}
\frac{1}{n} L_{r,\ep}(W^n|\mathrm{P}_{X_1 X_2}^n)
\end{align*}
and
\begin{align*}
C_{r}^{\mathtt{ICS}}(W|\bPP{X_1X_2}) := \lim_{\ep \to 0}
C_{r,\ep}^{\mathtt{ICS}}(W|\bPP{X_1X_2}).
\end{align*}
The single-letter expression of this general problem is characterized
as follows.

%% \begin{theorem}[\cite{YasAreGoh15}]\label{t:interactive_channel_simulation}
%% Given a pmf $\bPP{X_1, X_2}$, a discrete memoryless channel $W:
%% \cX_1\times \cX_2\to \cY_1 \times \cY_2$, and $r\geq 1$, we have
%% \begin{align} \label{eq:interactive-channel-simulation}
%% C_{r}^{\mathtt{ICS}}(W|\bPP{X_1X_2}) = \min_{V^r}\bigg[ I(V^r \wedge
%%   X_1|X_2) + I(V^r \wedge X_2|X_1) \bigg],
%% \end{align}
%% where the minimization is taken over auxiliary random variables $V^r =
%% (V_1,\ldots,V_r)$ satisfying
%% \begin{align*}
%% &V_i \mc (X_1,V^{i-1}) \mc X_2,~~~\mbox{for odd } i, \\ & V_i \mc
%%   (X_2,V^{i-1}) \mc X_1,~~~\mbox{for even } i, \\ & Y_1 \mc (X_1,V^r)
%%   \mc (X_2,Y_2), \\ & Y_2 \mc (X_2,V^r) \mc (X_1,Y_1).
%% \end{align*}
%% Moreover, it may be assumed that $|\cV_1| \le
%% |\cX_1||\cX_2||\cY_1||\cY_2|+3$ and $|\cV_i| \le
%% |\cX_1||\cX_2||\cY_1||\cY_2| \prod_{j=1}^{i-1} |\cV_j| +2$ for $2 \le
%% i \le r$.
%% \end{theorem}
%% By regarding $\Pi = V^r$ as a transcript, we can also describe the
%% expression in \eqref{eq:interactive-channel-simulation} as the
%% minimization of the internal information cost $\ICi(\pi| X_1, X_2)$
%% over all protocols that simulate the output $(Y_1,Y_2)$ of the
%% channel.
\begin{theorem}\label{t:interactive_channel_simulation}
  Given a pmf $\bPP{X_1, X_2}$, a discrete memoryless channel
  $W:\cX_1\times \cX_2\to \cY_1 \times \cY_2$, and $r\geq 1$, we have
\begin{align} \label{eq:interactive-channel-simulation}
  C_{r}^{\mathtt{ICS}}(W|\bPP{X_1X_2}) =\min_{\pi} \ICi(\pi| X_1,
  X_2),
\end{align}
where \htc{$\ICi$ denoted the internal information complexity defined
  in Section~\ref{s:comm_prot} and} the minimum is taken over all
$r$-round private coin protocols $\pi$ with output $(Y_1, Y_2)$
\htc{(see Section~\ref{s:comm_prot} for the definiton of output of
  protocol.)}.
\end{theorem}
The result above is from \cite{YasAreGoh15}, but we have restated it
using the notion of internal information cost.

Specializing to $r=1$ and $X_2 = Y_1 = \emptyset$ leads to the reverse
Shannon theorem of the previous section.  Another important special
case of the interactive channel simulation problem is the function
computation problem obtained by setting $Y_1 = Y_2= g(X_1,X_2)$ for a
function $g$ of $(X_1, X_2)$.  The function computation problem has a
rich history, starting from the pioneering work of
Yao~\cite{Yao79}. For completeness, we present a brief introduction of
this rather broad area of communication complexity. An interested
reader can see~\cite{KushilevitzNisan97} for a comprehensive treatment
of the classical formulation. Over the last decade or so, starting
with~\cite{ChakrabartiSWY01, BarakBCR10}, an information theoretic
approach has been taken for communication complexity problems;
see~\cite{Braverman12ii} for a short review. We present a quick
overview of the area.  Denote by $L_{r,\ep}(g|\bPP{X_1X_2})$ the
quantity $L_{r,\ep}(W|\bPP{X_1X_2})$ for the channel $W(y_1, y_2|x_1,
x_2) = \indicator(y_1 = y_2 = g(x_1, x_2))$. Note that for this
special case the bound on approximation error reduces to the
requirement
\[
\bPr{\hat{Y}_1= \hat{Y}_2 =g(X_1, X_2)}\geq 1- \ep.
\]
The quantity $L_{r,\ep}(W|\bPP{X_1X_2})$ is referred to as the
$r$-round (distributional) communication complexity of $g$.  When the
observations are iid random variables $(X_1^n,X_2^n)$ and the function
to be computed is given by $g^n(x_1^n,x_2^n) =
(g(x_{1,1},x_{2,1}),\ldots,g(x_{1,n},x_{2,n}))$, the asymptotic
optimal rate is called the {\em amortized communication complexity} of
$g$:
\begin{align*}
C(g|X_1,X_2) := \inf_{r\ge1} \lim_{\varepsilon \to 0}
\limsup_{n\to\infty} \frac{1}{n} L_{r,\ep}(g^n|\mathrm{P}_{X_1X_2}^n).
\end{align*} 
As a corollary of Theorem \ref{t:interactive_channel_simulation}, we
can characterize the amortized communication complexity. To state the
result, we define the {\em information complexity} of a function $g$,
denoted $\mathtt{IC}(g|X_1, X_2)$ as the infimum of internal
information complexity (see Section~\ref{s:comm_prot} for definition)
$\ICi(\pi|X_1, X_2)$ over all private coin protocols $\pi$ that
compute $g$ exactly, namely protocols with output $(O_1, O_2)$ such
that $H(g(X_1, X_2)|O_1) = H(g(X_1, X_2)|O_2)=0$.
\begin{corollary}[\cite{MaIsh11, BraRao14}]
For a given function $g:\cX_1\times \cX_2\to \cY$, the amortized
communication complexity is given by
\begin{align*}
C(g|X_1,X_2) = \mathtt{IC}(g|X_1, X_2).
\end{align*}
\end{corollary}  
This result is a special case of
Theorem~\ref{t:interactive_channel_simulation}, but was obtained
earlier in \cite{MaIsh11, BraRao11, BraRao14} (see also
\cite{OrlRoc01})\footnote{The interactive function computation problem
  can be also regarded as a special case of the interactive
  rate-distortion problem \cite{Kas85}.}. In view of our foregoing
presentation, it is not surprising that amortized communication
complexity can be characterized in terms of an information theoretic
quantity $\mathtt{IC}(g|X_1, X_2)$. Interestingly, information
complexity $\mathtt{IC}(g|X_1, X_2)$ also gives a handle over the
worst-case (over all input distributions $\bPP{X_1 X_2}$)
communication complexity of computing a single instance of a function
$g$. In fact, using a subadditivity property of information
complexity, it was shown in~\cite{BarakBCR10, BarakBCR13} that the
worst-case communication complexity for computing $n$-instances of a
function $g$ grows at least as $\tilde{O}(\sqrt{n})$. A formal
description of this result or other similar results
(cf.~\cite{Braverman12,BraRaoWeiYeh13}) is beyond the scope of this
article. A key tool is the simulation of interactive protocols, which
we review next.

%%%%%%%%%%%%%%%
\subsection{Protocol simulation}
A special case of the interactive channel simulation problem is when
the channel to be simulated has the structure of an interactive
communication protocol. \htc{Note that if we consider the ``data
  exchange protocol,'' namely the protocol where each party simply
  communicates its input to the other party, a simulation of this
  protocol can be used to simulate any channel. But the general
  problem of simulating a given interactive communication protocol is
  far better understood than that of simulating a given arbitrary
  channel.} This problem, termed the {\it interactive protocol
  simulation}, plays a central role in the information theoretic
method for deriving lower bounds on communication complexity of
function computation.

Given a private coin protocol $\pi$ with input $(X_1, X_2)$, let
$W_\pi: \cX_1\times \cX_2 \to \{0,1\}^*$
% \textchange{[SW: do we allow infinite rounds protocol?]: discussed}
denote the channel $\bPP{Y_1Y_2| X_1 X_2}$ with $Y_1 = Y_2 = \Pi$. The
interactive protocol simulation problem entails the interactive
simulation of the channel $W_\pi$ using as few bits of interactive
communication as possible. We denote this minimum communication by
$L_\ep(\pi| \bPP{X_1X_2})$, defined as
\[
L_\ep(\pi| \bPP{X_1X_2}) = \lim_{r\rightarrow \infty}L_{r,\ep}(W_\pi|
\bPP{X_1X_2}).
\]
When the goal is to simulate several independent copies of the same
protocol, we are interested in the amortized communication complexity
given by
\[
C_\ep(\pi| X_1,X_2) = \lim_{n\rightarrow \infty} \frac 1n L_\ep(\pi^n|
\bPP{X_1X_2}^n),
\]
where $\pi^n$ denotes the same protocol applied to each coordinate
using independent private randomness. Denote the limit of
$C_\ep(\pi|\bPP{X_1X_2})$ as $\ep$ goes to $0$ by $C(\pi|
\bPP{X_1X_2})$.  Specializing
Theorem~\ref{t:interactive_channel_simulation} to the case of
protocols as outputs, we get the following result.
%% \textchange{[SW: Does Theorem~\ref{t:interactive_channel_simulation} apply for unbounded rounds 
%% or transcript of unbounded length?]:discussed}
\begin{corollary}\label{c:BR11}
For a private coin protocol $\pi$ with input $(X_1, X_2)$, we have
\[
C(\pi|X_1, X_2) = \ICi(\pi| X_1, X_2).
\]
\end{corollary}
This result can be obtained using the analysis in \cite{BraRao14}. In
fact, a more refined asymptotic behavior was obtained recently in
\cite{TyagiVVW17}, which we summarize below.  To describe this result,
we need the notion of {\it information complexity density} defined in
\cite{TyagiVVW17}.
\begin{definition}
The {information complexity density} of a private-coin protocol $\pi$
is given by the function
\begin{align*}
\lefteqn{ \ic(\tau; x_1, x_2) } \\
&= \log
\frac{\bP{\Pi|X_1X_2}{\tau|x_1,x_2}}{\bP{\Pi|X_1}{\tau|x_1}} + \log
\frac{\bP{\Pi|X_1X_2}{\tau|x_1,x_2}}{\bP{\Pi|X_2}{\tau|x_2}},
\end{align*}
\end{definition}
Denote by $\icp$ the random variable denoting $\ic(\tau; x_1, x_2)$
when $(\tau, x_1,x_2)$ are generated randomly from $\bPP{\Pi
  X_1X_2}$. Note that $\ICi(\pi|X_1, X_2) = \bEE{\icp}$, and denote by
$\varp$ the variance $\mathrm{Var}\left[ \icp \right]$. The following
result yields a more refined asymptotic behavior of $L_\ep(\pi^n|
\bPP{X_1X_2}^n)$.
\begin{theorem}\label{t:second_order}
For every $0 < \ep < 1$ and every protocol $\pi$ with $\mathtt{V}(\pi)
> 0$,
\begin{align*}
\lefteqn{ L_\ep(\pi^n|\bPP{X_1X_2}^n) } \\
&= n \ICi(\pi| X_1, X_2) + \sqrt{n \varp}
Q^{-1}(\ep) + \order(\sqrt{n}),
\end{align*}
where $Q(x)$ is equal to the probability that a standard normal random
variable exceeds $x$.
\end{theorem}
As a corollary, we obtain the strong converse, namely
$C_\ep(\pi|X_1,X_2) = C(\pi| X_1,X_2)$ for all $0< \ep <1$. Note that
such a strong converse is unavailable for the interactive channel
simulation problem described in the previous section; for the special
case of function computation, the strong converse was recently proved
in \cite{TyaWat18}.

Unlike the general problem of interactive channel simulation, where
only asymptotic results are available, several schemes for simulating
a single instance of a protocol are available. In fact, the asymptotic
results stated above are obtained using more general single-shot
schemes and converse bounds. In the remainder of this section, we
describe these single-shot schemes. For applications of these results
to the function computation problem, see review
articles~\cite{Braverman12ii, Weinstein15}. We fix a private coin
protocol $\pi$ with internal information cost $\ICi(\pi| X_1, X_2) =
\mathtt{I}$ and length $|\pi|=\mathtt{C}$.  We shall evaluate the
communication cost of our simulation protocols in terms of its
dependence on $\mathtt{I}$ and $\mathtt{C}$. \htc{Broadly speaking, the schemes
  we sketch below rely on two ideas: Correlated sampling seen in
  Section~\ref{s:correlated_sampling} and a guess-and-check strategy.
  In particular, upon generating transcripts $\tau$ till round
  $r$, parties use the conditional probabilities for the communication
  of round $r+1$ (given $\Pi^r=\tau$) with correlated sampling to get
  the next round of communication. The second idea is used to guess
  the transcripts ahead where the parties form a guess-list of most likely
  communication in the next few rounds and verify their guess by
  exchanging random hashes. All the schemes below apply extensions
  of these two basic ideas in different ways.} 

\paragraph{Round-by-round simulation} We begin with schemes that
follow the protocol tree closely and simulate the interactive protocol
in a round-by-round fashion. For such protocols, it suffices to
describe the simulation of a protocol with $1$-round of communication;
multiple rounds are simulated by applying this simulation protocol
separately to each round.  

The first such simulation scheme is from \cite{BraRao14} and achieves
asymptotically the optimal rate of Corollary~\ref{c:BR11}. Note that
the transcript $\Pi$ of a $1$-round protocol $\pi$ satisfies the
Markov relation $\Pi \mc X_1 \mc X_2$. To simulate such a protocol, it
suffices to output estimates $(\Pi_1, \Pi_2)$ such that $\Pi_i$ has
distribution close to $\bPP{\Pi| X_i}$, $i=1,2$ and the probability
$\bPr{\Pi_1 = \Pi_2}$ is close to $1$. Assuming that $\cP_1$ initiates
the communication protocol $\pi$, $\cP_1$ knows the actual
distribution of the transcript $\bPP{\Pi|X_1X_2}$ since 
$\bPP{\Pi|X_1} = \bPP{\Pi|X_1X_2}$. On the other hand, $\cP_2$ only
has an estimate of this distribution given by
$\bPP{\Pi|X_2}$. Therefore, the goal of simulating a $1$-round
protocol can be described in an abstract fashion as follows: $\cP_1$
and $\cP_2$ know distributions $\dP$ and $\dQ$, respectively, and seek
to produce samples $Y_1\sim \dP$ and $Y_2\sim \dQ$ such that $\bPr{Y_1
  \neq Y_2}$ is small. This is very similar to the goal in the
correlated sampling problem described in
Section~\ref{s:correlated_sampling}, except that we are allowed to use
interactive communication to reduce the probability of disagreement to
an arbitrarily small quantity. Accordingly, the scheme proposed in
\cite{BraRao14} builds on correlated sampling of
\cite{Holenstein09} (reviewed in the proof of
Theorem~\ref{t:correlated_sampling}) and uses interactive
communication to ensure that the parties agree on the same index
$i$. Specifically, using the shared randomness to generate the iid
sequence $\{(A_i,
B_i)\}_{i=1}^\infty$ with $A_i$ uniform on $\cX$ and $B_i$ uniform on
$[0,1]$, $\cP_1$ finds the first 
index $i$ such that $\dP(A_i)>B_i$ and sends a random hash of this
index. Then, $\cP_2$ finds the first index $j$ such that $\dQ(A_j)>
B_j$ and checks if its hash matches the hash sent by $\cP_1$. If it
matches, it sends back an ACK signal; else, it sends back a NACK
signal and increments each $\dQ(x)$ by a factor of $2$. In the next
round of communication, $\cP_2$ searches for the least index $j$ using
this updated $\dQ$. Since we have relaxed the criterion for an
acceptable $j$, more such indices are now feasible. To compensate for
that, $\cP_1$ sends some more bits of random hash for $i$, and $\cP_2$
seeks the least $j$ satisfying $\dQ(A_j)> B_j$ and checks if all its
random hashes match those received from $\cP_1$ till this point. The
parties proceed interactively till a match is found.  

The analysis in \cite{BraRao14} shows that this scheme uses roughly
$D(\dP\|\dQ) + \cO(\sqrt{D(\dP\| \dQ)})$ bits of
communication. Substituting $\dP = \bPP{\Pi|X_1X_2}$  and $\dQ =
\bPP{\Pi|X_2}$ and taking expectation with respect to $\bPP{X_1X_2}$,
the overall communication is roughly  $D(\bPP{\Pi|X_1X_2}\|
\bPP{\Pi|X_2}| \bPP{X_1X_2}) = I(\Pi\wedge X_1|X_2) \text{ bits}$. 
Using the same scheme for each round, the leading term in
communication cost equals $I$, although the number of rounds of
interaction is much larger than the number of rounds of interaction in
the original protocol.  
For the amortized case, while the internal information cost of each
round grows linearly in $n$, the number of rounds remains
constant. Thus, the asymptotic rate equals the internal information
cost of $\pi$. 

Next, we describe the round-by-round simulation scheme of
\cite{TyagiVVW17} which is asymptotically optimal even up to the second
order term and attains the rate claimed in
Theorem~\ref{t:second_order}. As before, it suffices to describe
simulation of a single round. The simulation protocol builds upon the
information reconciliation step described in the context of SK
agreement in Section~\ref{subsection:SK-no-constraint}. Specifically,  
$\cP_1$ generates a transcript $\Pi$ using $\bPP{\Pi|X_1}
=\bPP{\Pi|X_1X_2}$ and sends a random hash to $\cP_2$ which uses it to
find a matching entry in a ``typical'' guess-list it forms using
$X_2$. However, this simple protocol is modified in two ways. First,
$\cP_1$ simulates $\Pi$ using shared randomness in such a manner that
a part of the random hash that needs to be sent is realized from the
shared randomness itself and need not be sent. Second, instead of
working with the original distributions $\bPP{\Pi|X_1}$ and
$\bPP{\Pi|X_2}$ to form the guess-lists, the parties use
``spectrum-slicing'' techniques introduced in \cite{Han03} (see \cite{TyagiVVW17} for details) to
search in a more greedy fashion by giving priority to more likely
transcripts. 
As in the scheme of \cite{BraRao14}, the protocol entails
several rounds of interaction for simulating each round of the $\pi$;
in the amortized setting, $\cO(n^{1/4})$ rounds of interaction are used
for simulating each round of $\pi^n$, \textchangeSW{which enables us to derive the
optimal second order term. Note that the scheme of \cite{BraRao14}
uses $\cO(\sqrt{n})$ rounds of interaction in the amortized setting.} 

\paragraph{Simulation using $\sqrt{\mathtt{I} \mathtt{C}}$ bits} 
Chronologically the first protocol simulation scheme, given in the seminal work
\cite{BarakBCR13}, requires $\cO(\sqrt{\mathtt{I} \mathtt{C}}\log \mathtt{C})$ bits
of communication. This scheme, too, builds upon the correlated
sampling of \cite{Holenstein09}. However, the usage of correlated
sampling is different from that in \cite{BraRao14}. It is now used for
simulating, without communication, a ``guess'' for the overall
transcript of the protocol at each party. Denote by $p_v(x_1)$ and  
$q_v(x_2)$, respectively, the probabilities $\bP{\Pi_v|X_1}{1|x_1}$
and $\bP{\Pi_v|X_2}{1|x_2}$ where $\Pi_v$ denotes the random output of
the protocol once it reaches the node $v$ in the protocol tree. For
input $(x_1, x_2)$, the parties begin by using correlated sampling to
generate bits $(B_{1}(v), B_2(v))$ using shared randomness with
$\bPr{B_1(v) = p_v(x_1)}$, $\bPr{B_2(v)=q_v(x_2)}$, and $\bPr{B_1(v)
  \neq B_2(v)}= |p_v(x_1)-q_v(x_2)|$. Then, starting from the root,
the parties follow their generated bits $B_1(v)$ and $B_2(v)$, with
$1$ denoting the right-child and $0$ the left-child, to identify paths
from the root to a leaf. This is, in essence, tantamount to both
parties guessing the transcript $\Pi$ but using correlated sampling to
ensure that the marginals of the bits are as prescribed by the
protocol. Next, the parties use a randomized algorithm suggested in
\cite{FeigeRPU94} to identify the highest node $v$ where the guessed
paths diverge. The entire process is then repeated by both parties
using the guess of the party controlling $v$ for that node and
repeating the process above with $v$ in place of the root. The
randomized algorithm for finding the first node of divergence takes no
more than $\log C$ bits of communication. The communication cost for
the protocol is dominated by the number of times we need to apply this
protocol, namely the number of places along the correct path where the
guesses diverge. The expected number of this guesses is shown in
\cite{BarakBCR13} to be bounded above by roughly $\sqrt{\mathtt{I} \mathtt{C}}$.

\paragraph{Simulation using $2^{\cO(\mathtt{I})}$ bits} The final simulation scheme we
describe is from \cite{Braverman12}, though a similar
scheme appears in a slightly restricted context in
\cite{OrlEl84ii}. Unlike the previous scheme, the scheme of
\cite{Braverman12} does not invest communication to sync midway the guesses of the transcript formed by the two parties. Instead, the parties simply form guess-lists of likely 
transcripts $\tau$ that have a significant probability of appearing
given their respective inputs and use random hash to find the
intersection of their guess-lists.  The proposed scheme is a variant
of that given in \cite{BraRao14} and uses a modified version of
correlated sampling. Specifically, the shared randomness is used to
generate the iid sequence $\{(A_i, B_i, C_i)\}_{i=1}^\infty$ where
$A_i$ is uniform over the leaves of the  protocol tree and $B_i$ and
$C_i$ are independent and uniformly distributed over $[0,1]$. 
Note that by the {rectangle property} of interactive communication 
(see Section~\ref{s:comm_prot}), 
the probability of a transcript $p(\tau|x_1, x_2)$
equals $f_\tau(x_1)g_\tau(x_2)$ where the first factor is known to
$\cP_1$ and the second to $\cP_2$. Similarly,
 $p(\tau|x_1)=f_\tau(x_1)\hat{g}_\tau(x_1)$ 
and $p(\tau|x_2)=\hat{f}_\tau)$. 
Furthermore, the summation of $\bEE{f_\Pi(X_1)/\hat{f}_\Pi(X_2)}$ and $\bEE{g_\Pi(X_2)/\hat{g}_\Pi(X_1)}$ equals $I$. Thus, if $\cP_1$ and $\cP_2$, respectively, form guess-lists 
$\cA=\{\tau: f_\tau(X_1)> B_i, \hat{g}_\tau(X_1)\gtrsim 2^I C_i\}$ and $\cB=\{\tau: \hat{f}_\tau(X_1)\gtrsim 2^I B_i, g_\tau(X_2)> C_i\}$. 
It can be seen that the intersection of two guess-lists, with large probability, contains a unique element which has the distribution $\bPP{\Pi|X_1X_2}$, and it can be found by communicating 
roughly $2^{\cO(\mathtt{I}/\varepsilon)}$ bits. Details can be found in \cite[Lemma 5.2]{Braverman12}. Note that this protocol is simple, namely the communication from both parties is simultaneous.

\begin{remark} A general scheme that includes all the schemes above as special cases and their 
unified analysis is unavailable. It is rather intriguing that the only feature of the structure of the protocol tree that enters the communication cost is its depth $\mathtt{C}$. 
Furthermore, the more closely our simulation protocol follows the protocol tree, the higher the number of rounds of interaction it requires and the more the communication cost depends on 
$\mathtt{C}$. In particular, the simple communication protocol of \cite{Braverman12} has communication cost that does not depend on $\mathtt{C}$ at all, 
but depends exponentially on $\mathtt{I}$. In fact, a recent result \cite{GanorKR14ii} exhibits a protocol for which this dependence is optimal. 
On the other hand, it remains open if the simulation scheme of \cite{BarakBCR13} is optimal for any specific example. 
\end{remark}

%%%%%%%%%%%%%%%%%%%%%%%%%%%%%%%%%%%%%%%%%%%%%%%
\section{Applications and extensions}\label{s:app} 
The problems we have reviewed have direct applications in areas ranging from information theory, cryptography, distributed control and coordination, communication, and theoretical computer science. For instance, generating common randomness and secret keys from correlated observations is a standard primitive in cryptography. Similarly, problems requiring distributed simulation of random variables appear in quantum computing as well as other realms in theoretical computer science. In this concluding section, rather than discussing these direct applications, we point the reader to two perhaps not-so-straightforward applications of correlated sampling discussed in Section~\ref{s:correlated_sampling}. We begin with {\em locality sensitive hashing}, a basic building block for modern data mining techniques. Next, we discuss the {\em parallel repetition theorem}, which is a standard tool for establishing hardness of approximation results in computational complexity theory. Finally, we close with a brief discussion on extensions of the models covered in this article.
%%%%%%%%%%%%%%%
\subsection{Locality sensitive hashing}
In the mid-nineties, research on computer systems and web
search~\cite{Manber94,Broder97} led to a new challenge: that of
designing ``hash'' functions that were actually sensitive to the
topology on the input domain, and preserved distances approximately during hashing. (Roughly, for a hash function $h$, the distance between $h(x)$ and $h(y)$ should depend on the distance between $x$ and $y$.) Constructions of such
 hash functions led to efficient methods to
detect similarity of files in distributed file systems and proximity
of documents on the web. Remarkably these methods closely resemble the
process of correlated sampling (and predated the first protocols for
correlated sampling). We describe the problem and results below.

Recall that a metric space $\mathcal{M} = (\cX,d(\cdot,\cdot))$ is given
by a set $\cX$ and a distance measure $d:\cX \times \cX \to \mR^{\geq 0}$
which satisfies the axioms of being a {\em metric}, i.e., (1) $d(x,y)
= 0$ if and only if $x = y$, (2) $d(x,y) = d(y,x)$ and (3) $d(x,y) +
d(y,z) \geq d(x,z)$. 

\begin{definition}[Basic Locality Sensitive Hashing]
Given a metric space $\mathcal{M} = (\cX,d)$ and set $\cS$, a family of
function $\mathcal{H} \subseteq \{h:\cX \to \cS\}$ is said to be a 
  basic {\em locality sensitive hash} (LSH) family if there exists an
increasing invertible function $\alpha:\mR^{\geq 0} \to [0,1]$ such that
for all $x,y \in \cX$, we have $\Pr_{h\sim \cH}[h(x)\neq h(y)]:=
  \frac 1 {|\cH|}\sum_{h \in \cH}\indicator(h(x) \neq h(y))\le \alpha(d(x,y))$.
\end{definition}
 
To contrast this with usual hash function families (for instance a
$2$-UHF defined in Definition~\ref{d:2UHF}) note that in the latter
the goal is to map a (large) domain $\cD$ to a (small) range 
$\cS$ such that the probability of a collision among any pair of
elements $x \ne y \in \cD$ is small. In contrast, with LSH families, we
wish for the probability of a collision to be small only when $d(x,y)$
is large, and we do want a high probability of collision when $d(x,y)$
is small. This requirement makes constructions non-trivial, but also lends itself
to a new family of applications. Of course, to get good applications,
we still want $S$ and $\mathcal{H}$ to be small, and in addition, we
want $\alpha^{-1}(\cdot)$ to be as numerically stable as possible.  

Given such a family, obviously we can estimate the probability of a
hash collision easily by sampling hash functions from $\mathcal{H}$
independently and uniformly. Then by inverting $\alpha$ we can also
get a good estimate of $d(x,y)$. The gain in this process is the
communication: If $x$ and $y$ are large ``files'' sitting on distinct
servers, the time it takes to estimate the distance between them no
longer scales with $\cX$, the size of the domain; but rather with $\log
|\cS|$ the size of the range of the hash families. A second advantage,
leading to many of the applications in modern web search, is that
LSHs reduce the task of ``nearest neighbor search'' (classically
considered complex) to the task of ``exact membership search'' (a very
well-studied and well-solved problem in  the design of data
structures).  

Returning to our setting, it turns out that ``correlated
sampling'' can  be interpreted as giving an LSH family for a
particular metric space. Let $\Omega$ be a finite set and let $\cX$ be
the space of all probability distributions over $\Omega$. Let $d(\dP,\dQ)$
denote the total variation distance between the distributions $\dP$ and
$\dQ$. Then the correlated sampling protocol from
Theorem~\ref{t:correlated_sampling} can be interpreted as providing a
family of hash function $\mathcal{H}$ mapping $\cX$ to $\Omega$ as
captured by the following theorem. 

\begin{theorem}[\cite{Broder97,Holenstein09}]
Let $\mathcal{M} = (\cX,d)$ be the metric space of probability
distributions over $\Omega$ under  total variation distance. Then
there exists a basic LSH $\mathcal{H} \subseteq \{h:\cX\to\Omega\}$  
such that $\Pr_{h \sim \mathcal{H}} [ h(x) \neq h(y) ] \le \alpha(d(x,y))$, for the function $\alpha(\theta) = 2\theta/(1+\theta)$.
\end{theorem} 
Note that the function $\alpha(\cdot)$ has inverse
$\alpha^{-1}(\tau) = \tau/(2-\tau)$ which is numerically stable.
We remark that Broder~\cite{Broder97} gives a slightly different
solution for the setting when $\dP$ and $\dQ$ 
are flat distributions, i.e., uniform distributions over subsets of 
$\Omega$. 

While our solution above does not attempt to make $\mathcal{H}$ small,
this has been the subject of a large body of work and has led to major
progress on ``nearest neighbor search''. We point the reader to
~\cite{AndoniIndyk16} for a survey of this area.  

%%%%%%%%%%%%%%%
\subsection{The parallel repetition theorem}
We now turn to a more sophisticated application of the technique of
correlated sampling, to a notion of profound importance in
computational complexity and to the study of {\em probabilistically
checkable proofs}, and to the related study of {\em complexity of
approximating optimization problems}.

The parallel repetition problem considers the amortized value of a
$2$-player game; we start by defining the latter. A $2$-player
game $\cG$ is specified by (1) four finite sets $\cX,\cY, \cA$ and
$\cB$, (2) a distribution $\bPP{XY}$ on $\cX\times \cY$ and (3) a value function  
$V:\cX \times \cY \times \cA \times \cB \to \{0,1\}$. The value of the
game $\cG$ denoted $\omega(\cG)$, is the maximum over all functions
$f:\cX\to\cA$ and $g:\cY\to\cB$ of $\bEE{V(X,Y,f(X),g(Y))}$.

The game $\cG$ captures the interaction between two
cooperating, noninteracting provers (players) $\cP_1$ and $\cP_2$ and
a verifier $\mathcal{V}$. The players are supposed to help $\cV$ verify computationally complex 
statements with ``easy to verify'' proofs. For instance to verify that
a graph $H = (S,E \subseteq S\times S)$ is three colorable, the
verifier might ask the two provers to provide consistent coloring of
the vertices of the graph, one that color endpoints differently. This
problem can be realized as the game $\cG^{\rm color}_H$ with $\cX = \cY =
S$, $\cA = \cB = \{R,B,G\}$, and value function
\[
V(X,Y,a,b) =
\begin{cases}
1, \quad \{X=Y \Leftrightarrow a=b\}
\\
0, \quad \text{otherwise}.
\end{cases}
\]
The distribution of the inputs $(X,Y)$ is given by
$\dP = \frac12 (\dP_1 + \dP_2)$ \textchangeSW{where $\dP_1$ samples $(X,X)$
  for $X$ distributed uniformly on $S$  and $\dP_2$ samples pair $(X,Y)$
  distributed uniformly on the edge set $E$.  In order to attain
  $\omega(\cG^{\rm color}_H) = 1$, the verifiers must answer the same
  color  when the prover asks the same vertices $(X,X) \sim \dP_1$ and
  the verifiers must answer different colors  when the prover asks
  adjacent vertices $(X,Y) \sim \dP_2$, which is possible if and only if
  $H$ is $3$-colorable, though non $3$-colorable graphs may 
have value tending to $1$ as $|H| \to \infty$. }

A fundamental question in computational complexity in the nineties was:
Does the value of a $2$-prover game tend to zero when the game is
repeated? To elaborate on this question, let us first define the $n$-fold
product $\cG^{\otimes n}$ of a game $\cG$. The $n$-fold product is
another two player game with (1) the four finite sets being $\cX^n$,
$\cY^n$, $\cA^n$ and $\cB^n$, (2) The distribution $\dP^n$ being the
$n$-fold product of $\dP$ and (3) the function $V^n : \cX^n \times cY^n
\times \cA^n \times \cB^n \to [0,1]$ being given by  
\begin{align*}
& V((X_1,\ldots,X_n,Y_1,\ldots,Y_n,a_1,\ldots,a_n,b_1,\ldots,b_n)  \\
& =\prod_{i=1}^n V(X_i,Y_i,a_i,b_i).
\end{align*}

If the value of the underlying game $\cG$ is $\alpha$ obtained by
functions $(f,g)$, then using the functions $f^n(X_1,\ldots,X_n)
=(f(X_1),\ldots,f(X_n))$ and $g^n(X_1,\ldots,X_n) = 
(g(X_1),\ldots,g(X_n))$ yields functions attaining a value of
$\alpha^n$, and thus, $\omega(\cG^{\otimes n}) \geq
\omega(\cG)^n$. However, this inequality is not tight, and indeed, there
exist games $\cG$ where $\omega(\cG^{\otimes 2}) = \omega(\cG) < 1$ --
so the value of the twice-repeated game does not change at
all. (The reader should verify that 
$\omega(\cG^{\otimes n}) \leq \omega(\cG)$ for all games $\cG$ and all $n \geq 1$.)
In view of this counterexample it becomes clear that even the
question ``does the value of the game $\cG^{\otimes n}$ tend to zero as
$n\to \infty$?'' does not  have an obvious answer. This question was
settled affirmatively by Verbitsky~\cite{Verbitsky96} though with a
very non-explicit bound on the rate at which $\omega(\cG^{\otimes n})$
goes to zero.  

Later in a remarkable result Raz~\cite{Raz98} showed that for every game $\cG$ with value less than $1$ there is a quantity $\tilde{\omega} = \tilde{\omega}(\omega(\cG), \cA, \cB) < 1$ such that 
$\omega(\cG^{\otimes n}) \leq \tilde{\omega}^n$. While even the fact
that exponential shrinkage in $n$ was new, the applications in
computational complexity needed the fact that the growth depended only
on $\cA$, $\cB$ and $\omega(\cG)$ and not on $\cX$ or $\cY$, which
Raz's parallel repetition theorem stated below establishes. 

\begin{theorem}[\cite{Raz98}]
\label{t:parallel}
For every $a,b \in \Z^+$ and $\omega \in (0,1)$, there exists an
$\tilde{\omega} \in (0,1)$ such that for every game $\cG =(\cX,\cY,\cA,\cB,\dP_{XY},V)$ with $\omega(\cG) \leq \omega$ and 
$|\cA| \leq a$ and $|\cB| \leq b$ it is the case that for every $n$,
$\omega(\cG^{\otimes n}) \leq \tilde{\omega}^n$.  
\end{theorem}
While Raz's proof is itself information-theoretic, the connection to
information-theoretic tools, and in particular to correlated sampling,
became more explicit in a later elegant work of
Holenstein~\cite{Holenstein09}.  
We point out some highlights from this work below. Our writeup
being based on the notes of Barak~\cite{Barak07};  
we point the reader to the original writeups~\cite{Raz98,Holenstein09}
and the lecture notes~\cite{Barak07} for further details.

A theorem such as Theorem~\ref{t:parallel} is proved by a reduction
argument where we roughly use a strategy for the $n$-fold game to
obtain a strategy for a single instance of the game \htc{-- the embedding
technique that appears several times in this article.} Specifically, we
assume $\omega(\cG^{\otimes n}) > \tilde{\omega}^n$ and let this value
be attained by functions $(F, G)$. Some technical
manipulations using a subadditivity property of the Kullback-Leibler
divergence allows us to obtain a coordinate $i$ where the players
with probability significantly greater than $\omega$, when conditioned on
some event $E_i$. (Roughly, $E_i$ is the event that the functions $(F,G)$ lead to a
win on all coordinates except $i$ in
$\cG^{\otimes n}$.) The key idea is to embed a single instance of the
game in this coordinate, thereby getting a value more than $\omega$
for it which is a contradiction. We could hope that such an embedding
could be implemented by generating the other inputs for the $n$-fold
game using the shared randomness. However, a technical difficulty
emerges since the conditioning on $E_i$ may render the inputs across
different coordinates dependent. A variant of the correlated sampling
result \eqref{e:corr_samp} comes in handy here and allows us to show
that the hypothetical distribution for which we have our bound on the
value can be simulated using a single instance of inputs and shared
randomness. While a full description of the proof is beyond the scope
of this article, the summary above brings out out the connection to
correlated sampling here.

%% For $\vec{X} = (X_1,\ldots,X_n)$ and $\vec{Y} = (Y_1,\ldots,Y_n)$, we say that $(F,G)$ win the game if $V^n(\vec{X},\vec{Y},F(\vec{X}),G(\vec{Y}) = 1$, and we say that they win the $i$th coordinate if $V(X_i,Y_i,F(\vec{X})_i,G(\vec{Y})_i)=1$. Let $W_j$ be the event that $F,G$ win in coordinates $\{1,\ldots,j-1\}$. $f$ and $g$ will be probabilistic strategies (constructed with shared randomness) that will aim, given $(x,y) \sim D$ to create inputs $\vec{X}$ and $\vec{Y}$ such that (1) $X_j = x$ and $Y_j = y$ and (2) $(\vec{X},\vec{Y})$ are sampled by a distribution close to $D^n|W_j$. Note that the conditioning on $W_j$ makes these distributions non-product form, but an intricate idea employing correlated sampling allows $\cP_1$ and $\cP_2$ to sample from these distributions without interacting (though with some error). (For technical reasons this sampling works only when the probability of $W_j$ is not too small, but this is ok since the ``too small'' threshold is $2^{-\Omega(n)}$.) But now one can appeal to the single-shot upper bound to note that the winning probability of $f,g$ may be at most
%% $\omega$ plus the error in the sampling procedure. This allows one to apply a chain rule to conclude that the probability of winning in the $j$th coordinate is at most some $\omega' < 1$ and thus that probability of winning in $j$ coordinates is at most $(\omega')^j$, which leads to the exponential decay in the value of the game.

%%%%%%%%%%%%%%%%%%%
\subsection{Extensions}\label{s:m_party}
In this article, we have restricted ourselves to two-party formulations with
classical correlation.  Many of the problems presented have natural
extensions to the multiparty case.  A multiparty version of the
problem of CR generation via channel was studied in
\cite{VenAna98,VenAna00}. The problem of SK agreement for multiple
parties was initiated in \cite{CsiNar04} and further studied in
\cite{CsiNar08, CsiNar13, Cha08, TyaWat17}.  Multiparty CR generation
and SK agreement with communication constraints are not as well
understood, but initial results are available. An extension of the
result in \cite{Tya13} was studied in \cite{MukKasSan16}; however, a
single-letter characterization of the communication rate required to
attain the secret key capacity is not available.  It seems difficult
to derive multiparty counterparts of the results in Theorem
\ref{t:CR_cap} and Theorem \ref{theorem:SK-Rate-Tradeoff}. In a
similar vein, there is no consensus on a useful definition of
multiparty information complexity that bears an asymptotic operational
significance and facilitates single-shot bounds. The definitions vary
depending on communication models and tasks; for instance, see
\cite{KerRosUrr16} and references therein.

As an extension in another direction, it is of interest to consider
correlation generation problems when either a quantum resource is
available or when the target correlation itself is quantum. A
systematic study of entanglement generation was initiated in
\cite{BenDiVSmoWot96}; see \cite{HorHorHorHor09} for a comprehensive
review.  Quantum entanglement as a resource has several applications;
for instance, see \cite{BuhCleMasWol10} for an application to
communication complexity.
\textchangeSW{In the context of physics, there has been a long-standing debate on
what kind of correlations are physically allowed. Such questions are related closely to the ones considered in this article, and 
in the past few decades, information theoretic approach
has contributed richly to this research; see \cite{BruCavPirScaWeh14}
and references therein.}

\textchangeSW{We close by observing that we have reviewed utility of
  common randomness only in the context of
information theory and computer science. However, common randomness is also useful
for problems in other fields such as distributed control, distributed optimization, distributed consensus, and distributed game theory.
For instance, in {distributed zero-sum games} where two players separately choose their strategies, the Nash equilibrium may not exist in general unless the parties share sufficient amount of common randomness to coordinate their strategies \cite{AnaBor07}.}

%%%%%%%%%%%%%%%
\section*{Acknowledgment}
The authors would like to thank the former Editor-in-Chief Prakash Narayan and anonymous reviewers
for careful reading of the manuscript and providing many valuable comments, which substantially
improved the presentation of the paper. \textchangeHT{We especially thank a reviewer who identified a gap in one of our proof-sketches  and provided Example~\ref{example:RS}.}
Thanks to Noah Golowich for detecting an error in a previous version of our characterization in Theorem~\ref{t:CR_cap} and suggesting a fix.

M.S. was supported in part by a Simons Investigator Award and NSF Award CCF 1715187.
H.T. was supported in part by the Department of Science and Technology, India under the grant EMR/2016/002569. 
S.W. was supported in part by the Japan Society for the Promotion of Science KAKENHI under Grant 16H06091.

%%%%%%%%%%%%%%%%%%%%%%%%%%%%%%%%%%%%%%%%%%%%%%%
\bibliography{IEEEabrv,references}
\bibliographystyle{IEEEtranS}
%%%%%%%%%%%%%%%%%%%%%%%%%%%%%%%%%%%%%%%%%%%%%%%%%%%%%%%

 %%%%%%%%%%%%%%%%%%%%%%%%%%%%%%%%%%%%%%%%%
\end{document}